\documentclass[superscriptaddress,aps,preprintnumbers,pacs, nofootinbib, twocolumn]{revtex4}
\usepackage{eurosym}
\usepackage{graphicx}
\usepackage{amsmath}
\usepackage{epstopdf}

\def\be{\begin{equation}}
\def\ee{\end{equation}}
\def\bea{\begin{eqnarray}}
\def\eea{\end{eqnarray}}
\def\beaa{\begin{eqnarray*}}
\def\eeaa{\end{eqnarray*}}

\begin{document}

\title{Jacobi stability analysis of scalar field models with minimal coupling to gravity in a cosmological background}
\author{Bogdan D\u{a}nil\u{a}}
\email{bogdan.danila22@gmail.com}
\affiliation{Astronomical Observatory, 19 Ciresilor Street, Cluj-Napoca, Romania,}
\author{Tiberiu Harko}
\email{t.harko@ucl.ac.uk}
\affiliation{Department of Physics, Babes-Bolyai University, Kog\u{a}lniceanu Street 400084,
Cluj-Napoca, Romania,}
\affiliation{Department of Mathematics, University College London, Gower Street, London
WC1E 6BT, United Kingdom,}
\author{Man Kwong Mak}
\email{mankwongmak@gmail.com}
\affiliation{The Open University of Hong Kong, School of Science and Technology, Homantin, Kowloon, Hong Kong, Hong Kong SAR, P. R. China}
\affiliation{Departamento de F\'{\i}sica, Facultad de Ciencias Naturales, Universidad de
Atacama, Copayapu 485, Copiap\'o, Chile,}
\author{Praiboon Pantaragphong}
\email{kppraibo@kmitl.ac.th}
\affiliation{Mathematics Department, King Mongkut's Institute of Technology, Ladkrabang
BKK 10520, Thailand,}
\author{Sorin V. Sabau}
\email{sorin@tokai.ac.jp}
\affiliation{School of Science, Department of Mathematics, Tokai University, Sapporo
005-8600, Japan}

\begin{abstract}
We perform the study of the stability of the cosmological scalar field
models, by using the Jacobi stability analysis, or the Kosambi-Cartan-Chern
(KCC) theory. In the KCC approach we describe the time evolution of the
scalar field cosmologies in geometric terms, by performing a "second
geometrization", by considering them as paths of a semispray. By
introducing a non-linear connection and a Berwald type connection associated
to the Friedmann and Klein-Gordon equations, five geometrical invariants can
be constructed, with the second invariant giving the Jacobi stability of the
cosmological model. We obtain all the relevant geometric quantities, and we
formulate the condition of the Jacobi stability for scalar field cosmologies
in the second order formalism. As an application of the developed methods we
consider the Jacobi stability properties of the scalar fields with
exponential and Higgs type potential. We find that the Universe dominated by
a scalar field exponential potential is in Jacobi unstable state, while the
cosmological evolution in the presence of Higgs fields has alternating
stable and unstable phases. By using the standard first order formulation of
the cosmological models as dynamical systems we have investigated the
stability of the phantom quintessence and tachyonic scalar fields, by
lifting the first order system to the tangent bundle. It turns out that in
the presence of a power law potential both these models are Jacobi unstable
during the entire cosmological evolution.
\end{abstract}

\pacs{}
\maketitle

\section{Introduction}

A large number of cosmological observations, obtained initially from distant
Type Ia Supernovae, have convincingly proven that the Universe has undergone
a late time accelerated expansion \cite{1n,2n,3n,4n}. In order to explain
these observations a deep change in our paradigmatic understanding of the
cosmological dynamics is necessary, and many ideas have been put forward to
address them. The "standard" explanation of the late time acceleration is
based on the assumption of the existence of a mysterious component, called
dark energy, which is responsible for the observed characteristics of the
late time evolution of the Universe.

On the other hand, the combination of the results of the observations of
high redshift supernovae, and of the WMAP and of the recently released
Planck data, indicate that the location of the first acoustic peak in the
power spectrum of the Cosmic Microwave Background Radiation is consistent
with the prediction of the inflationary model for the density parameter $%
\Omega $, according to which $\Omega =1 $. The cosmological observations
also provide strong evidence for the behavior of the parameter $w=p/\rho $
of the equation of state of the cosmological fluid, where $p$ is the
pressure and $\rho $ is the density, as lying in the range $-1\leq w=p/\rho
<-1/3$ \cite{acc}.

In order to explain the observed cosmological dynamics, it is assumed
usually that the Universe is dominated by two main components: cold
(pressureless) dark matter (CDM), and dark energy (DE) with negative
pressure, respectively. CDM contributes $\Omega _{m}\sim 0.3$ \cite{P2}, and
its introduction is mainly motivated by the necessity of theoretically
explaining the galactic rotation curves and the large scale structure
formation. On the other hand, DE represents the major component of the
Universe, providing $\Omega _{DE}\sim 0.7$. Dark energy is the major factor
determining the recent acceleration of the Universe, as observed from the
study of the distant type Ia supernovae \cite{acc}. Explaining the nature
and properties of dark energy has become one of the most active fields of
research in cosmology and theoretical physics, with a huge number of
proposed DE models( for reviews see, for instance, \cite{PeRa03,
Pa03,Od,LiM, Mort}).

One interesting possibility for explaining DE are cosmological models
containing a mixture of cold dark matter and quintessence, representing a
slowly-varying, spatially inhomogeneous component \cite{8n}. From a
theoretical as well as a particle physics point of view the idea of
quintessence can be implemented by assuming that it is the energy associated
with a scalar field $Q$, having a self-interaction potential $V(Q)$. When
the potential energy density $V(Q)$ of the quintessence field is greater
than the kinetic one, then it follows that the pressure $p=\dot{Q}%
^{2}/2-V(Q) $ associated to the quintessence $Q$-field is negative. The
properties of the quintessential cosmological models have been actively
considered in the physical literature (for a recent review see \cite{Tsu}).
As opposed to the cosmological constant of standard general relativity, the
equation of state of the quintessence field changes dynamically with time
\cite{11n}. Alternative models, in which the late-time acceleration can be
driven by the kinetic energy of the scalar field, called $k-$essence models,
have also been proposed \cite{kessence}.

Scalar fields $\phi $ that are minimally coupled to gravity via a negative
kinetic energy, can also explain the recent acceleration of the Universe.
Interestingly enough, they allow values of the parameter of the equation of
state with $w<-1$. These types of scalar fields, known as phantom fields,
have been proposed in \cite{phan1}. For phantom scalar fields the energy
density and pressure are given by $\rho _{\phi}=-\dot{\phi}^2/2+V\left(\phi
\right)$ and $p _{\phi}=-\dot{\phi}^2/2-V\left(\phi \right)$, respectively.
The interesting properties of phantom cosmological models for dark energy
have been investigated in detail in \cite{phan2,phan3, phan4}. Recent
cosmological observations show that at some moment during the cosmological
evolution the value of the parameter $w$ may have crossed the standard value
$w= -1$, corresponding to the general relativistic cosmological constant $%
\Lambda $. This cosmological situation is called \textit{the phantom divide
line crossing} \cite{phan4}. In the case of scalar field models with cusped
potentials, the crossing of the phantom divide line was investigated in \cite%
{phan3}. Another alternative way of explaining the phantom divide line
crossing is to model dark energy by a scalar field, which is non-minimally
coupled to gravity \cite{phan3}.

Scalar fields are also assumed to play a fundamental role in the evolution
of the very early Universe, playing a major role in the inflationary
scenario \cite{1nn, 2nn}. Originally, the idea of inflation was proposed to
provide solutions to the singularity, flat space, horizon, and homogeneity
problems, to the absence of magnetic monopoles, as well as to the problem of
large numbers of particles \cite{Li90, Li98}. However, presently it is
believed that the most important feature of inflation is the generation of
both initial density perturbations and of the background cosmological
gravitational waves. These important cosmological parameters can be
determined in many different ways, like, for example, through the study of the
anisotropies of the microwave background radiation, the analysis of the
local (peculiar) velocity galactic flows, of the clustering of galaxies, and
the determination of the abundance of gravitationally bound structures of
different types, respectively \cite{Li98}.

In many inflationary models the dynamical evolution of the early Universe is
driven by a single scalar field, called the inflaton, with the inflaton
rolling in some underlying self-interaction potential \cite{1nn,2nn,Li90,
Li98}. One common approximation in the study of the inflationary evolution
is the slow-roll approximation, which can be successfully used in two
separate contexts. The first situation is in the study of the classical
inflationary dynamics of expansion in the lowest order approximation. Hence
this implies that the contribution of the kinetic energy of the inflaton
field to the expansion rate is ignored. The second situation is represented
by the calculation of the perturbation spectra. The standard expressions
deduced for these spectra are valid in the lowest order in the slow roll
approximation \cite{Co94}. Finding exact inflationary solutions of the
gravitational field equations for different types of scalar field potentials
is also of great importance for the understanding of the dynamics of the
early Universe. Such exact solutions have been found for a large number of
inflationary potentials. Moreover, the potentials allowing a graceful exit
from inflation have been classified \cite{Mi}.

Hence, the theoretical investigation of the scalar field models is an
essential task in cosmology. Among the various methods used to study the
properties of scalar fields the methods based on the applications of the
mathematical formalism of the qualitative study of dynamical systems is of
considerable importance.

The usefulness of dynamical systems formulation of physical models is mainly
determined by their powerful predictive power. This predictive power is
essentially determined by the stability of their solutions. In a realistic
physical system, due to the limited precision of the measurements, some
uncertainties in the initial conditions always exist. Therefore a physically
meaningful mathematical model must also offer detailed and useful
information on the evolution of the deviations of the possible trajectories
of the dynamical system from a given reference trajectory. Hence an
important requirement in mathematical modelling is the understanding of the
local stability of the physical and cosmological processes. This information
on the system behavior is as important as the understanding of the late-time
deviations. The global stability of the solutions of the dynamical systems
described by systems of non-linear ordinary differential equations is
analyzed in the framework of the well-known mathematical theory of Lyapounov
stability. In this mathematical approach the fundamental quantities that
measure exponential deviations from a given trajectory are the so-called
Lyapunov exponents \cite{1,2}. It is usually very difficult to determine the
Lyapounov exponents analytically for a given dynamical system, and thus one
must resort to numerical methods. On the other hand, the important problem
of the local stability of the solutions of dynamical systems, described by
ordinary differential equations, is less understood.

Cosmological models have been intensively investigated by using methods from
dynamical systems and Lyapounov stability theory \cite{W,C,Liap,cosm3,cosm4}%
. In particular, phase space analysis proved to be a very useful method for
the understanding of the cosmological evolution. When studying the evolution
of cosmological models, the dynamical equations can be represented by an
autonomous dynamical system, described by a set of coupled - usually
strongly non-linear - differential equations for the physical parameters.
This representation allows the study of the Lyapunov stability of the model,
without explicitly solving the field equations for the basic variables.
Furthermore, the importance of the Lyapunov analysis is related to the fact
that stationary points of the dynamical system correspond to exact or
approximate analytic solutions of the field equations. Thus the dynamical
system formulation provide a useful tool for obtaining exact or approximate
solutions of the field equations in cosmologically interesting situations.

Even that the mathematical methods of the Lyapounov stability analysis are
well established, the study of the stability of the dynamical systems from
different points of view is extremely important. The comparison of the
results of the alternative approach with the corresponding Lyapunov
exponents analysis can provide a deeper understanding of the stability
properties of the system. An alternative, and very powerful method for the
study of the systems of the ordinary differential equations is represented
by the so-called Kosambi-Cartan-Chern (KCC) theory, which was initiated in
the pioneering works of Kosambi \cite{Ko33}, Cartan \cite{Ca33} and Chern
\cite{Ch39}, respectively. The KCC theory was inspired and influenced by the
geometry of the Finsler spaces (for a recent review of the KCC theory see
\cite{rev}). From a mathematical point of view the KCC theory is a
differential geometric theory of the variational equations for the
deviations of the whole trajectory with respect to the nearby ones \cite%
{An00}. In the KCC geometrical description of the systems of ordinary
differential equations one associates a non-linear connection, and a Berwald
type connection to the system of equations. With the use of these geometric
quantities five geometrical invariants are obtained. The most important
invariant is the second invariant, also called the curvature deviation
tensor, which gives the Jacobi stability of the system \cite{rev, An00,
Sa05,Sa05a}. The KCC theory has been applied for the study of different
physical, biochemical or technical systems (see \cite{Sa05, Sa05a, An93,
KCC, KCC1}.

An alternative geometrization method for dynamical systems, with
applications in classical mechanics and general relativity, was proposed in
\cite{Pet10} and \cite{Kau}, and further investigated in \cite{Pet0,Pet1}.
The Henon-Heiles system and Bianchi type IX cosmological models were also
investigated within this framework. In particular, in \cite{Pet0} a
theoretical approach based on the geometrical description of dynamical
systems and of their chaotic properties was developed. For the base manifold
a Finsler space was introduced, whose properties allow the description of a
wide class of physical systems, including those with potentials depending on
time and velocities, for which the Riemannian approach is unsuitable.

It is the purpose of the present paper to consider a systematic
investigation of the Jacobi stability properties of the flat homogeneous and
isotropic general relativistic cosmological models. By starting from the
standard Friedmann equations we perform, as a first step in our analysis, a
"second geometrization" of these equations, by associating to them a
non-linear connection, and a Berwald connection, respectively. This
procedure allows to obtain the so-called KCC invariants of the Friedmann
equations. The second invariant, called the curvature deviation tensor,
gives the Jacobi stability properties of the cosmological model. The KCC
theory can be naturally applied to systems of second order ordinary
differential equations. The Friedmann equations can be formulated as second
order differential equations, similarly to the Klein-Gordon equation
describing the scalar field. Therefore the KCC theory can be applied to
matter and scalar field dominated cosmological models. We obtain the general
condition for the Jacobi stability of scalar fields, which is described by
two inequalities involving the second and the first derivative of the scalar
field potential, the energy density of the field, as well as the time
derivative of the field itself. The geodesic deviation equations describing
the time variation of the deviation vector are also obtained. As an
application of the developed formalism we investigate the stability
properties of the scalar field cosmological models with exponential and
Higgs type potentials, respectively. It turns out that the exponential
potential scalar field is Jacobi unstable during its entire evolution, while
the time evolution of the scalar field cosmological models with Higgs
potential show a complicated dynamics with alternating stable and unstable
Jacobi phases. The Jacobi stability properties of the Higgs type models are
determined by the numerical value of the ratio of the self-coupling constant
and the square of the mass of the Higgs particle.

The Lyapunov stability properties of the scalar field cosmological models
are usually investigated by reformulating the evolution equation as a set of
three first order ordinary differential equations. In order to apply the KCC
theory to such systems they must be lifted to the tangent bundle. From
mathematical point of view this requires to take the time derivative of the
first order equations, so that their "second geometrization" can be easily
performed. We consider in detail the Jacobi stability properties of the
phantom quintessence and tachyon scalar field cosmological models. We study
in detail the Jacobi stability condition of these models, and we find that
they are Jacobi unstable during the entire expansionary cosmological
evolution.

The present paper is organized as follows. We review the basic ideas and the
mathematical formalism of the KCC theory in Section~\ref{kcc}. The Jacobi
stability analysis of the homogenous isotropic flat cosmological models by
using the second order formulation of the dynamics is performed in Section~%
\ref{sect3}. We consider both the cases of the matter dominated and scalar
field dominated cosmological models. As an application of the developed
formalism we investigate in detail the Jacobi stability of the scalar fields
with exponential potential, and Higgs potential, respectively. The Jacobi
stability of the first order dynamical system formulation of scalar field
cosmological models is considered in Section~ref{sect4}, in which the KCC
geometrization of the phantom quintessence and tachyonic scalar field models
is analyzed in detail. We discuss and conclude our results in Section~\ref%
{sect5}. The KCC geometric quantities giving the geometric description of
the phantom quintessence and tachyon scalar field cosmologies are presented
in Appendix \ref{appA} and Appendix~\ref{appB}, respectively.

\section{Kosambi-Cartan-Chern (KCC) theory and Jacobi stability}

\label{kcc}

In the present Section we briefly summarize the basic concepts and results
of the KCC theory (for a detailed presentation see \cite{rev} and \cite{An00}%
).

\subsection{Dynamical systems as paths of a semispray}

In the following we denote by $\mathcal{M}$ a real, smooth $n$-dimensional
manifold, and by $T\mathcal{M}$ its tangent bundle. On an open connected
subset $\Omega $ of the Euclidian $(2n+1)$ dimensional space $R^{n}\times
R^{n}\times R^{1}$ we introduce a $2n+1$ dimensional coordinates system $%
\left(x^i,y^i,t\right)$, $i=1,2,...,n$, where $\left( x^{i}\right) =\left(
x^{1},x^{2},...,x^{n}\right) $, $\left( y^{i}\right) =\left(
y^{1},y^{2},...,y^{n}\right) $ and $t$ is the time $t$. The coordinates $y^i$
are defined as
\begin{equation}
y^{i}=\left( \frac{dx^{1}}{dt},\frac{dx^{2}}{dt},...,\frac{dx^{n}}{dt}%
\right) .
\end{equation}

We assume that the time $t$ is an absolute invariant, and therefore the only
admissible change of coordinates will be
\begin{equation}
\tilde{t}=t,\tilde{x}^{i}=\tilde{x}^{i}\left( x^{1},x^{2},...,x^{n}\right)
,i\in \left\{1 ,2,...,n\right\} .  \label{ct}
\end{equation}

\textbf{Definition \cite{Punzi}.} \textit{A deterministic dynamical systems
is a formal set of rules that describe the evolution of points in some set $%
S $ with respect to an external, discrete, or continuous time parameter $t$
running in another set $T$.}

In a rigorous mathematical formulation, a dynamical system is a map \cite%
{Punzi}
\begin{equation}
\phi:T \times S \rightarrow S, (t,x)\mapsto \phi (t,x),
\end{equation}
which satisfies the condition $\phi (t , \cdot) \circ \phi (s , \cdot)=\phi
(t+s , \cdot)$ for all times $t ,s\in T$. In order to model realistic
dynamical systems or physical processes additional structures must be added
to the above definition.

In many cases the equations of motion of a dynamical system can be derived
from a Lagrangian $L$ via the Euler-Lagrange equations,
\begin{equation}
\frac{d}{dt}\frac{\partial L}{\partial y^{i}}-\frac{\partial L}{\partial
x^{i}}=F_{i},i=1,2,...,n,  \label{EL}
\end{equation}%
where $F_{i}$, $i=1,2,...,n$, is the external force. For a regular
Lagrangian $L$, the Euler-Lagrange equations given by Eq.~(\ref{EL}) are
equivalent to a system of second-order differential equations \cite{MHSS}
\begin{equation}
\frac{d^{2}x^{i}}{dt^{2}}+2G^{i}\left( x^{j},y^{j}\right) =0,i\in \left\{
1,2,...,n\right\} ,  \label{EM}
\end{equation}%
where each function $G^{i}\left( x^{j},y^{j},t\right) $ is $C^{\infty }$ in
a neighborhood of some initial conditions $\left( \left( x\right)
_{0},\left( y\right) _{0},t_{0}\right) $ in $\Omega $.

A vector field $S$ on $T\mathcal{M}$ of the form
\begin{equation}
S=y^i\frac{\partial }{\partial x^i}-2G^i\left(x^i,y^i\right)\frac{\partial }{%
\partial y^i},
\end{equation}
is called a \textit{semispray} \cite{S0,S1,S2}. The functions $G^i( x^i,
y^i) $ are \textit{the local coefficients} of the semispray, and they are
defined on domains of local charts. In the particular case in which the
coefficients $G^i = G^i \left(x^i, y^i\right)$ are homogeneous of degree two
in $y^i$, the vector field $S$ is called a \textit{spray}.

\textbf{Definition.} A \textit{path}  of the semispray
$S$ is defined as a curve $c:t\rightarrow x^{i}(t)$ on $\mathcal{M}$, with
the property that its lift $c^{\prime }:t\rightarrow \left( x^{i}(t),\dot{x}%
^{i}(t)\right) $ to $T\mathcal{M}$ is an integral curve of $S$, that is, a
curve satisfying the equation
\begin{equation}
\frac{d^{2}x^{i}}{dt^{2}}+2G\left( x^{i},y^{i}\right) =0.  \label{EM1}
\end{equation}

Conversely, for any system of ordinary differential equations of the form (%
\ref{EM1}), which is globally defined, the functions $G^{i}$ define a
semispray on $T\mathcal{M}$ \cite{S1,S2}.

More generally, one can start from \textit{an arbitrary system of
second-order differential equations of the form (\ref{EM})}, where no
\textit{a priori} given Lagrangian function is assumed, and study the
behavior of its trajectories \textit{by analogy with the trajectories of the
Euler-Lagrange equations.}

\subsection{The KCC geometrization of dynamical systems}

To associate a geometrical structure to the dynamical system defined by
Eqs.~(\ref{EM}), we introduce first a nonlinear connection $N$ on $M$, with
coefficients $N_{j}^{i}$, defined as \cite{MHSS}
\begin{equation}
N_{j}^{i}=\frac{\partial G^{i}}{\partial y^{j}}.
\end{equation}

The nonlinear connection can be understood in terms of a dynamical covariant
derivative $\nabla ^N$ \cite{Punzi}: for two vector fields $v$, $w$ defined
over M, we introduce the covariant derivative $\nabla ^N$ as
\begin{equation}  \label{con}
\nabla _v^Nw=\left[v^j\frac{\partial }{\partial x^j}w^i+N^i_j(x,y)w^j\right]%
\frac{\partial }{\partial x^i}.
\end{equation}

For $N_{j}^{i}(x,y)=\Gamma _{jl}^{i}(x)y^{l}$, Eq.~(\ref{con}) reduces to
the definition of the covariant derivative for the special case of a linear
connection.

For a non-singular coordinate transformations given by Eq. (\ref{ct}), the
KCC-covariant differential of an arbitrary vector field $\theta ^{i}(x)$ on
the open subset $\Omega \subseteq R^{n}\times R^{n}\times R^{1}$ is defined
as \cite{Sa05,rev}
\begin{equation}
\frac{D\theta ^{i}}{dt}=\frac{d\theta ^{i}}{dt}+N_{j}^{i}\theta ^{j}.
\label{KCC}
\end{equation}

For $\theta ^{i}=y^{i}$ we obtain
\begin{equation}
\frac{Dy^{i}}{dt}=N_{j}^{i}y^{j}-2G^{i}=-\epsilon ^{i},
\end{equation}
where the contravariant vector field $\epsilon ^{i}$ on $\Omega $ is called
the first KCC invariant.

\subsubsection{The curvature deviation tensor}

Let us now vary the trajectories $x^{i}(t)$ of the system (\ref{EM}) into
nearby ones according to
\begin{equation}
\tilde{x}^{i}\left( t\right) =x^{i}(t)+\eta \xi ^{i}(t),  \label{var}
\end{equation}
where $\left| \eta \right| $ is a small parameter and $\xi ^{i}(t)$ are the
components of some contravariant vector field defined along the path $%
x^{i}(t)$. Substituting Eqs. (\ref{var}) into Eqs. (\ref{EM}) and taking the
limit $\eta \rightarrow 0$ we obtain the variational equations \cite%
{An93,An00,Sa05,Sa05a}
\begin{equation}
\frac{d^{2}\xi ^{i}}{dt^{2}}+2N_{j}^{i}\frac{d\xi ^{j}}{dt}+2\frac{\partial
G^{i}}{\partial x^{j}}\xi ^{j}=0.  \label{def}
\end{equation}

By using the KCC-covariant differential we can write Eq. (\ref{def}) in the
covariant form
\begin{equation}
\frac{D^{2}\xi ^{i}}{dt^{2}}=P_{j}^{i}\xi ^{j},  \label{JE}
\end{equation}
where we have denoted
\begin{equation}
P_{j}^{i}=-2\frac{\partial G^{i}}{\partial x^{j}}-2G^{l}G_{jl}^{i}+ y^{l}%
\frac{\partial N_{j}^{i}}{\partial x^{l}}+N_{l}^{i}N_{j}^{l},
\end{equation}
and we have introduced the Berwald connection $G_{jl}^{i}$, defined as \cite%
{rev, An00, An93,MHSS,Sa05,Sa05a}
\begin{equation}
G_{jl}^{i}\equiv \frac{\partial N_{j}^{i}}{\partial y^{l}}.
\end{equation}

The tensor $P_{j}^{i}$ is called the second KCC-invariant, or \textit{the
deviation curvature tensor}, and Eq.~(\ref{JE}) is called \textit{the Jacobi
equation}, respectively. In both Riemann or Finsler geometry, when the
system (\ref{EM}) describes the geodesic equations in the given geometry,
Eq. (\ref{JE}) represents the Jacobi field equation.

The trace $P$ of the curvature deviation tensor is obtained as
\begin{equation}
P=P_{i}^{i}=-2\frac{\partial G^{i}}{\partial x^{i}}-2G^{l}G_{il}^{i}+ y^{l}%
\frac{\partial N_{i}^{i}}{\partial x^{l}}+N_{l}^{i}N_{i}^{l}+\frac{\partial
N_{i}^{i}}{\partial t}.
\end{equation}

The third, fourth and fifth invariants of the system (\ref{EM}) are defined
as \cite{An00}
\begin{equation}  \label{31}
P_{jk}^{i}\equiv \frac{1}{3}\left( \frac{\partial P_{j}^{i}}{\partial y^{k}}-%
\frac{\partial P_{k}^{i}}{\partial y^{j}}\right) ,P_{jkl}^{i}\equiv \frac{%
\partial P_{jk}^{i}}{\partial y^{l}},D_{jkl}^{i}\equiv \frac{\partial
G_{jk}^{i}}{\partial y^{l}}.
\end{equation}

The third invariant $P_{jk}^{i}$ can be interpreted as a torsion tensor. The
fourth and fifth invariants $P_{jkl}^{i}$ and $D_{jkl}^{i}$ are called the
Riemann-Christoffel curvature tensor, and the Douglas tensor, respectively
\cite{rev, An00}. In Berwald spaces these tensors always exist. Generally,
they can be used to describe the geometrical properties of systems of
second-order differential equations.

\subsection{The definition of Jacobi stability}

In many physical applications the behavior of the trajectories of the
dynamical system (\ref{EM}) in a vicinity of a point $x^{i}\left(
t_{0}\right) $ is of major interest. In the following for simplicity we take
$t_{0}=0$. The trajectories $x^{i}=x^{i}(t)$ can be considered as curves in
the Euclidean space $\left( R^{n},\left\langle .,.\right\rangle \right) $,
where $\left\langle .,.\right\rangle $ defines the canonical inner product
of $R^{n}$. As for the deviation vector $\xi $ we assume that it satisfies
the natural initial conditions $\xi \left( 0\right) =O$ and $\dot{\xi}\left(
0\right) =W\neq O$, where $O\in R^{n}$ is the null vector \cite{rev, An00,
Sa05,Sa05a}.


We describe the focusing tendency of the trajectories around $t_{0}=0$ as
follows: if $\left| \left| \xi \left( t\right) \right| \right| <t^{2}$, $%
t\approx 0^{+}$, the trajectories are bunching together. On the other hand,
if $\left| \left| \xi \left( t\right) \right| \right| >t^{2}$, $t\approx
0^{+}$, the trajectories are dispersing \cite{rev, An00, Sa05,Sa05a}.
Alternatively, the focusing tendency of the trajectories can be described in
terms of the deviation curvature tensor: the trajectories of the system of
second order differential equations (\ref{EM}) are bunching together for $%
t\approx 0^{+}$ if and only if the real parts of the eigenvalues of the
deviation curvature tensor $P_{j}^{i}\left( 0\right) $ are strictly
negative. On the other hand the trajectories are dispersing if and only if
the real parts of the eigenvalues of the tensor $P_{j}^{i}\left( 0\right) $
are strictly positive \cite{rev, An00, Sa05,Sa05a}.

Based on the above intuitive considerations we can define the Jacobi
stability for a second order system of differential equations as follows
\cite{rev, An00,Sa05,Sa05a}:

\textbf{Definition:} \textit{If the system of second order differential
equations (\ref{EM}) satisfies the initial conditions
\begin{equation*}
\left| \left| x^{i}\left( t_{0}\right) -\tilde{x}^{i}\left( t_{0}\right)
\right| \right| =0, \left| \left| \dot{x}^{i}\left( t_{0}\right) -\tilde{x}%
^{i}\left( t_{0}\right) \right| \right| \neq 0,
\end{equation*}
with respect to the norm $\left| \left| .\right| \right| $ induced by a
positive definite inner product, then we call the trajectories of (\ref{EM})
as Jacobi stable if and only if the real parts of the eigenvalues of the
deviation tensor $P_{j}^{i}$ are strictly negative everywhere. Otherwise,
the trajectories are called Jacobi unstable.}

\begin{figure}[htp]
\includegraphics[width=8.0cm]{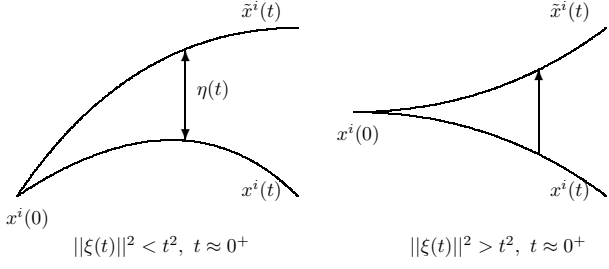}
\caption{Behavior of the trajectories near zero.}
\label{pict1}
\end{figure}

The focussing/dispersing behavior of the trajectories of a system of second
order ordinary differential equations near the origin is represented in Fig.~%
\ref{pict1}.

\subsection{The deviation curvature tensor for two and three dimensional dynamical
systems}

In the important two dimensional case the curvature deviation tensor can be
written in a matrix form as
\begin{equation}
P_{j}^{i}=\left(
\begin{array}{c}
P_{1}^{1}\;\;\;\;P_{2}^{1} \\
P_{1}^{2}\;\;\;P_{2}^{2}%
\end{array}%
\right) ,
\end{equation}%
with the eigenvalues given by
\begin{equation}
\lambda _{\pm }=\frac{1}{2}\left[ P_{1}^{1}+P_{2}^{2}\pm \sqrt{\left(
P_{1}^{1}-P_2^2\right) ^{2}+4P_{2}^{1}P_{1}^{2}}\right] .
\end{equation}

The eigenvalues of the curvature deviation tensor can be obtained as
solutions of the quadratic equation
\begin{equation}  \label{eql}
\lambda ^2-\left(P_1^1+P_2^2\right)\lambda
+\left(P_1^1P_2^2-P_2^1P_1^2\right)=0.
\end{equation}

A very powerful algebraic method to obtain the signs of the eigenvalues of
the curvature deviation tensor is represented by the Routh-Hurwitz criteria
\cite{RH}. According to these criteria all of the roots of the polynomial $%
P(\lambda)$ are negatives or have negative real parts if the determinants of
all Hurwitz matrices $\mathrm{det}\;H_j$, $j=1,2,..,n$, are strictly
positive. For $n=2$, corresponding to the case of Eq.~(\ref{eql}), the
Routh-Hurwitz criteria takes the simple form
\begin{equation}
P_1^1+P_2^2<0,\;\;P_1^1P_2^2-P_2^1P_1^2>0.
\end{equation}

The curvature properties along a given geodesic are described by the
eigenvalues of the deviation curvature tensor $\lambda _{\pm}$, which are
invariant functions on the tangent space. Moreover, once they are known, we
can introduce two quantities that can characterize the way the geodesic
explores the base manifold. They are defined as the (half) of the Ricci
curvature scalar along the flow, $\kappa $, and the anisotropy $\theta $,
given by
\begin{equation}
\kappa =\frac{1}{2}\left(\lambda _{+}+\lambda _{-}\right)=\frac{P}{2}=\frac{%
P_1^1+P_2^2}{2},
\end{equation}
and
\begin{equation}
\theta =\frac{1}{2}\left(\lambda _{+}-\lambda _{-}\right)=\frac{\sqrt{\left(
P_{1}^{1}-P_2^2\right) ^{2}+4P_{2}^{1}P_{1}^{2} }}{2},
\end{equation}
respectively.

In the case of a three-dimensional dynamical system with $n=3$, the
characteristic equation of the matrix of the curvature deviation tensor
becomes
\begin{eqnarray}
&&\lambda ^3- \left(P^1_1+P^2_2+P^3_3\right)\lambda ^2-\Bigg[-P^1_1
\left(P^2_2+P^3_3\right)+  \notag \\
&&P^1_2 P^2_1+P^1_3 P^3_1-P^2_2 P^3_3+P^2_3 P^3_2\Bigg]\lambda -  \notag \\
&&\Bigg[P^1_1 \left(P^2_2 P^3_3- P^2_3 P^3_2\right)-P^1_2 \left(P^2_1 P^3_3-
P^2_3 P^3_1\right)+  \notag \\
&& P^1_3 \left(P^2_1 P^3_2- P^2_2 P^3_1\right)\Bigg]=0.
\end{eqnarray}
Therefore the conditions of the Jacobi stability for a three-dimensional
system of second order ordinary differential equations can be formulated as
follows:
\begin{equation}  \label{3dim1}
\Sigma =P^1_1+P^2_2+P^3_3<0,
\end{equation}
\begin{equation}  \label{3dim2}
\Phi =-P^1_1 \left(P^2_2+P^3_3\right)+P^1_2 P^2_1+P^1_3 P^3_1-P^2_2
P^3_3+P^2_3 P^3_2<0,
\end{equation}
\begin{eqnarray}  \label{3dim3}
&&\Psi=P^1_1 \left(P^2_2 P^3_3- P^2_3 P^3_2\right)-P^1_2\left( P^2_1 P^3_3-
P^2_3 P^3_1\right)+  \notag \\
&& P^1_3 \left(P^2_1 P^3_2- P^2_2 P^3_1\right)<0,
\end{eqnarray}
\begin{eqnarray}  \label{3dim4}
&&\Omega=\left(P^1_1+P^2_2+P^3_3\right)\Bigg[-P^1_1
\left(P^2_2+P^3_3\right)+P^1_2 P^2_1+  \notag \\
&& P^1_3 P^3_1- P^2_2 P^3_3+P^2_3 P^3_2\Bigg]- \Bigg[P^1_1 \left( P^2_2
P^3_3- P^2_3 P^3_2\right)-  \notag \\
&& P^1_2 \Big(P^2_1 P^3_3- P^2_3 P^3_1\Big)+ P^1_3 \left(P^2_1 P^3_2- P^2_2
P^3_1\right)\Bigg]>0.  \notag \\
\end{eqnarray}

\section{Jacobi stability analysis of isotropic matter dominated and scalar
field cosmologies}

\label{sect3}

In the present Section we use the KCC approach for the study of the
dynamical properties of the matter dominated and scalar field cosmologies.
We explicitly obtain the non-linear and Berwald connections, and the
deviation curvature tensors. The eigenvalues of the deviation curvature
tensor are also obtained, and we study their properties in the equilibrium
points of the matter dominated model. The study of the sign of the
eigenvalues allows us to obtain the Jacobi stability properties of the fixed
points of the matter field dominated cosmological models. Next, we proceed
to a detailed analysis of the scalar field cosmologies in the framework of
the KCC theory. A full "second geometric" description is introduced, and the
time evolution of the relevant physical and geometrical parameters is
obtained. In particular, the nature of the Jacobi stability is analyzed in
detail. In the present study we restrict our analysis to the case of the
flat Friedmann-Robertson-Walker metric, given by
\begin{equation}
ds^{2}=dt^{2}-a^{2}(t)\left( dx^{2}+dy^{2}+dz^{2}\right) ,
\end{equation}%
where $a$ is the scale factor.

\subsection{Jacobi stability of matter dominated cosmological models}

For a Universe filled with pressureless dust and radiation only, the
cosmological expansion is described mathematically by the Friedmann
equations, which take the well-known form
\begin{equation}  \label{Fr1}
3H^2=\rho _m+\rho _r+\Lambda, \dot{H}=-\frac{1}{2}\left(\rho _m+\frac{4}{3}%
\rho _r\right),
\end{equation}
\begin{equation}  \label{Fr2}
\dot{\rho }_m+3H\rho _m=0,\dot{\rho}_r+4H\rho _r=0
\end{equation}
where a dot denotes the derivative with respect to the time $t$. In Eqs.~(%
\ref{Fr1}) and (\ref{Fr2}) $H=\dot{a}/a$ denotes the Hubble function, $\rho
_m$ is the baryonic matter energy density, $\rho _r$ is the energy density
of the radiation, while $\Lambda $ is the cosmological constant.

\subsubsection{Friedmann equations as an autonomous dynamical system}

In order to reformulate the cosmological evolutions equations as a dynamical
system, we need to introduce first the density parameters $\left( \Omega
_{m},\Omega _{r},\Omega _{\Lambda }\right)$ of the matter, radiation and
cosmological constant, defined as
\begin{equation}
\Omega _{m}=\frac{\rho _{m}}{3H^{2}},\Omega _{r}=\frac{\rho _{r}}{3H^{2}}%
,\Omega _{\Lambda }=\frac{\Lambda }{3H^{2}}.
\end{equation}%
The density parameters satisfy the normalization relation
\begin{equation}
\Omega _{m}+\Omega _{r}+\Omega _{\Lambda }=1.
\end{equation}%
As the basic variables $(x,y)$ in the phase space we adopt the quantities $%
x\equiv \Omega _{r}$, and $y\equiv \Omega _{\Lambda }$, respectively \cite%
{cosm3}. Then the density parameter of the mater is given by $\Omega
_{m}=1-x-y$, and the range of the variables $\left(x,y,\Omega _m\right)$ is $%
0\leq x\leq 1$, $0\leq y\leq 1$, and $0\leq \Omega _{m}\leq 1$. To describe
the cosmological dynamics we define the physically significant phase space
as $\Phi =\left\{ (x,y):x+y\leq 1,0\leq x\leq 1,0\leq y\leq 1\right\} $.
Next we take the time derivatives of $x $ and $y$ with respect to the time $%
t $, and, after introducing the new time variable $\tau =\ln a(t)$, we can
formulate the Friedmann equations as an autonomous dynamical system given by
\cite{cosm3}
\begin{equation}
\frac{dx}{d\tau }=-x(1-x+3y),  \label{ca1}
\end{equation}%
\begin{equation}
\frac{dy}{d\tau }=(3+x-3y)y.  \label{ca2}
\end{equation}%
The critical points of the system (\ref{ca1}) and (\ref{ca2}) in the phase
space region defined by $\Psi $ are obtained by solving the algebraic
equations $x(1-x+3y)=0$ and $(3+x-3y)y=0 $, respectively, and are given by
\begin{eqnarray}
&&P_{dS}=\{x=0,y= 1\},P_r=\{x= 1,y= 0\},  \notag \\
&&P_m=\{x =0,y= 0\}.
\end{eqnarray}

The Lyapunov stability properties of the system follows from the study of
the Jacobian matrix
\begin{equation}
J=%
\begin{pmatrix}
-1+2x-3y & -3x \\
y & 3+x-6y%
\end{pmatrix}%
.
\end{equation}

The critical point of the system (\ref{ca1}) and (\ref{ca2}) have a clear
cosmological interpretation. Thus, the critical point $P_{dS} = (0, 1)$,
with $\Omega _{\Lambda} = 1$, has $a(t) \propto \exp\left( \sqrt{\Lambda/3}%
t\right)$, and is associated to an accelerated de Sitter type expansion,
being a future attractor \cite{cosm4}. The critical point $P_r = (1, 0)$,
with $\Omega _r = 1$ and $a(t)\propto \sqrt{t}$, corresponds to the
radiation-dominated era in the cosmological evolution of the Universe, and
is a source point, or a past attractor. Finally, the critical point $P_m (0,
0)$, with $\Omega _m = 1$ and $a(t) \propto t^{2/3}$, corresponds to the
decelerating matter-dominated phase of the cosmological expansion. It turns
out that $P_m$ is a saddle critical point \cite{cosm4}.

\subsubsection{The KCC geometrization and the Jacobi stability of the
Friedmann equations}

In order to apply the KCC theory to the cosmological dynamical system given
by equations (\ref{ca1}) and (\ref{ca2}), we first relabel the variables as $%
x\equiv x^1$ and $y\equiv x^2$. We also denote $y^1=dx^1/d\tau$ and $%
y^2=dx^2/d\tau$, respectively. Hence we obtain
\begin{equation}  \label{ca3}
\frac{dx^1}{d\tau}=-x^1\left(1-x^1+3x^2\right)=f\left(x^1,x^2\right),
\end{equation}
\begin{equation}  \label{ca4}
\frac{dx^2}{d\tau}=\left(3+x^1-3x^2\right)x^2=g\left(x^1,x^2\right).
\end{equation}

Next, we take the derivative with respect to $\tau $ of Eqs.~(\ref{ca3}) and
(\ref{ca4}), respectively, thus obtaining the following lift on the tangent
bundle of the cosmological dynamical system,
\begin{equation}
\frac{d^2x^1}{d\tau ^2}=\left(-1+2x^1-3x^2\right)y^1-3x^1y^2,
\end{equation}
\begin{equation}
\frac{d^2x^2}{d\tau ^2}=x^2y^1+\left(3+x^1-6x^2\right)y^2.
\end{equation}

By comparison with Eqs.~(\ref{EM}) we obtain immediately
\begin{eqnarray}
&&G^{i}\left( x^{1},x^{2},y^{1},y^{2}\right) =  \notag \\
&&-\frac{1}{2}%
\begin{pmatrix}
\left( -1+2x^{1}-3x^{2}\right) y^{1}\;\;\;\;\;\;\;\;\;-3x^{1}y^{2} \\
\;\;\;\;x^{2}y^{1}\;\;\;\;\;\;\;\;\;\;\;\;\;\;\;\;\;\;\;\;\left(
3+x^{1}-6x^{2}\right) y^{2}%
\end{pmatrix}%
.
\end{eqnarray}

Hence the components of the non-linear connection $\left( N\right)
=N_{i}^{j} $ associated to the matter dominated cosmological dynamical
system in the presence of a cosmological constant are obtained as
\begin{equation}
\left( N\right) =N_{i}^{j}=-\frac{1}{2}%
\begin{pmatrix}
-1+2x^{1}-3x^{2} & -3x^{1} \\
x^{2} & 3+x^{1}-6x^{2}%
\end{pmatrix}%
.
\end{equation}

For the components of the deviation tensor $P=P_{i}^{j}$ we obtain
\begin{equation}
P=\frac{1}{2}%
\begin{pmatrix}
H_{f}\cdot y & H_{g}\cdot y%
\end{pmatrix}%
^{t}+\frac{1}{4}J^{2}\left( f,g\right) ,
\end{equation}
where
\begin{equation}
H_{f}=%
\begin{pmatrix}
f_{11} & f_{12} \\
f_{21} & f_{22}%
\end{pmatrix}%
,
\end{equation}
is the Hessian of $f$, and $H_g$ is the Hessian of $g$. Explicitly, the
curvature deviation tensor for the Friedmann cosmological dynamical system
can be obtained as
\begin{widetext}
\be
\left.P\right|_{\left(y^1=0,y^2=0\right)}=\left(
\begin{array}{cc}
 \frac{1}{4} (-2 x^1+3 x^2+1)^2+\frac{3 x^1 x^2}{2} & \frac{3}{2} x^1 (x^1-6 x^2+3)+\frac{3}{4} x^1 (-2 x^1+3 x^2+1)
   \\
 (x^1-6 x^2+3) x^2+\frac{1}{2} (-2 x^1+3 x^2+1) x^2 & (x^1-6 x^2+3)^2+\frac{3 x^1 x^2}{2}
\end{array}
\right).
\ee
\end{widetext}

By evaluating $P$ at the critical points gives
\begin{equation}
\left.P\right|_{\left(y^1=0,y^2=0,x^1=x^1_{cr},x^2=x^2_{cr}\right)}=\frac{1}{%
4}A^2,
\end{equation}
where $A=\left.J\right|_{\left(y^1=0,y^2=0,x^1=x^1_{cr},x^2=x^2_{cr}\right)}$%
.

\subsubsection{Jacobi stability of the critical points of the matter
dominated cosmological models}

In order to obtain the Jacobi stability of the critical points of the
cosmological model described by Eqs.~(\ref{ca1}) and (\ref{ca2}), we need to
compute the numerical values of the curvature deviation tensor at the
critical points. At the critical point $P_{dS}=\left(x^1=0,x^2=1\right)$ the
curvature deviation tensor takes the form
\begin{equation}
\left.P\right|_{\left(y^1=0,y^2=0,x^1=0,x^2=1\right)}= \left(
\begin{array}{cc}
4 & 0 \\
-1 & 9%
\end{array}
\right)
\end{equation}
and has the eigenvalues $\lambda _1=4$ and $\lambda _2=9$. Hence it follows
that the critical point $P_{dS}$ of the standard $\Lambda $CDM cosmological
model is \textit{Jacobi unstable}. For the critical point $P_r=
\left(x^1=1,x^2=0\right)$, we find
\begin{equation}
\left.P\right|_{\left(y^1=0,y^2=0,x^1=1,x^2=0\right)}= \left(
\begin{array}{cc}
\frac{1}{4} & \frac{21}{4} \\
0 & 16%
\end{array}
\right),
\end{equation}
with the corresponding eigenvalues of the curvature deviation tensor
obtained as $\lambda _1=16$ and $\lambda _2=1/4$. Hence, the critical point $%
P_r$ is also \textit{Jacobi unstable}. Finally, for the last critical point $%
P_m=(0,0)$ we obtain
\begin{equation}
\left.P\right|_{\left(y^1=0,y^2=0,x^1=0,x^2=0\right)}= \left(
\begin{array}{cc}
\frac{1}{4} & 0 \\
0 & 9%
\end{array}
\right),
\end{equation}
with the eigenvalues $\lambda _1=9$, $\lambda _2=1/4$. Hence, we obtain the
result that \textit{all the critical points} of the dynamical system
equivalent to the cosmological Friedmann equations are \textit{Jacobi
unstable}.

\subsection{The cosmological evolution equations in the presence of
minimally coupled scalar fields}

Let us consider a rather general class of scalar field models, minimally
coupled to the gravitational field, for which the Lagrangian density in the
Einstein frame reads
\begin{equation}
L=\frac{1}{2\kappa}\sqrt{\left| g\right| }\left\{ R+\kappa \left[ g^{\mu \nu
}\left( \partial _{\mu }\phi \right) \left( \partial _{\nu }\phi \right)
-2V\left( \phi \right) \right] \right\} \,,
\end{equation}
where $R$ is the curvature scalar, $\phi $ is the scalar field, $V\left(
\phi \right) $ is the self-interaction potential and $\kappa =8\pi G/c^{4}$
is the gravitational coupling constant, respectively. In the following, we
use natural units with $c=8\pi G=\hbar =1$, and we adopt as our signature
for the metric $\left( +1,-1,-1,-1\right) $, as is common in particle
physics.

For a flat FRW scalar field dominated Universe the evolution of a
cosmological model is determined by the system of the field equations
\begin{eqnarray}
3\left( \frac{\dot{a}}{a}\right) ^{2} &=&\rho _{\phi }=\frac{\dot{\phi}^{2}}{%
2}+V\left( \phi \right) ,  \label{H} \\
2\frac{\ddot{a}}{a}+\left( \frac{\dot{a}}{a}\right) ^{2} &=&-p_{\phi }=-%
\frac{\dot{\phi}^{2}}{2}+V\left( \phi \right) ,  \label{H1}
\end{eqnarray}%
and the evolution equation for the scalar field
\begin{equation}
\ddot{\phi}+3\frac{\dot{a}}{a}\dot{\phi}+V^{\prime }\left( \phi \right) =0,
\label{phi}
\end{equation}%
where the overdot denotes the derivative with respect to the time-coordinate
$t$, and the prime denotes the derivative with respect to the scalar field $%
\phi $, respectively. By substituting $\dot{a}/a$ from Eq. (\ref{H}) into
Eqs. (\ref{H1}) and (\ref{phi}) we can reformulate the dynamics of the
scalar field cosmological models in terms of two second order non-linear
ordinary differential equations, given by
\begin{equation}
\ddot{a}+\frac{1}{3}\left[ \dot{\phi}^{2}-V(\phi )\right] a=0,  \label{Ff1}
\end{equation}%
and
\begin{equation}
\ddot{\phi}+\sqrt{3}\sqrt{\frac{\dot{\phi}^{2}}{2}+V\left( \phi \right) }%
\dot{\phi}+V^{\prime }\left( \phi \right) =0,  \label{Ff2}
\end{equation}%
respectively.

\subsubsection{The non-linear and Berwald connections, and the KCC
invariants of the scalar field cosmological models}

In the following we introduce a new notation for the dependent variables $a$
and $\phi $, and for their time derivatives, respectively, as
\begin{equation}
a=x^{1},\phi =x^{2},\dot{a}=y^{1},\dot{\phi}=y^{2}.
\end{equation}

The cosmological dynamics of scalar field dominated Universes can be
formulated as a second order differential system, given by two second order
differential equations of the form
\begin{equation}
\frac{d^{2}x^{i}}{dt^{2}}+2G^{i}\left( x^{i},y^{i}\right) =0,i=1,2.
\end{equation}

From Eqs. (\ref{Ff1}) and (\ref{Ff2}) it follows immediately that
\begin{equation}
G^{1}\left( a,\phi,\dot{a},\dot{\phi}\right) =\frac{1}{6}\left[ \dot{\phi}
^{2}-V( \phi) \right] a,  \label{G1}
\end{equation}%
and
\begin{equation}
G^{2}\left( a,\phi,\dot{a},\dot{\phi}\right) =\frac{V^{\prime }(\phi )}{2}+%
\frac{\sqrt{3}}{2} \dot{\phi} \sqrt{V(\phi )+\frac{\dot{\phi}^2}{2}} ,
\end{equation}
respectively. Therefore we first obtain the components of the non-linear
connection as
\begin{equation}
N_{1}^{1}=\frac{\partial G^{1}\left( a,\phi,\dot{a},\dot{\phi}\right) }{%
\partial \dot{a}}=0,  \label{N}
\end{equation}%
\begin{equation}
N_{2}^{1}=\frac{\partial G^{1}\left( a,\phi,\dot{a},\dot{\phi}\right() }{%
\partial \dot{\phi}}=\frac{a \dot{\phi} }{3},  \label{N1}
\end{equation}%
\begin{equation}
N_{1}^{2}=\frac{\partial G^{2}\left( a,\phi,\dot{a},\dot{\phi}\right) }{%
\partial \dot{a}}=0,  \label{N3}
\end{equation}%
\begin{eqnarray}  \label{N4}
N_{2}^{2}&=&\frac{\partial G^{2}\left( a,\phi,\dot{a},\dot{\phi}\right) }{%
\partial \dot{\phi}}=\sqrt{\frac{3}{2}}\frac{ \left(V(\phi )+\dot{\phi}
^2\right)}{\sqrt{2 V(\phi )+\dot{\phi} ^2}}=  \notag \\
. &&\frac{\sqrt{3}}{2}\frac{2\rho _{\phi}-V(\phi)}{\sqrt{\rho _{\phi}}}.
\end{eqnarray}

For the non-zero components of the Berwald connection, defined as
\begin{equation}
G_{jl}^{i}=\frac{\partial N_{j}^{i}}{\partial y^{l}},i,j,l=1,2,
\end{equation}
we obtain
\begin{eqnarray}
&&G_{22}^{1}=\frac{\partial N_{2}^{1}}{\partial y^{2}}=\frac{1}{3}a,
\label{G111}
\end{eqnarray}
\begin{eqnarray}  \label{G112}
&&G_{22}^{2}=\frac{\partial N_{2}^{2}}{\partial y^{2}}=\sqrt{\frac{3}{2}}%
\frac{ \dot{\phi} \left(3 V(\phi )+\dot{\phi} ^2\right)}{\left(2 V(\phi )+%
\dot{\phi} ^2\right)^{3/2}} .
\end{eqnarray}

The components of the first KCC invariant of the minimally coupled scalar
field model are obtained as
\begin{equation}
\epsilon ^{1}=2G^{1}\left( a,\phi,\dot{a},\dot{\phi}\right) -N_{2}^{1}\dot{%
\phi}=-\frac{1}{3}aV\left( \phi\right) ,
\end{equation}%
and
\begin{eqnarray}
\epsilon ^{2} &=&2G^{2}\left( a,\phi,\dot{a},\dot{\phi}\right) -N_{2}^{2}%
\dot{\phi}= V^{\prime }\left(\phi\right)+  \notag \\
&& \sqrt{\frac{3}{2}}\frac{ \dot{\phi} V\left(\phi\right)}{\sqrt{2
V\left(\phi\right)+\dot{\phi}^2}}=V^{\prime }\left(\phi\right)+\frac{\sqrt{3}%
}{2}\frac{\dot{\phi}V(\phi)}{\sqrt{\rho _{\phi}}},
\end{eqnarray}
respectively.

The components of the curvature deviation tensor for minimally coupled
scalar field cosmological models are given by
\begin{equation}
P_{1}^{1}=\frac{1}{3} \left[V(\phi )-\dot{\phi} ^2\right] ,
\end{equation}%
\begin{equation}
P_{2}^{1}=\frac{\dot{a} \dot{\phi} }{3}-\frac{1}{\sqrt{6}}\frac{a \dot{\phi}
V(\phi )}{\sqrt{2 V(\phi )+\dot{\phi} ^2}},
\end{equation}%
\begin{equation}
P_{1}^{2}=0,
\end{equation}
\begin{eqnarray}
P_{2}^{2}&=&-V^{\prime \prime }(\phi )-\frac{\sqrt{6} \dot{\phi} V^{\prime
}(\phi )}{\sqrt{2 V(\phi )+\dot{\phi} ^2}}+  \notag \\
&&\frac{9 V(\phi )^2}{2 \left[2 V(\phi )+\dot{\phi} ^2\right]}-\frac{3
V(\phi \ )}{2}.
\end{eqnarray}
For the trace of the curvature deviation tensor we obtain
\begin{eqnarray}
P&=&P_{1}^{1}+P_{2}^{2}=\frac{1}{6} \Bigg\{-2 \left[3 V^{\prime \prime
}(\phi )+\dot{\phi} ^2\right]- \\
&& \frac{6 \sqrt{6} \dot{\phi} V^{\prime }(\phi )}{\sqrt{2 V(\phi )+\dot{\phi%
} ^2}}+\frac{27 V^2(\phi )}{2 V(\phi )+\dot{\phi} ^2}-7 V(\phi )\Bigg\} ,
\end{eqnarray}
while for $\chi =P_{1}^{1}P_{2}^{2}-P_{2}^{1}P_{1}^{2}=P_{1}^{1}P_{2}^{2}$
we have
\begin{eqnarray}
\chi &=&\frac{1}{3} \left[V(\phi )-\dot{\phi} ^2\right] \times \Bigg[%
-V^{\prime \prime }(\phi )- \frac{\sqrt{6} \dot{\phi} V^{\prime }(\phi )}{%
\sqrt{2 V(\phi )+\dot{\phi} ^2}}+  \notag \\
&& \frac{9 V^2(\phi )}{2 \left(2 V(\phi )+\dot{\phi} ^2\right)}- \frac{3
V(\phi )}{2}\Bigg].
\end{eqnarray}

Therefore if the conditions $P_{1}^{1}+P_{2}^{2}<0$ and $%
\;P_{1}^{1}P_{2}^{2}-P_{2}^{1}P_{1}^{2}>0$ are simultaneously satisfied, the
scalar field cosmological model is \textit{Jacobi stable}. These conditions
allows us to formulate the following

\textbf{Jacobi stability condition of isotropic and homogeneous scalar field
cosmological models}. \textit{If the parameters $\left(V(\phi), \rho
_{\phi}, p_{\phi},\dot{\phi}\right)$ of a homogeneous scalar field in an
isotropic flat FRW geometry simultaneously satisfy the conditions
\begin{equation*}
V^{\prime \prime }(\phi)>\frac{9}{4}\frac{V^2(\phi)}{\rho _{\phi}}-\frac{3}{2%
}\rho _{\phi}-\frac{1}{2}V(\phi)-\sqrt{3}\frac{\dot{\phi}V^{\prime }(\phi)}{%
\sqrt{\rho _{\phi}}},
\end{equation*}
\begin{eqnarray*}
&&\left[2p_{\phi}+V(\phi)\right]\times  \notag \\
&&\left[V^{\prime \prime }(\phi)+\frac{\sqrt{3}\dot{\phi}V^{\prime }(\phi)}{%
\rho _{\phi}}+\frac{3V(\phi)}{2}-\frac{9V^2(\phi)}{4\rho _{\phi}}\right] >0,
\end{eqnarray*}
the corresponding cosmological model is Jacobi stable, and Jacobi unstable
otherwise.}

For the variational differential equations determining the deviation vector $%
\xi$, we obtain
\begin{equation}
3 \frac{d^2\xi ^1}{dt^2}+2 a \dot{\phi} \frac{d\xi^2}{dt}+\left[\dot{\phi}
^2-V(\phi )\right]\xi^1(t) -a V^{\prime 2}(t) =0,
\end{equation}
and
\begin{eqnarray}
&&\frac{d^2\xi ^2}{dt^2}+\frac{\sqrt{6} \left[V(\phi )+\dot{\phi} ^2\right]
}{\sqrt{2 V(\phi )+\dot{\phi} ^2}}\frac{d\xi ^2}{dt} +  \notag \\
&& \left[V^{\prime \prime }(\phi )+\frac{\sqrt{\frac{3}{2}} \dot{\phi}
V^{\prime }(\phi \ )}{\sqrt{2 V(\phi )+\dot{\phi} ^2}}\right]\xi ^2(t) =0,
\end{eqnarray}
respectively.

In the case of the isotropic cosmological scalar field models the third,
fourth and fifth KCC invariants, as defined by Eqs.~(\ref{31}), are
identically equal to zero.

\subsubsection{Applications: the case of the scalar field with exponential
self-interaction potential}

As an application of the KCC geometrization of the scalar field cosmological
models we will consider the case of a scalar field with an exponential
self-interaction potential of the form
\begin{equation}
V(\phi )=V_{0}e^{\pm \lambda \phi },
\end{equation}%
where $V_{0}$ and $\lambda $ are constants. The cosmological equations
describing the time evolution of this scalar field model are
\begin{equation}
\ddot{a}+\frac{1}{3}\left[ \dot{\phi}^{2}-V_{0}e^{\pm \lambda \phi }\right]
a=0,  \label{V1}
\end{equation}%
and
\begin{equation}
\ddot{\phi}+\sqrt{3}\sqrt{\frac{\dot{\phi}^{2}}{2}+V_{0}e^{\pm \lambda \phi }%
}\dot{\phi}\pm \lambda V_{0}e^{\pm \lambda \phi }=0,  \label{V2}
\end{equation}
respectively. By introducing a new time variable $\tau =\sqrt{V_{0}}t$, it
turns out that the system of Eqs. (\ref{V1}) and (\ref{V2}) can be written
as
\begin{equation}
\frac{d^{2}a}{d\tau ^{2}}+\frac{1}{3}\left[ \left( \frac{d\phi }{d\tau }%
\right) ^{2}-e^{\pm \lambda \phi }\right] a=0,  \label{V3}
\end{equation}
\begin{equation}
\frac{d^{2}\phi }{d\tau ^{2}}+\sqrt{3}\sqrt{\frac{1}{2}\left( \frac{d\phi }{%
d\tau }\right) ^{2}+e^{\pm \lambda \phi }}\frac{d\phi }{d\tau }\pm \lambda
e^{\pm \lambda \phi }=0.  \label{V4}
\end{equation}

The system of Eqs. (\ref{V3})-(\ref{V4}) must be integrated with the initial
conditions $a(0)=a_{0}$, $\dot{a}(0)=a_{0}H_{0}$, $\phi \left( 0\right)
=\phi _{0}$, and $\dot{\phi}\left( 0\right) =\dot{\phi}_{0}$, respectively.

\begin{figure*}[tb]
\centering
\includegraphics[width=8.15cm]{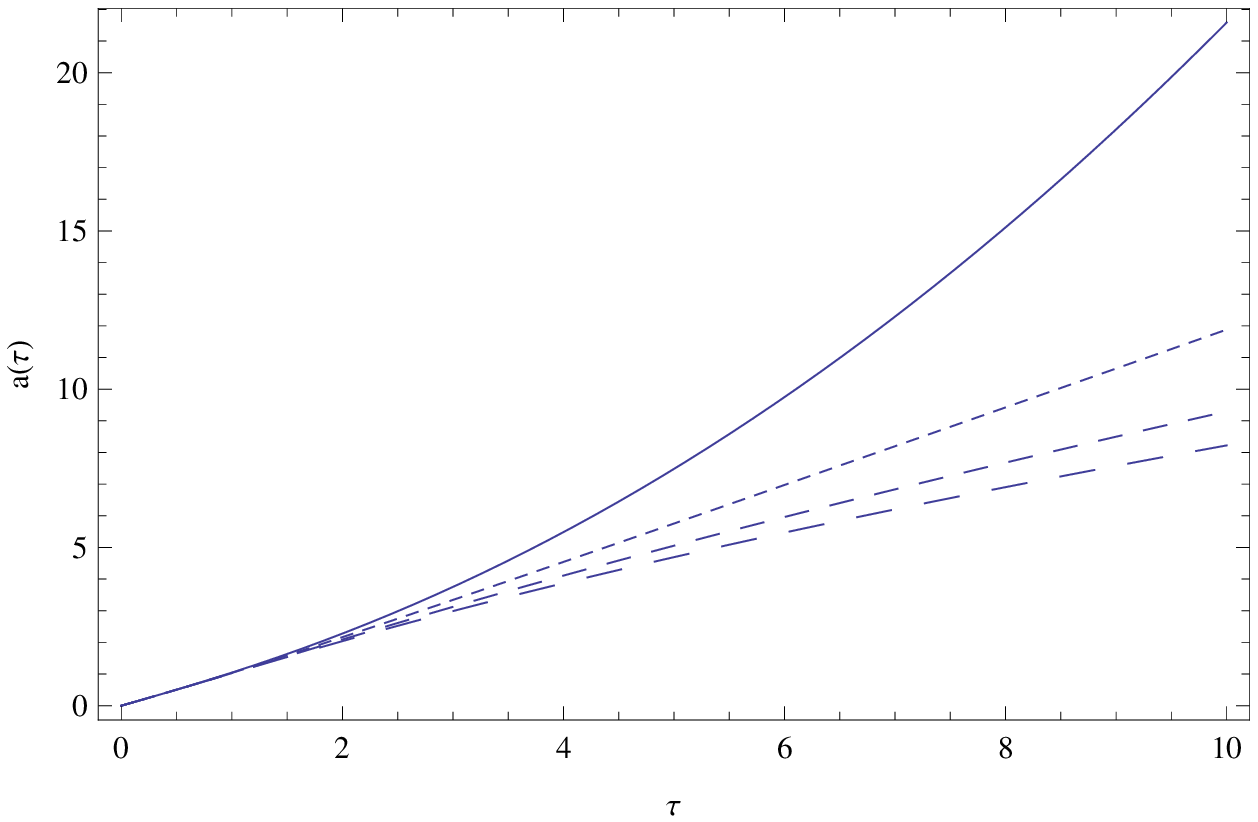} %
\includegraphics[width=8.15cm]{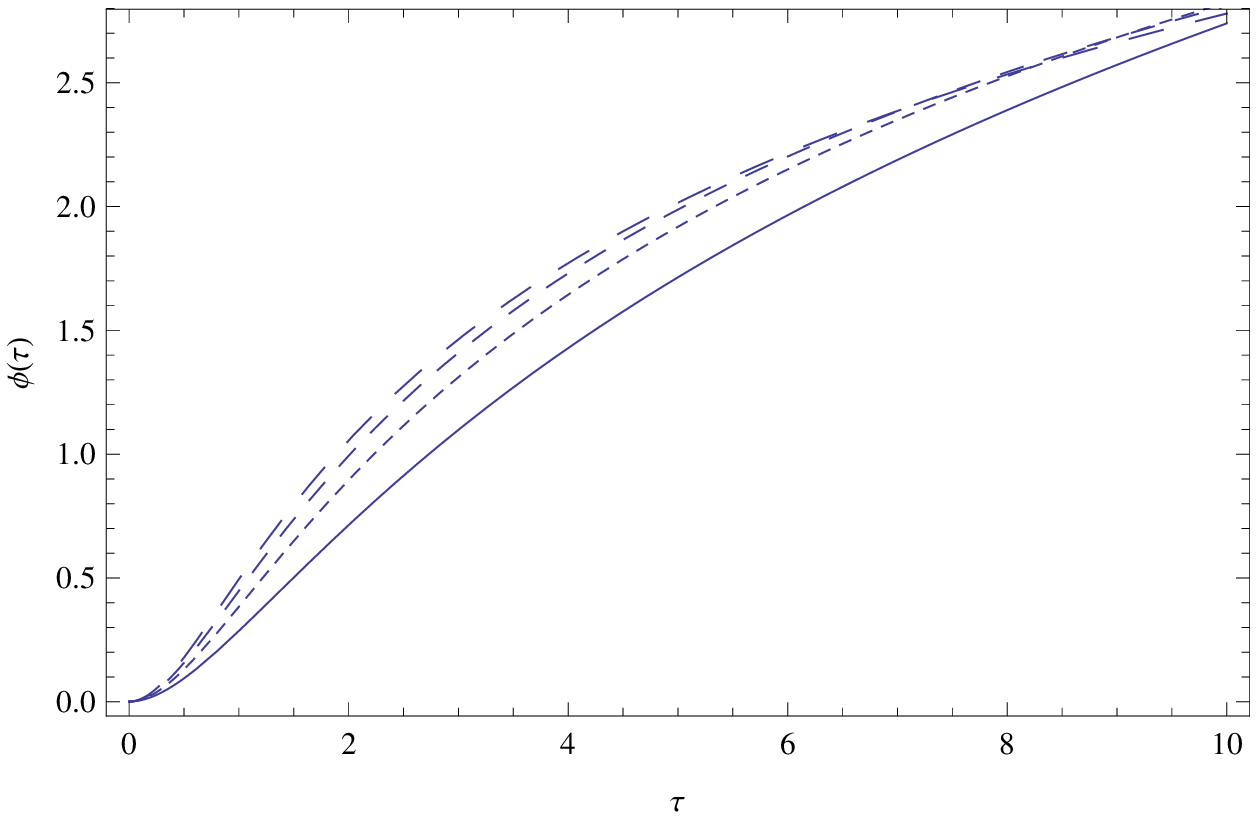}
\caption{ Variation of the scale factor of the Universe (left figure) and of
the scalar field (right figure) as a function of the dimensionless time $%
\protect\tau $ for the scalar field with exponential potential, for
different values of the parameter $\protect\lambda $: $\protect\lambda =-1$
(solid curve), $\protect\lambda =-\protect\sqrt{2}$ (dotted curve), $\protect%
\lambda =-\protect\sqrt{3}$ (short dashed curve), and $\protect\lambda =-2$
(dashed curve).}
\label{fig2}
\end{figure*}

The variations of the scale factor and of the scalar field of the
exponential potential scalar field filled Universe are presented, for
different values of the parameter $\lambda $, in Fig.~\ref{fig2}. In order
to numerically solve the field equations we have used the initial conditions
$a(0)=0$, $\phi (0)=0.001$, $\dot{a}(0)=1$, and $\dot{\phi}(0)=-0.001$,
respectively.

\begin{figure}[tb]
\centering
\includegraphics[width=8.15cm]{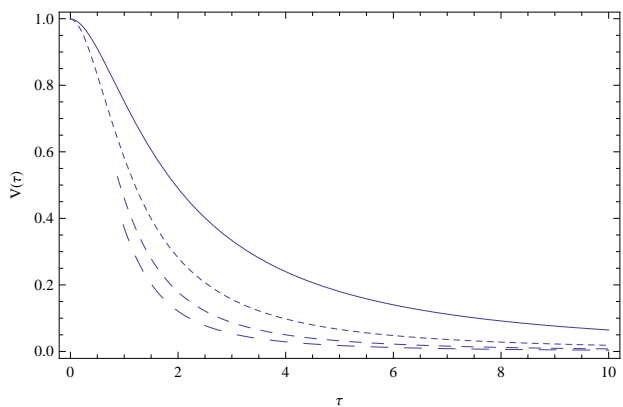}
\caption{ Variation of the scalar field potential $V=\exp (-\protect\lambda
\protect\phi )$ as a function of the dimensionless time $\protect\tau $ for
different values of the parameter $\protect\lambda $: $\protect\lambda =-1$
(solid curve), $\protect\lambda =-\protect\sqrt{2}$ (dotted curve), $\protect%
\lambda =-\protect\sqrt{3}$ (short dashed curve), and $\protect\lambda =-2$
(dashed curve).}
\label{fig3}
\end{figure}

As one can see from the figures, the scale factor is a monotonically
increasing function of time, while the scalar field also increases during
the cosmological evolution. The time variation of the scalar field potential
is presented in Fig.~\ref{fig3}.

\begin{figure*}[tb]
\centering
\includegraphics[width=8.15cm]{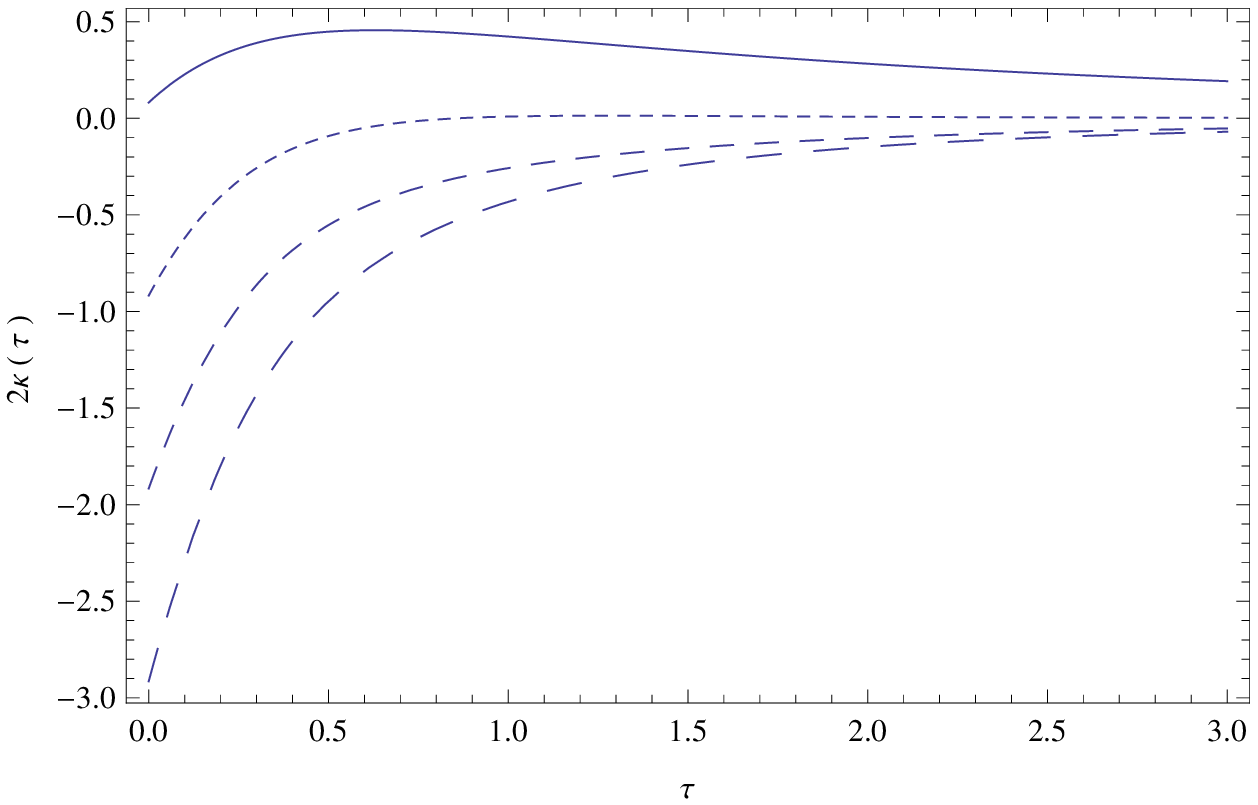} %
\includegraphics[width=8.15cm]{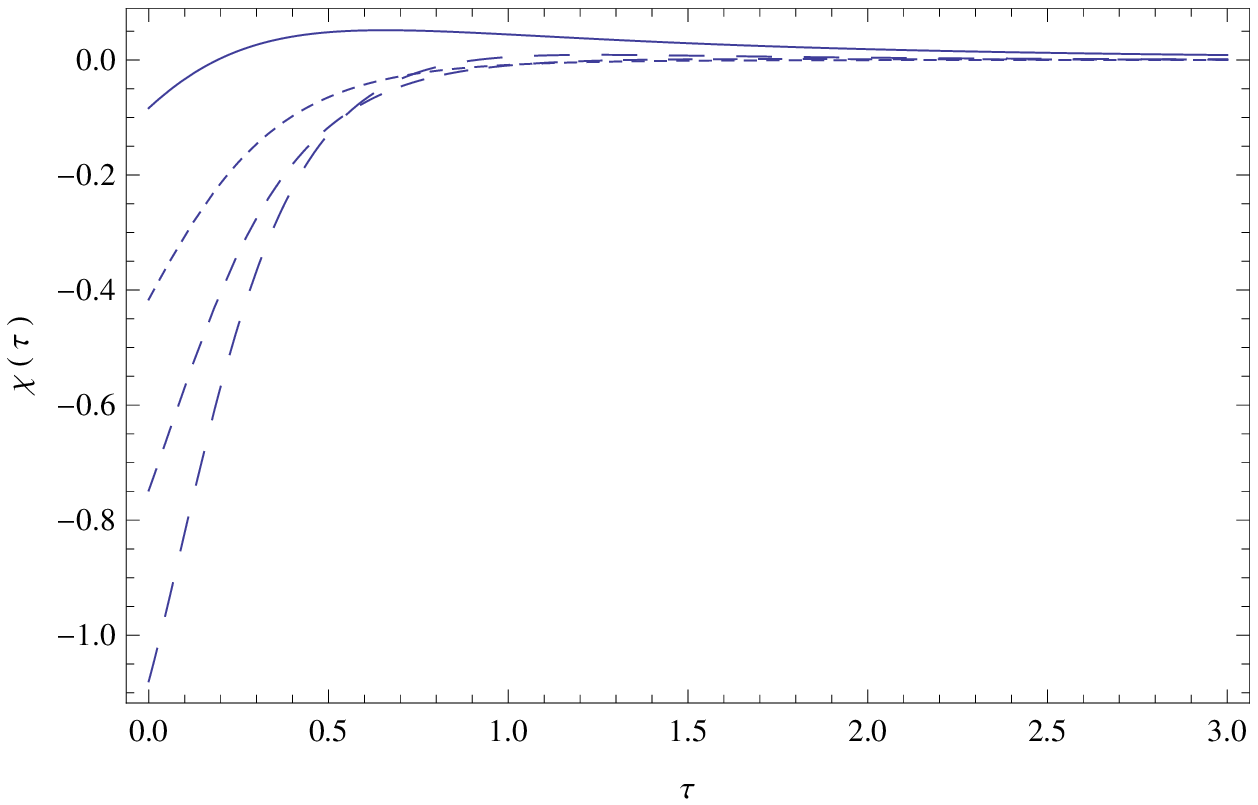}
\caption{ Time variations of the KCC invariants $2\protect\kappa %
=P_1^1+P_2^2 $ (left figure) and $\protect\chi =P_1^1P_2^2-P_2^1P_2^1$
(right figure) for a Universe filled with a scalar field with exponential
potential, for different values of the parameter $\protect\lambda $: $%
\protect\lambda =-1$ (solid curve), $\protect\lambda =-\protect\sqrt{2}$
(dotted curve), $\protect\lambda =-\protect\sqrt{3}$ (short dashed curve),
and $\protect\lambda =-2$ (dashed curve).}
\label{fig4}
\end{figure*}

The scalar field potential is a monotonically decreasing function of time,
which tends, in the large time limit, to zero. The time variations of the
KCC invariant $2\kappa =P_1^1+P_2^2$ and $\chi =P_1^1P_2^2-P_2^1P_2^1$ are
represented in Figs.~\ref{fig4}.

\begin{figure*}[tb]
\centering
\includegraphics[width=8.15cm]{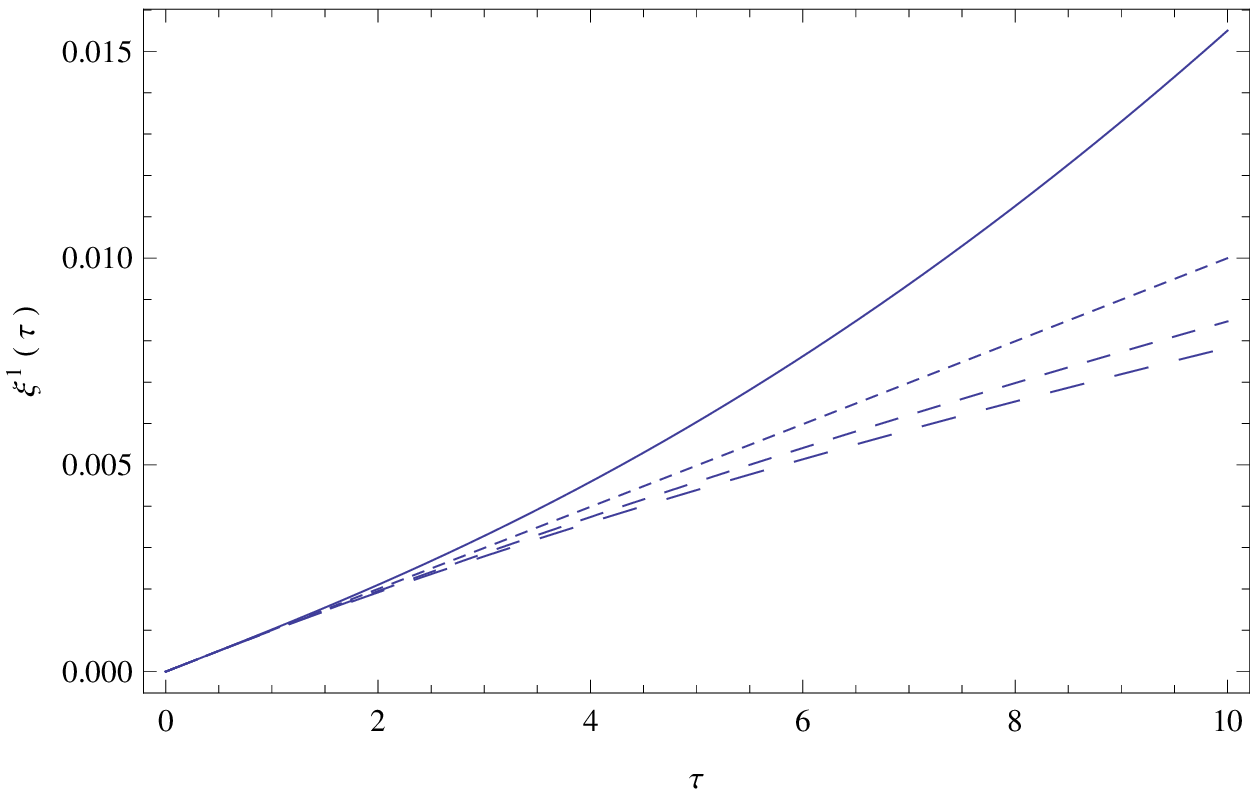} %
\includegraphics[width=8.15cm]{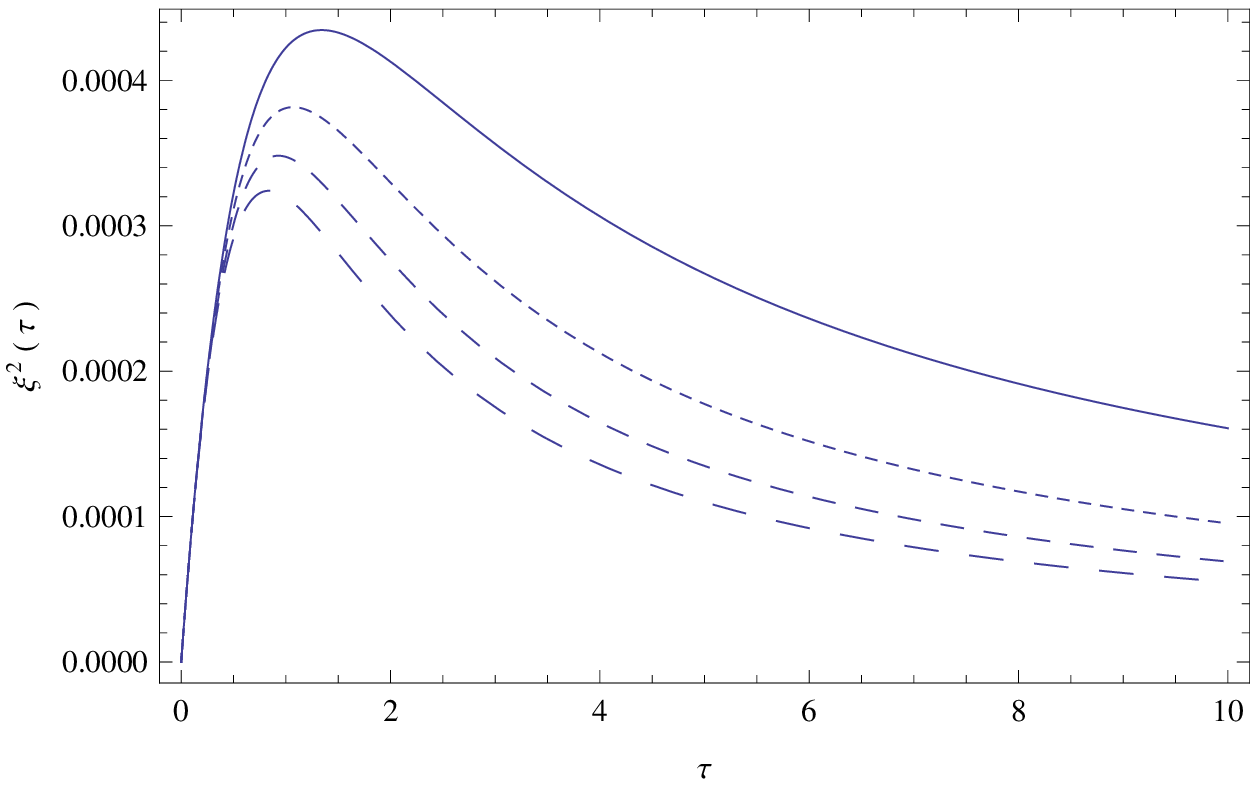}
\caption{ Time variations of the deviation vector components $\protect\xi ^1$
(left figure) and $\protect\xi ^2$ (right figure) for a Universe filled with
a scalar field with exponential potential, for different values of the
parameter $\protect\lambda $: $\protect\lambda =-1$ (solid curve), $\protect%
\lambda =-\protect\sqrt{2}$ (dotted curve), $\protect\lambda =-\protect\sqrt{%
3}$ (short dashed curve), and $\protect\lambda =-2$ (dashed curve).}
\label{fig5}
\end{figure*}

In order for the considered model of the Universe be stable, the invariants
must simultaneously satisfy the conditions $2\kappa <0$ and $\chi >0$,
respectively. As one can see from the Figures, from the chosen set of
parameters, these conditions are not satisfied for any interval of time
during the cosmological evolution. Therefore it follows that during its
entire evolution an exponential potential scalar field Universe is in a
\textit{Jacobi unstable} state. Finally, in Figs.~\ref{fig5} we present the
time variation of the components of the deviation vector $\xi ^i$, $i=1,2$.

The component $\xi ^1$ of the deviation vector increases exponentially in
time, indicating that the trajectories do diverge exponentially near the
origin. Thus, this result also confirms the presence of the Jacobi
instability for the exponential potential scalar field cosmology.

\subsubsection{Scalar fields with Higgs potential}

\begin{figure*}[tb]
\centering
\includegraphics[width=8.15cm]{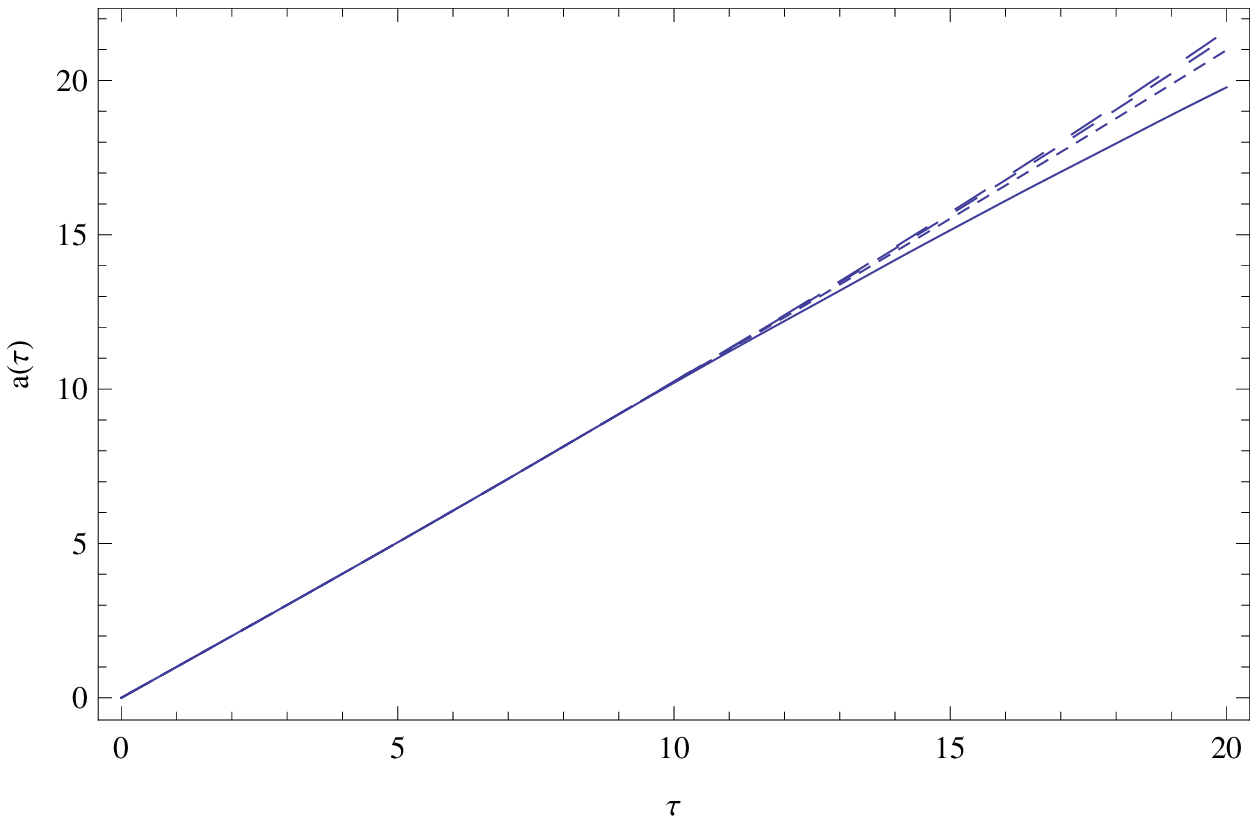} %
\includegraphics[width=8.15cm]{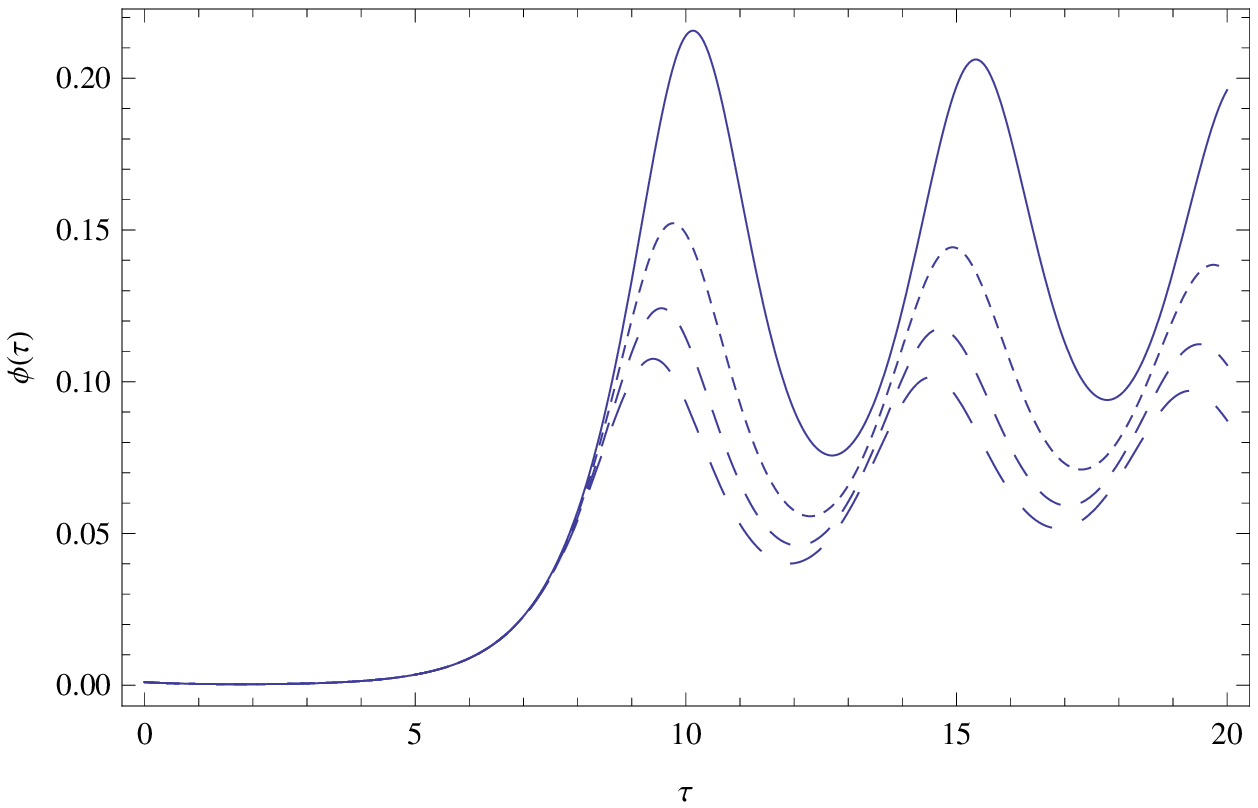}
\caption{ Variation of the scale factor of the Universe (left figure) and of
the scalar field (right figure) as a function of the dimensionless time $%
\protect\tau $ for the scalar field with Higgs type self-interaction
potential, for different values of the parameter $\protect\eta $: $\protect%
\eta =10$ (solid curve), $\protect\eta =20$ (dotted curve), $\protect\eta %
=30 $ (short dashed curve), and $\protect\eta =40$ (dashed curve).}
\label{fig6}
\end{figure*}

As a next case of the investigation of the Jacobi stability of a
cosmological scalar field model we consider that the Universe is filled with
a Higgs-like field, with self-interaction potential given by
\begin{equation}
V(\phi )=V_{0}+\frac{1}{2}M^{2}\phi ^{2}+\frac{\lambda }{4}\phi ^{4},
\end{equation}%
where $V_{0}$ is a constant, and $M^{2}<0$ is related to the mass of the
Higgs boson by the relation $m_{H}=V^{\prime \prime 2}$, where $%
v^{2}=-M^{2}/\lambda $ gives the minimum of the potential. The Higgs
self-coupling constant $\lambda \approx 1/8$ \cite{Higgs}, a value inferred
based on the determination of $m_{H}$ from accelerator experiments. By
introducing a new dimensionless time variable $\tau =Mt$, the basic
equations determining the cosmological evolution are given by
\begin{equation}
\frac{d^{2}a}{d\tau ^{2}}+\frac{1}{3}\left[ \frac{1}{2}\left( \frac{d\phi }{%
d\tau }\right) ^{2}-v_{0}+\frac{1}{2}\phi ^{2}-\eta \phi ^{4}\right] =0,
\end{equation}%
\begin{equation}
\frac{d^{2}\phi }{d\tau ^{2}}+\sqrt{3}\sqrt{\frac{1}{2}\left( \frac{d\phi }{%
d\tau }\right) ^{2}+v_{0}-\frac{1}{2}\phi ^{2}+\eta \phi ^{4}}\;\frac{d\phi
}{d\tau }-\phi +4\eta \phi ^{3}=0,
\end{equation}
where $v_{0}=V_{0}/M^{2}$ and $\eta =\lambda /4M^{2}>0$, respectively. The
time variations of the scale factor of the Higgs field filled Universe and
of the scalar field are represented in Fig.~\ref{fig6}. To numerically
integrate the gravitational field equations, we have fixed the value of the
constant $v_{0}$ to $v_{0}=0.005$, and we have adopted as initial conditions
the values the initial conditions $a(0)=a_{0}$, $\dot{a}(0)=a_{0}H_{0}$, $%
\phi \left( 0\right) =\phi _{0}$, and $\dot{\phi}\left( 0\right) =\dot{\phi}%
_{0}$, respectively.

\begin{figure}[tb]
\centering
\includegraphics[width=8.15cm]{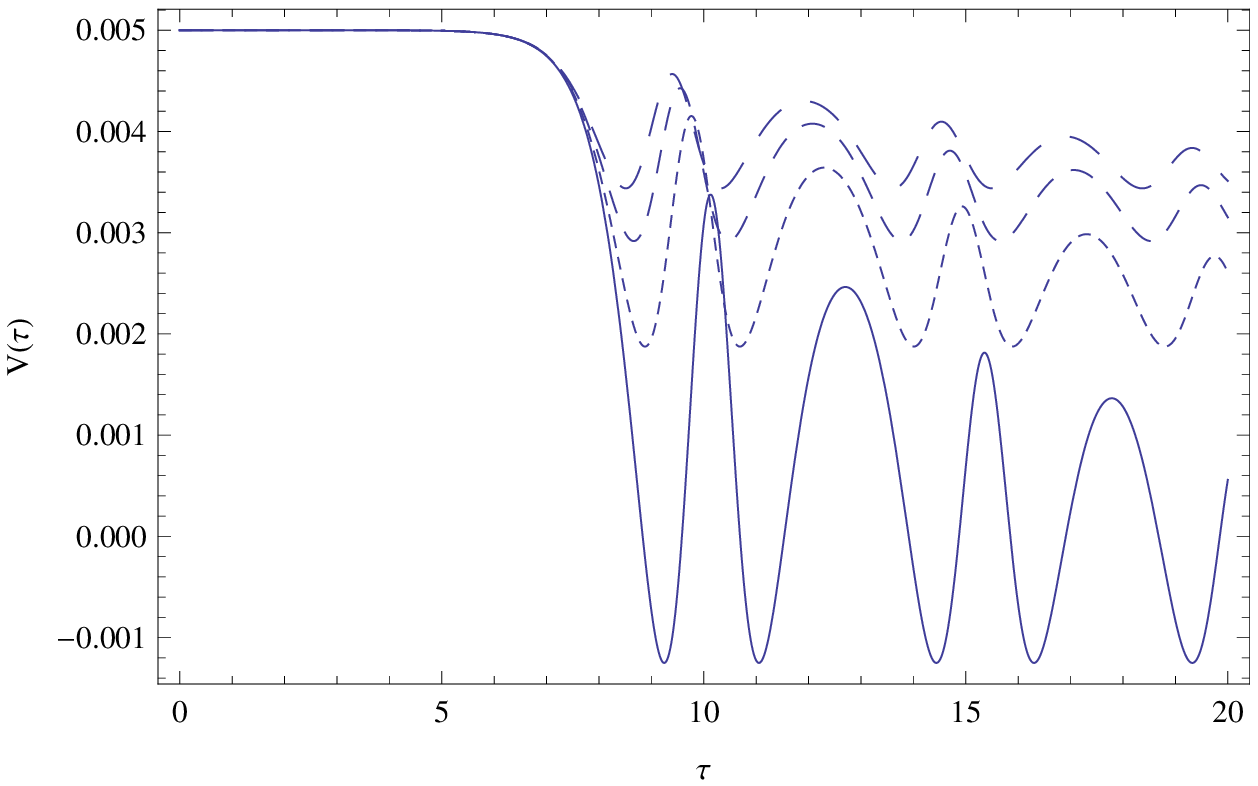}
\caption{ Variation of the Higgs type potential of the scalar field as a
function of the dimensionless time $\protect\tau $ for different values of
the parameter $\protect\eta $: $\protect\eta =10$ (solid curve), $\protect%
\eta =20$ (dotted curve), $\protect\eta =30$ (short dashed curve), and $%
\protect\eta =40$ (dashed curve).}
\label{fig7}
\end{figure}

The Universe filled with a Higgs type scalar field is expanding, with the
scale factor monotonically increasing in time. In the initial phases of the
expansion the scale factor can be approximated by a linearly increasing
function of time. The Higgs scalar field $\phi$ keeps a constant value in
the initial stages of the evolution, followed by a rapid increase associated
with an oscillatory behavior with a decreasing amplitude, associated with
the energy dissipation. The variation of the Higgs potential is represented
in Fig.~\ref{fig7}.

\begin{figure*}[tb]
\centering
\includegraphics[width=8.15cm]{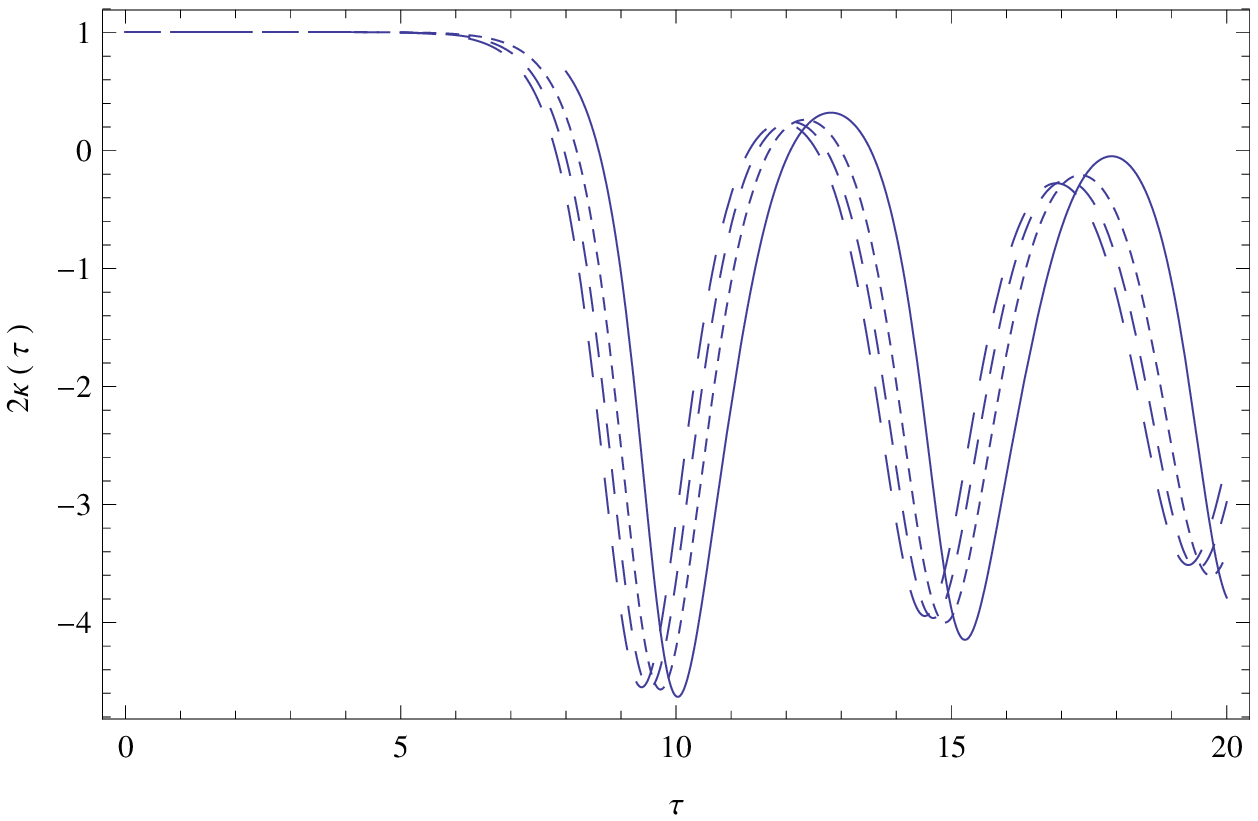} %
\includegraphics[width=8.15cm]{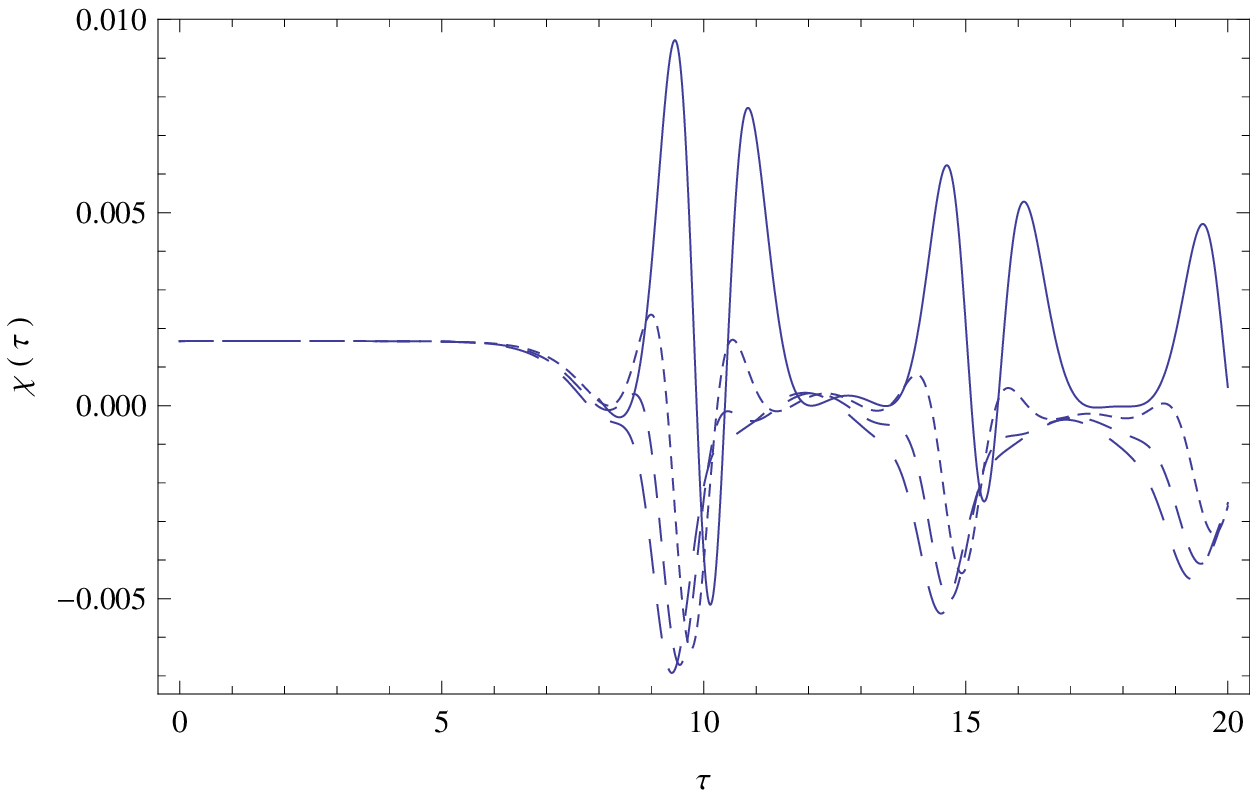}
\caption{ Time variations of the KCC invariants $2\protect\kappa %
=P_1^1+P_2^2 $ (left figure) and $\protect\chi =P_1^1P_2^2-P_2^1P_2^1$
(right figure) for a Universe filled with a Higgs type scalar field, for
different values of the parameter $\protect\eta $: $\protect\eta =10$ (solid
curve), $\protect\eta =20$ (dotted curve), $\protect\eta =30$ (short dashed
curve), and $\protect\eta =40$ (dashed curve).}
\label{fig8}
\end{figure*}

After an initial phase in which the potential is constant, rapid
oscillations with decreasing amplitude do follow, and the potential decays
in time. The variation of the KCC invariants $2\kappa $ and $\chi $ is
represented in Fig.~\ref{fig8}.

\begin{figure*}[tb]
\centering
\includegraphics[width=8.15cm]{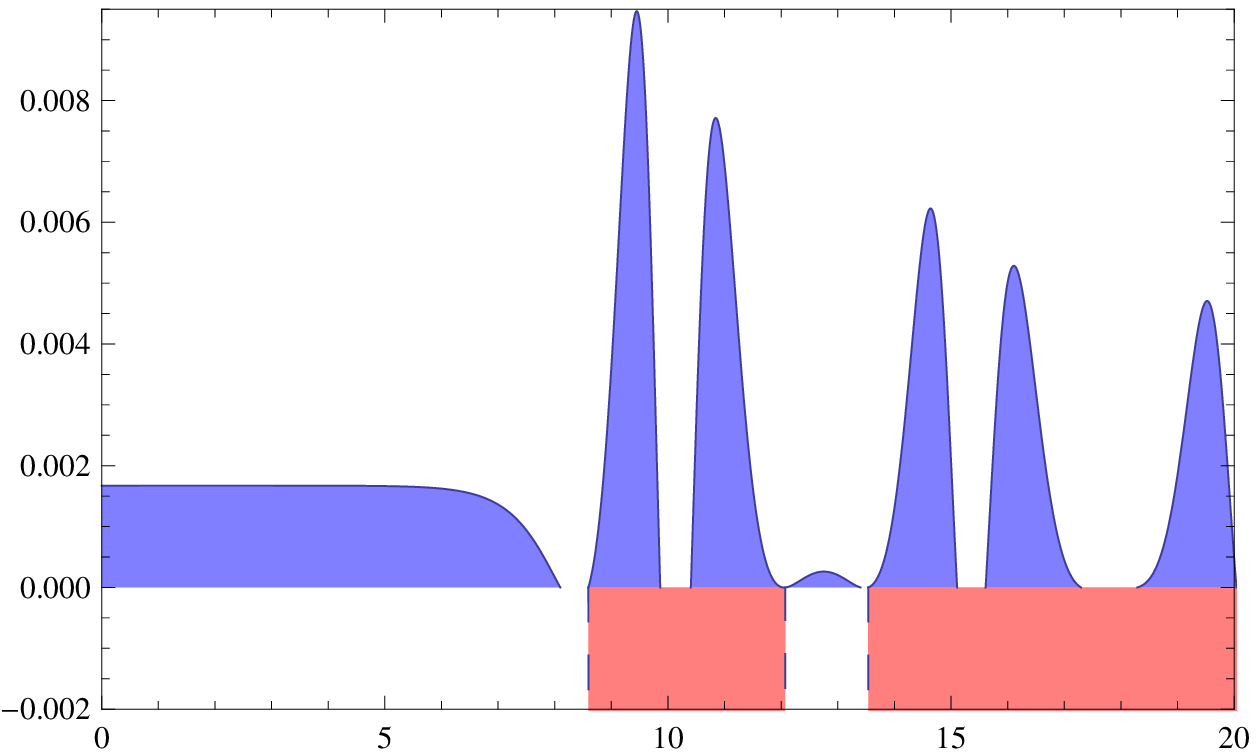} %
\includegraphics[width=8.15cm]{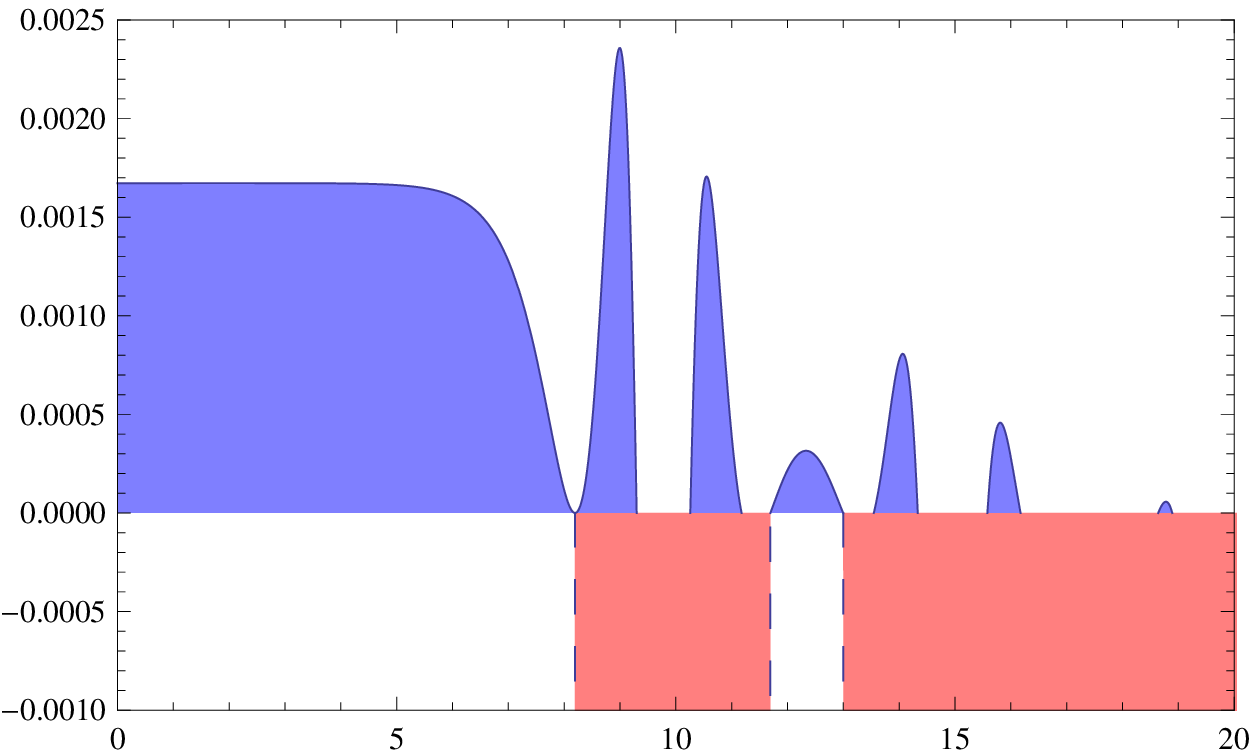} %
\includegraphics[width=8.15cm]{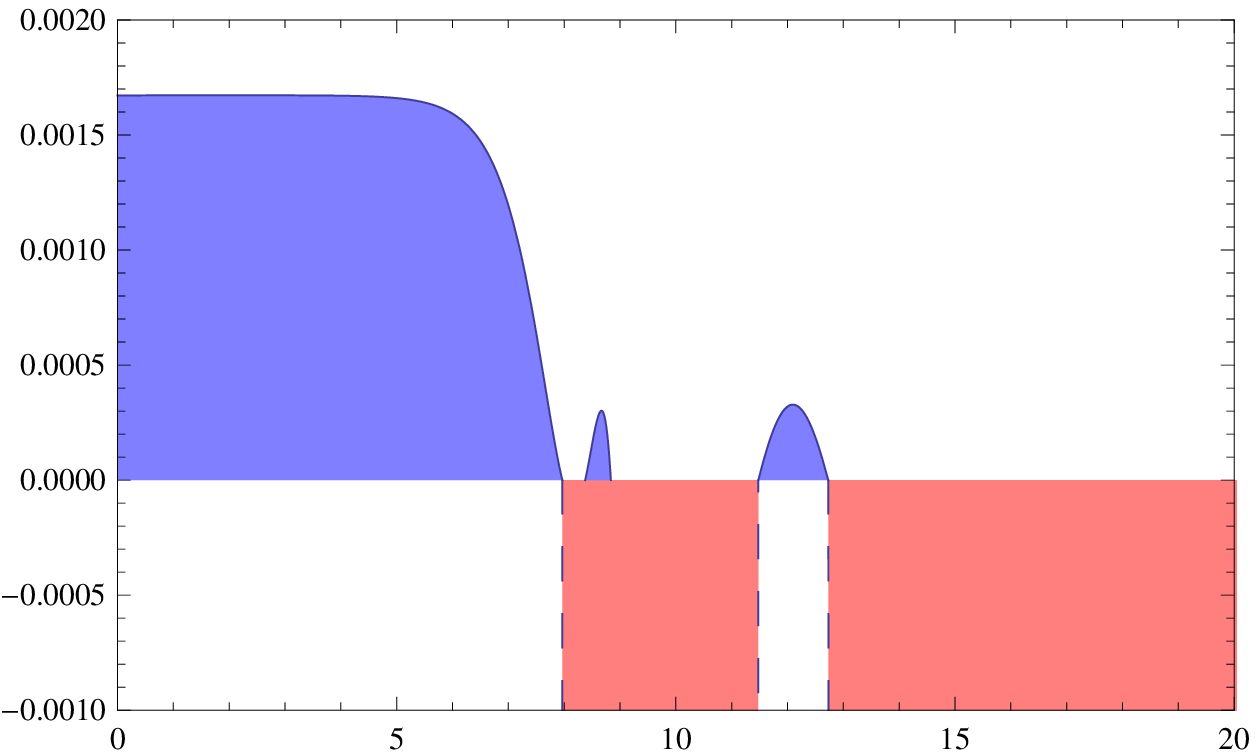} %
\includegraphics[width=8.15cm]{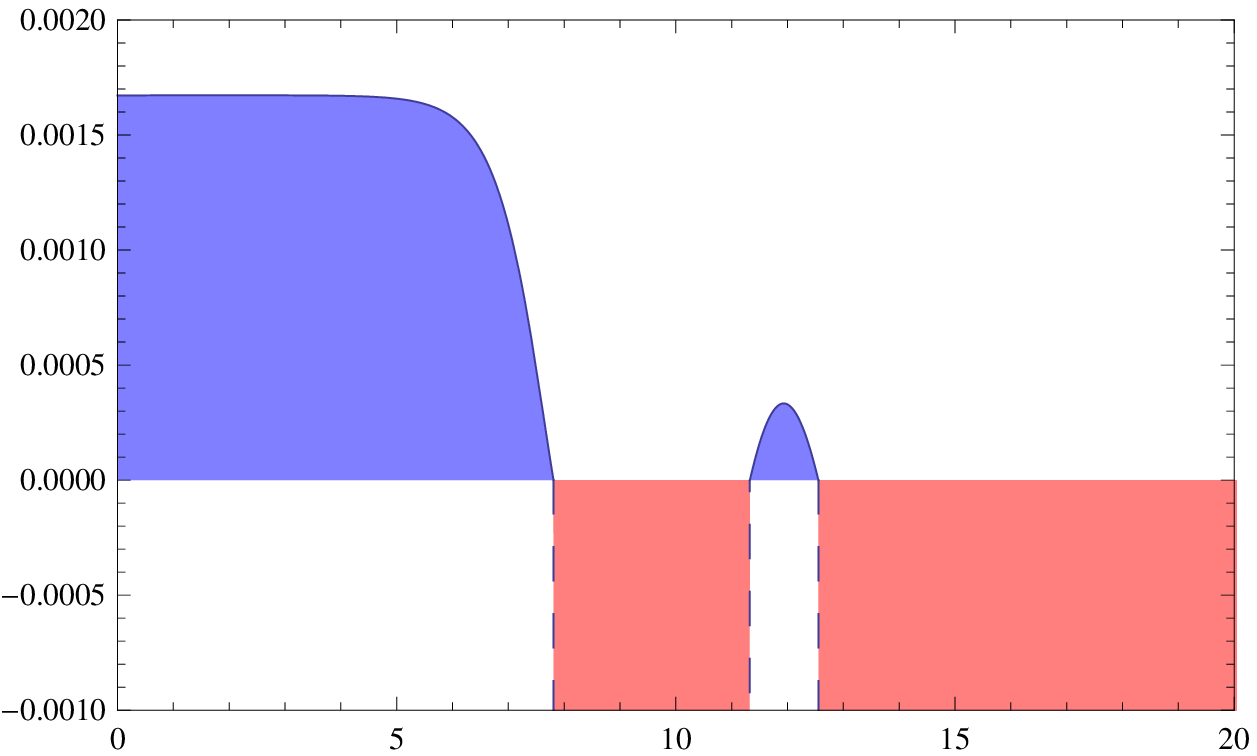}
\caption{ The Jacobi stability regions for the scalar field cosmological
model with Higgs potential for different values of $\protect\eta $: $\protect%
\eta=10$ (top left), $\protect\eta=20$ (top right), $\protect\eta=30$
(bottom left) and $\protect\eta=40$ (bottom right). The blue area shows the
time intervals with $\protect\chi>0$, and the black area shows the time
intervals with $\protect\kappa<0$. Where they overlap, both conditions $%
\protect\chi>0$ and $\protect\kappa<0$ are simultaneously satisfied, and the
evolution is Jacobi stable. }
\label{fig8a}
\end{figure*}

Jacobi stability of the cosmological model with Higgs type scalar field
requires the conditions $2\kappa <0$ and $\chi >0$ to be simultaneously
satisfied. The Jacobi stable/unstable regions formed during the cosmological
expansion of scalar field cosmologies with Higgs self-interacting potential
are presented in Fig.~\ref{fig8a}.

\begin{figure*}[tb]
\centering
\includegraphics[width=8.15cm]{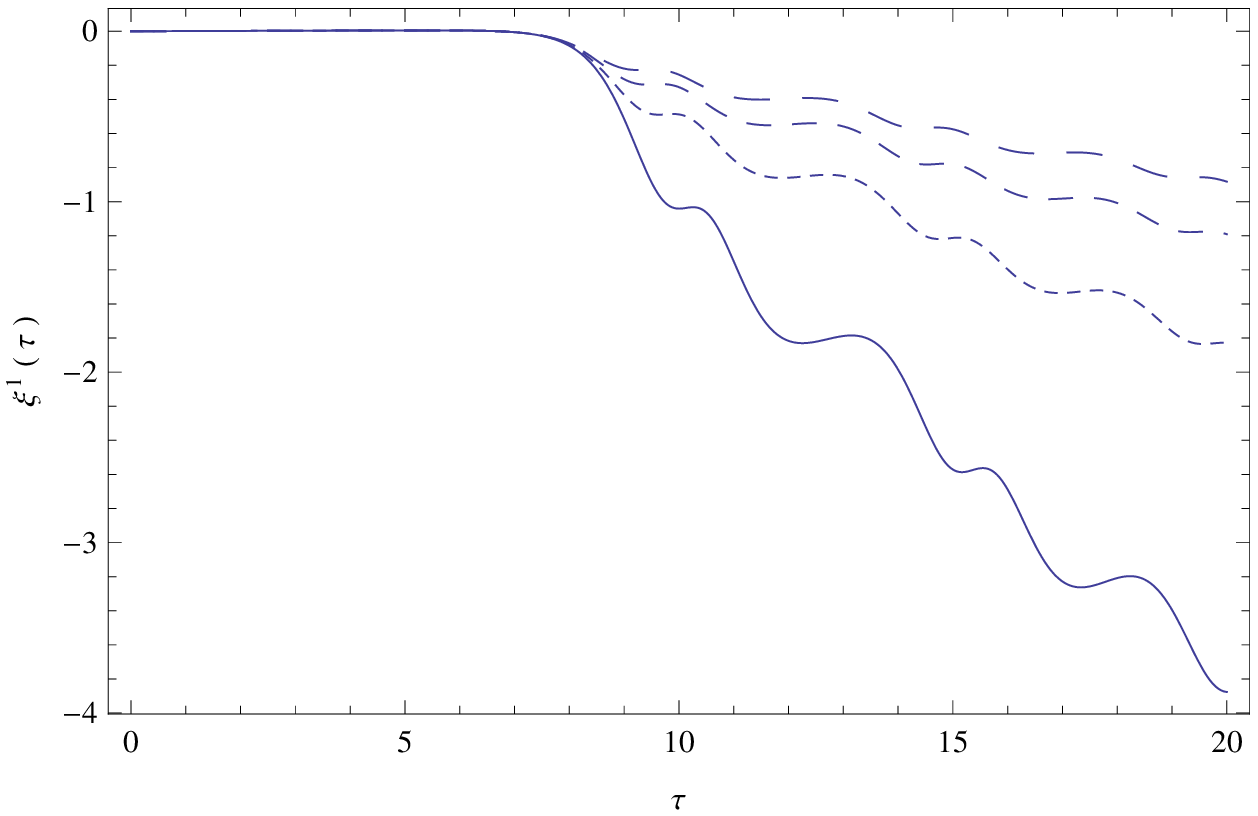} %
\includegraphics[width=8.15cm]{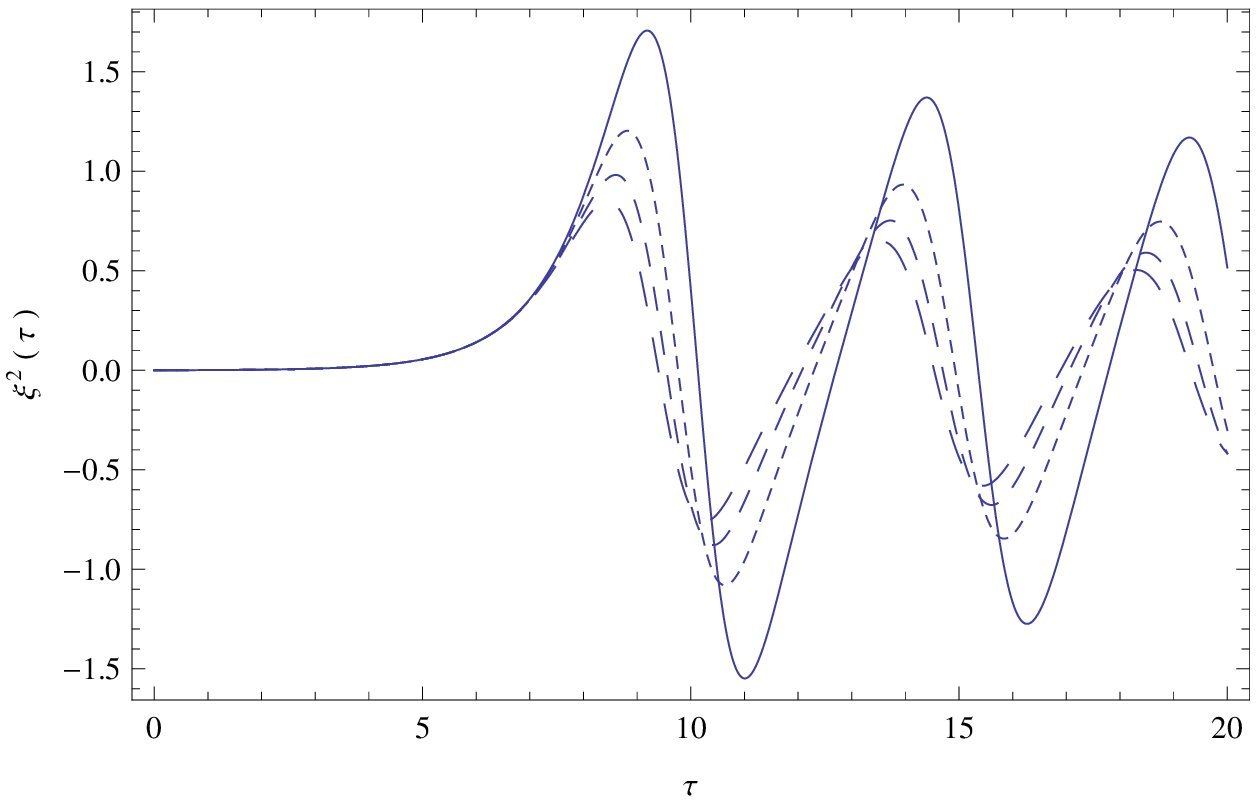}
\caption{ Time variations of the deviation vector components $\protect\xi ^1$
(left figure) and $\protect\xi ^2$ (right figure) for a Universe filled with
a Higgs type scalar field, for different values of the parameter $\protect%
\eta $: $\protect\eta =10$ (solid curve), $\protect\eta =20$ (dotted curve),
$\protect\eta =30$ (short dashed curve), and $\protect\eta =40$ (dashed
curve).}
\label{fig9}
\end{figure*}

In Figs.~\ref{fig9} we present the time variation of the components of the
deviation vector $\xi ^i$, $i=1,2$.

\section{Jacobi stability analysis of the phantom quintessence and tachyonic
cosmological models}

\label{sect4}

In the present Section we will consider the Jacobi stability analysis of the
standard dynamical system formulation of scalar field cosmological models.
In this formulation the Friedmann equations are rewritten in the equivalent
form of a three-dimensional first order dynamical system. Geometrically, we
can describe the solutions of a first order dynamical system as a flow $%
\varphi _{t}:D\subset R^{n}\rightarrow R^{n}$, or, more generally, $\varphi
_{t}:D\subset \mathcal{M}\rightarrow \mathcal{M}$, where $\mathcal{M}$ is a
smooth $n$-dimensional manifold. The canonical lift of $\varphi _{t}$ to the
tangent space $T\mathcal{M}$ is given by $\hat{\varphi}_{t}:T\mathcal{M}%
\rightarrow T\mathcal{M}, \hat{\varphi}_{t}\left( u\right) =\left( \varphi
_{t}\left( u\right) ,\dot{\varphi}_{t}\left( u\right) \right) $, where $\dot{%
\varphi}_{t}\left( u\right) =\frac{\partial \varphi \left( t,u\right) }{%
\partial t}$. Therefore, in order to apply the KCC theory to first order
dynamical systems we must first lift the equations to the tangent bundle.
Mathematically, this is equivalent to simply take the time derivative of the
dynamical system.

\subsection{First order dynamical system formulation of quintessence and
phantom quintessence scalar field cosmological models}

In the following we assume that the energy density and pressure of the
scalar field can be generally represented as
\begin{equation}
\rho _{\phi }=\zeta \frac{1}{2}\dot{\phi}^{2}+V(\phi ),p_{\phi }=\zeta \frac{%
1}{2}\dot{\phi}^{2}-V(\phi ),
\end{equation}%
where $\zeta =\pm 1$. The case $\zeta =+1$ corresponds to the quintessence
fields, while $\zeta =-1$ describes the phantom scalar fields. We assume
that the Universe is filled by ordinary matter with energy density $\rho
_{m} $ and pressure $p_{m}$, and scalar fields. Each component of the model
is characterized by their dimensionless density parameters $\Omega _{m}$ and
$\Omega _{\phi }$, defined as
\begin{equation}
\Omega _{m}=\frac{\rho _{m}}{3H^{2}},\Omega _{\phi }=\frac{\rho _{\phi }}{%
3H^{2}},
\end{equation}%
and satisfying the constraint $\Omega _{m}+\Omega _{\phi }=1$. The
cosmological equations describing this Universe model take the form
\begin{equation}
3H^{2}=\rho _{m}+\rho _{\phi },2\dot{H}=-\left( \rho _{m}+p_{m}+\rho _{\phi
}+p_{\phi }\right) ,
\end{equation}%
and
\begin{equation}
\ddot{\phi}+3H\dot{\phi}+\zeta V^{\prime }(\phi )=0,
\end{equation}
respectively. In the following we assume that the matter obeys a linear
barotropic equation of state of the form $p_{m}=\left( \gamma _{m}-1\right)
\rho _{m}$, where $\gamma _{m}$ is constant, and $1\leq \gamma _{m}\leq 2$.
From the second Friedmann equation we immediately obtain the relation
\begin{equation}
-\frac{2\dot{H}}{H^{2}}=3\gamma _{m}\Omega _{m}+\zeta \frac{\dot{\phi}^{2}}{%
H^{2}}.
\end{equation}%
As basic variables in the first order dynamical description of cosmological
dynamics we introduce the quantities $x$ and $y$, defined as
\begin{equation}
x=\frac{\dot{\phi}}{\sqrt{6}H},y=\frac{\sqrt{V}}{\sqrt{3}H}.  \label{87}
\end{equation}

In these variables the density parameter of the scalar field is given by $%
\Omega _{\phi }=x^{2}+y^{2}$, while the density parameter of the matter can
be written as $\Omega _{m}=1-\left( x^{2}+y^{2}\right) $. Moreover, the energy density and pressure of the scalar field are $p_{\phi}=3H^2\left(\zeta x^2+y^2\right)$ and $p_{\phi}=3H^2\left(\zeta x^2-y^2\right)$, respectively. Instead of the
ordinary time variable $t$ we introduce the new independent variable $\tau $%
, defined as $\tau =\ln a(t)$, giving $d/dt=H\left( d/d\tau \right) $. Then,
by taking the time derivative of $x$ and $y$ we obtain after some simple
transformations \cite{cosm4}
\begin{equation}
\frac{dx}{d\tau }=-3x(1-\zeta x^{2})+\frac{3\gamma _{m}}{2}x\left(
1-x^{2}-y^{2}\right) -\sqrt{\frac{3}{2}}\zeta \frac{V^{\prime }}{V}y^{2},
\end{equation}
\begin{equation}
\frac{dy}{d\tau }=\left[ \frac{3\gamma _{m}}{2}\left( 1-x^{2}-y^{2}\right)
+3\zeta x^{2}\right] y+\sqrt{\frac{3}{2}}\frac{V^{\prime }}{V}xy.
\end{equation}%
The above system is not a closed system since one more equation involving $%
V^{\prime }/V$ is still lacking. Therefore we introduce a third variable $z$%
, defined as $z=V^{\prime }/V$. By taking its derivative with respect to the
time $\tau $ we obtain
\begin{equation}
\frac{dz}{d\tau }=\sqrt{6}xz^{2}\left[ \Gamma (z)-1\right] ,
\end{equation}%
where $\Gamma =VV^{\prime \prime }/V^{\prime 2}$. In order to lift this
dynamical system to the tangent bundle we relabel the coordinates as
\begin{equation}
x=x^{1},y=x^{2},z=x^{3},y^{1}=\frac{dx^{1}}{d\tau },y^{2}=\frac{dx^{2}}{%
d\tau },y^{3}=\frac{dx^{3}}{d\tau }.
\end{equation}%
Then, by taking the derivatives with respect to $\tau $ of the first order
cosmological dynamical system we obtain the equivalent second order system
\begin{eqnarray}
&&\frac{d^{2}x^{1}}{d\tau ^{2}}=-\frac{3}{2} \Bigg\{ x' \Bigg(-\gamma_m+3 (\gamma_m-2 \zeta ) x^2+\gamma_m y^2+2\Bigg)- \notag \\
&&y \Bigg(6 \gamma_m x y'+\sqrt{6} \zeta  \left(2 z y'+y z'\right)\Bigg)\Bigg\}=-2G^{1},  \notag
\\
&&
\end{eqnarray}%
\begin{eqnarray}
&&\frac{d^{2}x^{2}}{d\tau ^{2}}=\frac{1}{2}\left[ \sqrt{6}%
x^{2}x^{3}-6(\gamma _{m}-2\zeta )x^{1}x^{2}\right] y^{1}+  \notag \\
&&\frac{1}{2}\left[ 3\gamma _{m}-3(\gamma _{m}-2\zeta )\left( x^{1}\right)
^{2}+\sqrt{6}x^{1}x^{3}-9\gamma _{m}\left( x^{2}\right) ^{2}\right] y^{2}+
\notag \\
&&\sqrt{\frac{3}{2}}x^{1}x^{2}y^{3}=-2G^{2},
\end{eqnarray}%
\begin{eqnarray}
&&\frac{d^{2}x^{3}}{d\tau ^{2}}=\sqrt{6}(x^{3})^{2}\left[ \Gamma (x^{3})-1%
\right] y^{1}+\sqrt{6}x^{3}\Bigg\{x^{1}x^{3}\Gamma ^{\prime }(x^{3})+  \notag
\\
&&2x^{1}\left[ \Gamma (x^{3})-1\right] \Bigg\}y^{3}=-2G^{3}.
\end{eqnarray}%
The KCC geometric quantities (non-linear connection, deviation curvature
tensor, and geodesic deviation equations) describing the Jacobi stability
properties of the phantom quintessence cosmological model are presented in
Appendix \ref{appA}.

\subsection{The phantom scalar field with power law potential}

We assume that the potential of the phantom quintessence scalar field is
given by a power law function, so that $V(\phi )=V_{0}\phi ^{\alpha }$,
where $\alpha $ is a constant. Then it follows immediately that $z=V^{\prime
}/V=\alpha /\phi $, $\Gamma =(\alpha -1)/\alpha =\mathrm{constant}$, and $%
\Gamma ^{\prime }(z)=0$. An important cosmological indicator is represented
by the parameter of the total equation of state of the matter $w$, which for
dust is defined as
\begin{equation}
w=\frac{p_{\phi }}{\rho _{\phi }+\rho _{m}}=\frac{\zeta x^{2}-y^{2}}{\left(\zeta -1\right)x^2+1 }.
\end{equation}

\begin{figure*}[tb]
\centering
\includegraphics[width=8.15cm]{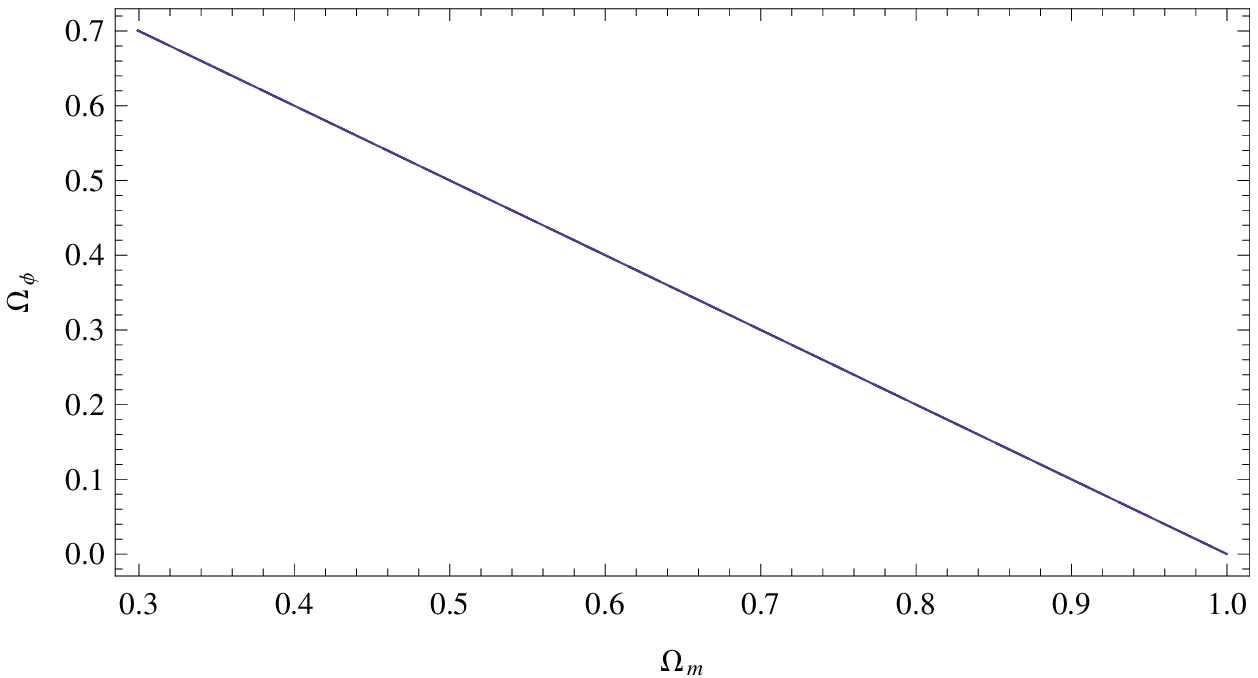} %
\includegraphics[width=8.15cm]{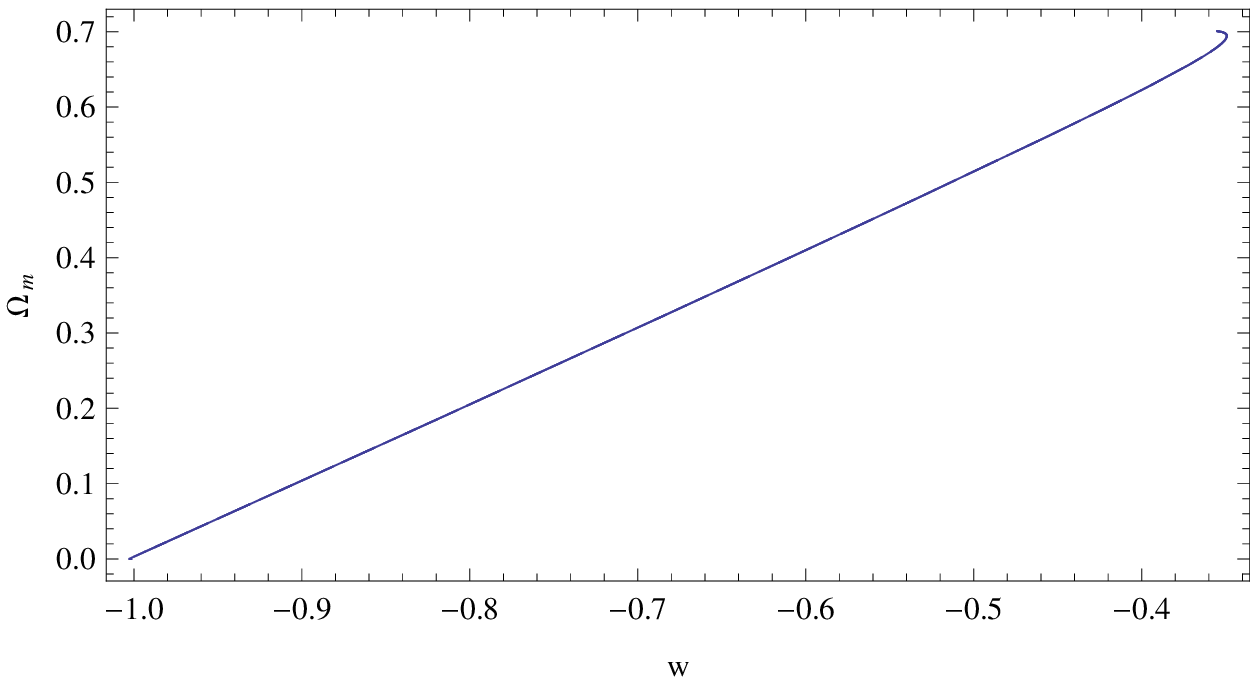}
\caption{ Variation of the density parameter of the phantom quintessence
field as a function of the density parameter of the dust cosmological matter
(left figure) and of the parameter $w$ of the total equation of state (right
figure) for different values of $\protect\alpha $: $\protect\alpha =0.8$
(solid curve), $\protect\alpha =1$ (dotted curve), $\protect\alpha =1.2$
(short dashed curve), and $\protect\alpha =1.4$ (dashed curve),
respectively. }
\label{fig10}
\end{figure*}

The variations of the density parameter of the phantom quintessence field $%
\Omega _{\phi}$ and of the parameter of the total equation of state are
represented, as a function of the dust ($\gamma _m=1$) cosmological matter
density parameter in Fig.~\ref{fig10}. The initial conditions used to
integrate the cosmological dynamical system are $x(0)=0.28$, $y(0)=0.47$,
and $z(0)=0.1$, respectively.

In this model the Universe starts its evolution in the large time limit from
a matter dominated phase, with $\Omega _m=1$. During the cosmological
expansion the role of the scalar field becomes dominant, and the Universe
reaches the present day in a state of accelerated expansion, with $\Omega
_{\phi}=0.75$, and $\Omega _m=0.25$, respectively. On the other hand, the de
Sitter phase with $w=-1$ is reached only for vanishing ordinary matter
density, when the Universe is fully dominated by the phantom quintessence
field. It is also important to note that the cosmological evolution is
basically independent on the numerical values of the exponent $\alpha $ in
the scalar field potential.

\begin{figure*}[tb]
\centering
\includegraphics[width=8.15cm]{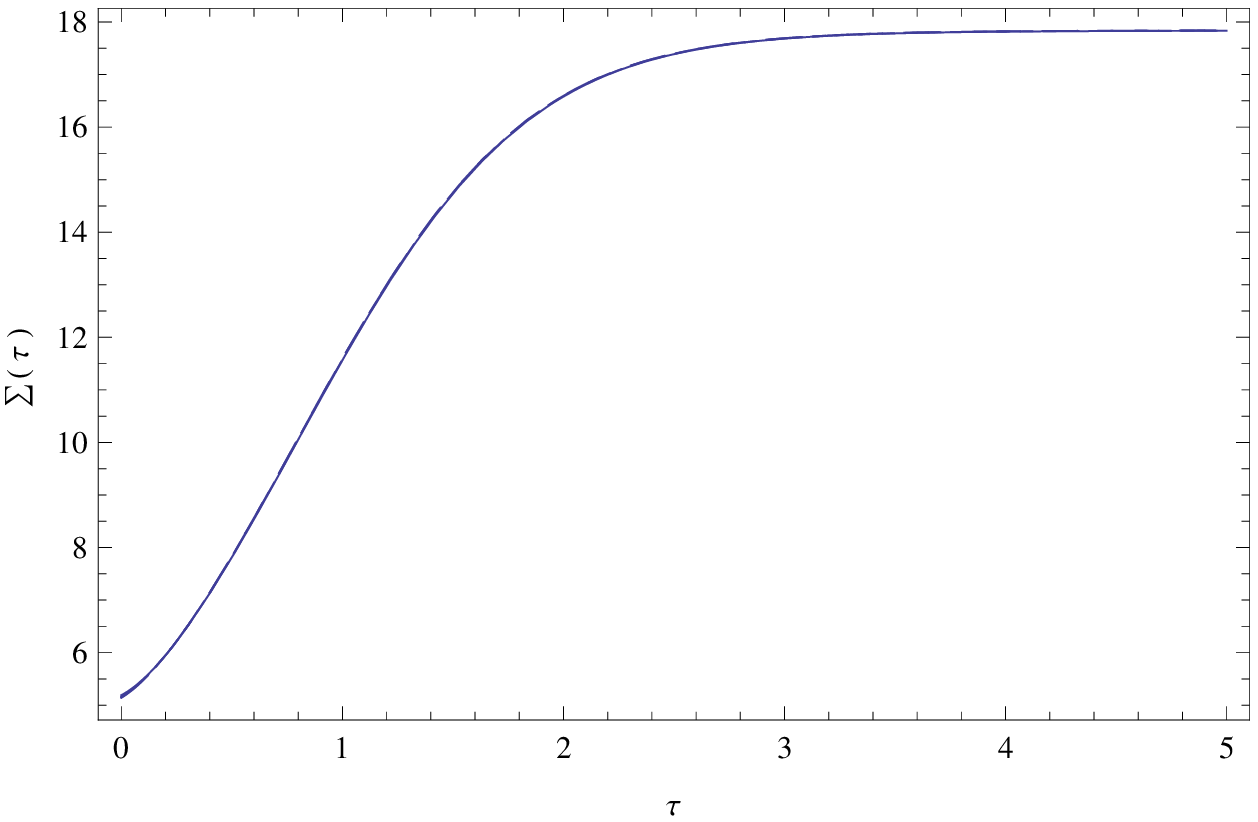} %
\includegraphics[width=8.15cm]{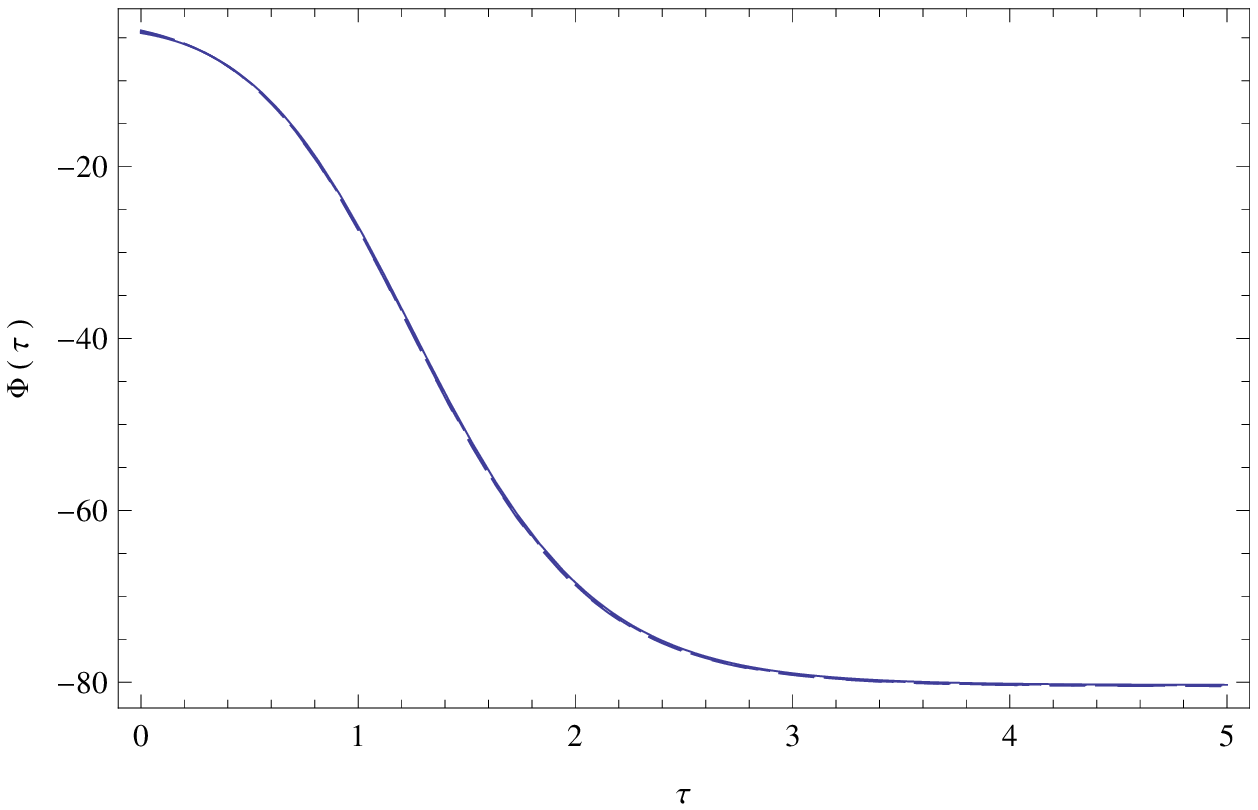} %
\includegraphics[width=8.15cm]{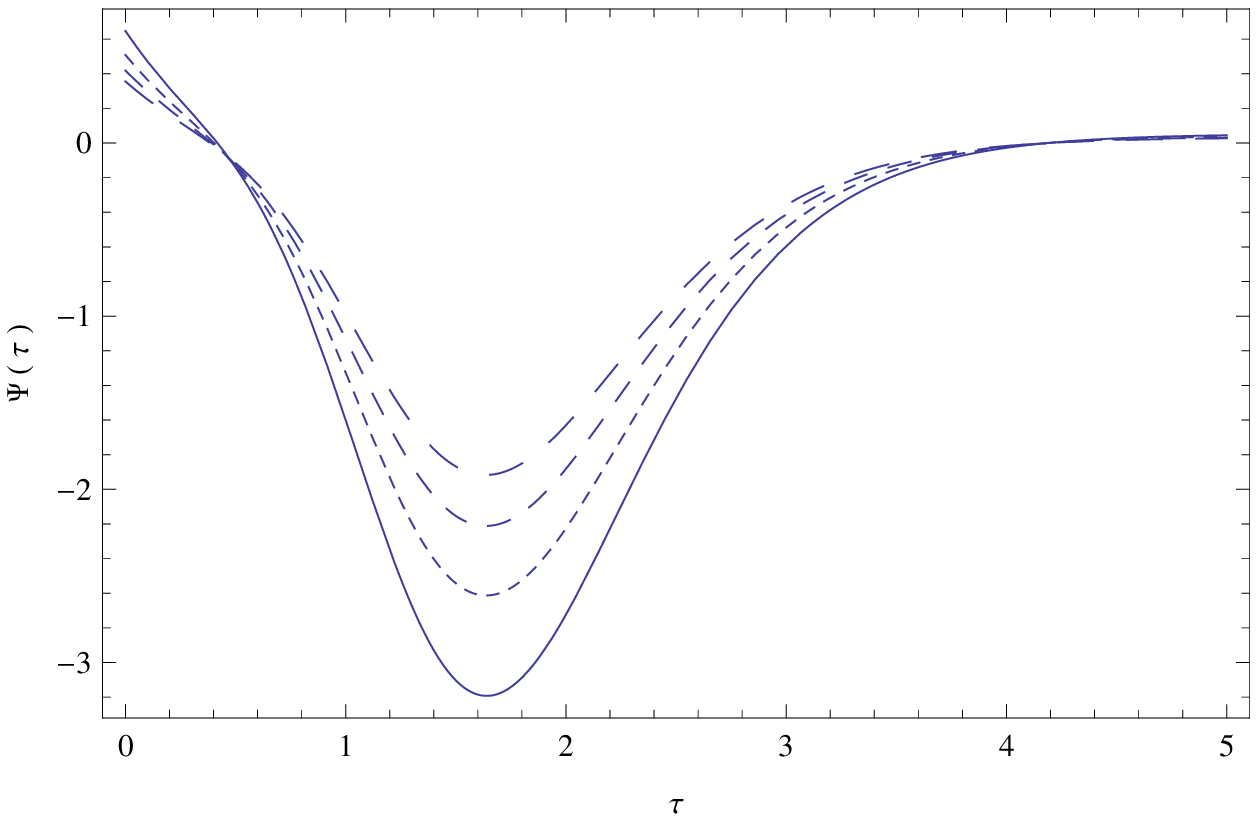} %
\includegraphics[width=8.15cm]{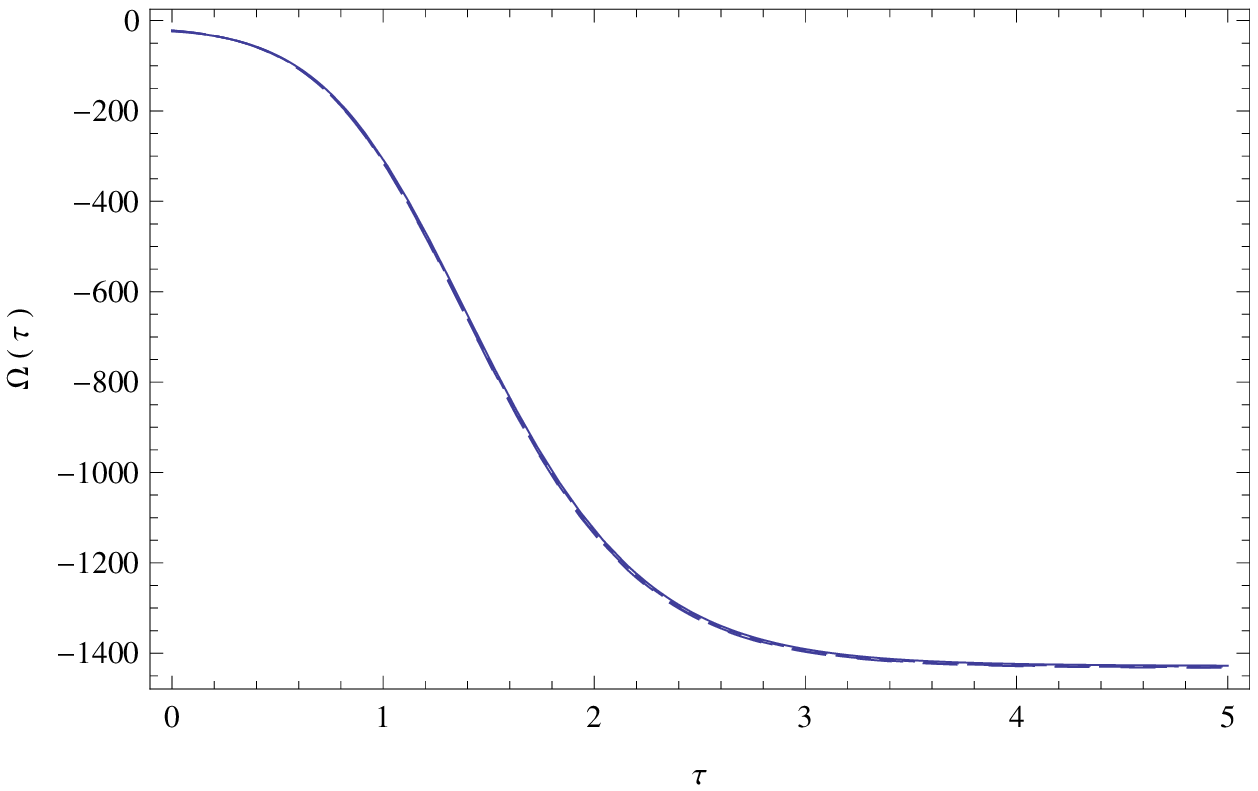}
\caption{ Time variation of the quantities $\left(\Sigma,
\Phi,\Psi,\Omega\right)$, giving the Jacobi stability of the phantom
quintessence field cosmological model with power law potential for different
values of $\protect\alpha $: $\protect\alpha =0.8$ (solid curve), $\protect%
\alpha =1$ (dotted curve), $\protect\alpha =1.2$ (short dashed curve), and $%
\protect\alpha =1.4$ (dashed curve), respectively.}
\label{fig11}
\end{figure*}

The conditions of the Jacobi stability of the phantom quintessence
cosmological model in its three-dimensional dynamic system representation
are given by the four conditions that must be satisfied by the quantities $%
\left(\Sigma, \Phi,\Psi,\Omega\right)$, which are functions of the
components of the deviation curvature tensor, and which are presented in
Eqs.~(\ref{3dim1})-(\ref{3dim4}), for different values of $\alpha $. The
time variations of $\left(\Sigma, \Phi,\Psi,\Omega\right)$ are represented
in Fig.~\ref{fig11}.

As one can see from Fig.~(\ref{fig11}), the Jacobi stability condition $%
\Omega >0$ is not satisfied during the entire time evolution of this
cosmological model. Therefore phantom quintessence scalar field cosmological
models with power law potential are \textit{Jacobi unstable} for all time
intervals.

\subsection{Jacobi stability of tachyon field cosmological models}

Tachyon scalar fields have been proposed as possible candidates to explain
both inflation and the late acceleration of the Universe \cite{tach1,tach2}.
It is also possible for the tachyonic field to trigger an the inflationary
expansion, and at a later time generate a non-relativistic fluid, which
could account for the existence of dark matter. A tachyonic scalar field is
described by the Lagrangian $L=-V(\phi) \sqrt{1-\dot{\phi}^2}$, and it can
be described in terms of an effective energy density and pressure, given by
\begin{equation}
\rho _{\phi}=\frac{V(\phi)}{\sqrt{1-\dot{\phi}^2}},p_{\phi}=-V(\phi)\sqrt{1-%
\dot{\phi}^2}.
\end{equation}
The Friedmann equations are given by
\begin{equation}
3H^2=\rho _{m}+\frac{V(\phi)}{\sqrt{1-\dot{\phi}^2}}, 2\dot{H}%
+3H^2=-p_m+V(\phi)\sqrt{1-\dot{\phi}^2},
\end{equation}
while the Klein-Gordon type equation satisfied by the scalar field is
\begin{equation}
\ddot{\phi}+3H\dot{\phi}\left(1-\dot{\phi}^2\right)+\frac{V^{\prime }(\phi)}{%
V(\phi)}\left(1-\dot{\phi}^2\right)=0.
\end{equation}

The parameter of the equation of state of the tachyon field is given by
\begin{equation}
w=\frac{p_{\phi}}{\rho _{\phi}}=-\left(1-\dot{\phi}^2\right).
\end{equation}
By adopting again for the matter the linear barotropic equation of state $%
p_m=\left(\gamma _m-1\right)\rho _m$, $1\leq \gamma _m\leq 2$, from the
Friedmann equation we obtain first
\begin{equation}
\frac{2\dot{H}}{H^2}=-3\gamma _m\Omega _m-\frac{V(\phi)}{\sqrt{1-\dot{\phi}^2%
}}\frac{\dot{\phi}^2}{H^2}.
\end{equation}
To formulate the cosmological evolution equation as a dynamical system we
introduce a set of variables $(\tau, x,y,z)$ defined as
\begin{equation}
\tau =\ln a, x=\dot{\phi},y=\frac{\sqrt{V}}{\sqrt{3}H},z=\frac{V^{\prime
}(\phi)}{V^{3/2}(\phi)}.
\end{equation}
In these variables the density parameters of the scalar field and of the
matter are given by
\begin{equation}
\Omega _{\phi}=\frac{y^2}{\sqrt{1-x^2}},\Omega _m=1-\frac{y^2}{\sqrt{1-x^2}}.
\end{equation}

Then, by taking the derivative of $x$ and $y$ with respect to $\tau $ we
obtain
\begin{equation}
\frac{dx}{d\tau }=-\sqrt{3}\left( 1-x^{2}\right) \left( \sqrt{3}x+yz\right) ,
\end{equation}%
\begin{equation}
\frac{dy}{d\tau }=\frac{\sqrt{3}}{2}%
y\left\{ xyz+\sqrt{3}\left[ \gamma _{m}+\frac{y^{2}}{\sqrt{1-x^{2}}}\left(
x^{2}-\gamma _{m}\right) \right] \right\} ,
\end{equation}
\begin{equation}
\frac{dz}{d\tau }=\frac{3\sqrt{3}}{2}xyz^{2}\left[ \Gamma (z)-1\right] ,
\end{equation}%
where $\Gamma (z)=(2/3)VV^{\prime \prime }/V^{\prime }{^{2}}$. By lifting
the cosmological field equations to the tangent bundle we obtain
\begin{eqnarray}
\frac{d^{2}x}{d\tau ^{2}} &=&2\sqrt{3}xx^{\prime }\left( \sqrt{3}x+yz\right)
-\sqrt{3}\left( 1-x^{2}\right) \Bigg(\sqrt{3}x^{\prime }+  \notag \\
&&y^{\prime }z+yz^{\prime }\Bigg)=-2G^{1},
\end{eqnarray}%
\begin{eqnarray}
\frac{d^{2}y}{d\tau ^{2}} &=&\frac{\sqrt{3}}{2}\Bigg\{3y\Bigg[\gamma
_{m}\left( -\frac{xy^{2}x^{\prime }}{\left( 1-x^{2}\right) ^{3/2}}-\frac{%
2yy^{\prime }}{\sqrt{1-x^{2}}}\right)  \notag \\
&&+2xy^{2}x^{\prime 2}yy^{\prime }\Bigg]+y^{2}zx^{\prime }+3y^{\prime }\times
\notag \\
&&\Bigg[\gamma _{m}\left( 1-\frac{y^{2}}{\sqrt{1-x^{2}}}\right) +x^{2}y^{2}%
\Bigg]+  \notag \\
&&2xyzy^{\prime 2}z^{\prime }\Bigg\}=-2G^{2},
\end{eqnarray}%
\begin{eqnarray}
\frac{d^{2}z}{d\tau ^{2}} &=&\frac{3}{2}\sqrt{3}z\Bigg\{(\Gamma (z)-1)\left[
y\left( zx^{\prime }+2xz^{\prime }\right) +xzy^{\prime }\right] +  \notag \\
&&xyzz^{\prime }\Gamma ^{\prime }(z)\Bigg\}=-2G^{3}.
\end{eqnarray}

The KCC geometric quantities (non-linear connection, deviation curvature
tensor, and geodesic deviation equations) describing the Jacobi stability
properties of the tachyon scalar field cosmological model are presented in
Appendix \ref{appB}.

\subsubsection{Power law potential tachyonic scalar field cosmological models%
}

\begin{figure*}[tb]
\centering
\includegraphics[width=8.15cm]{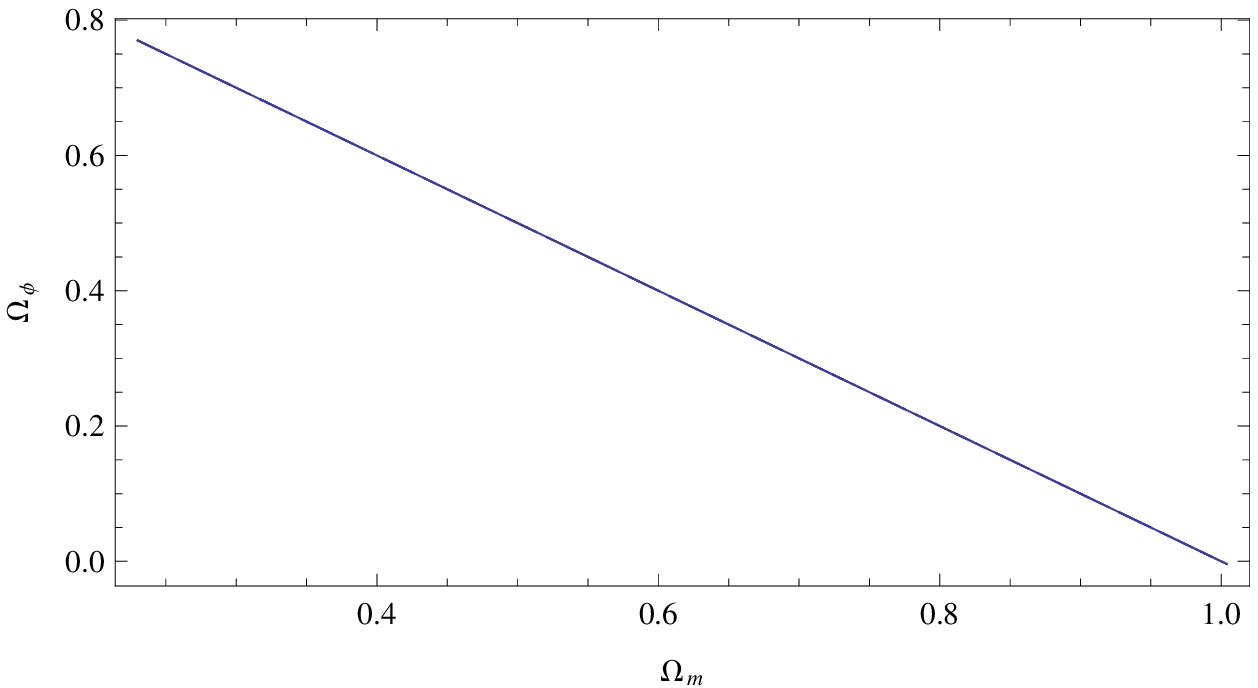} %
\includegraphics[width=8.15cm]{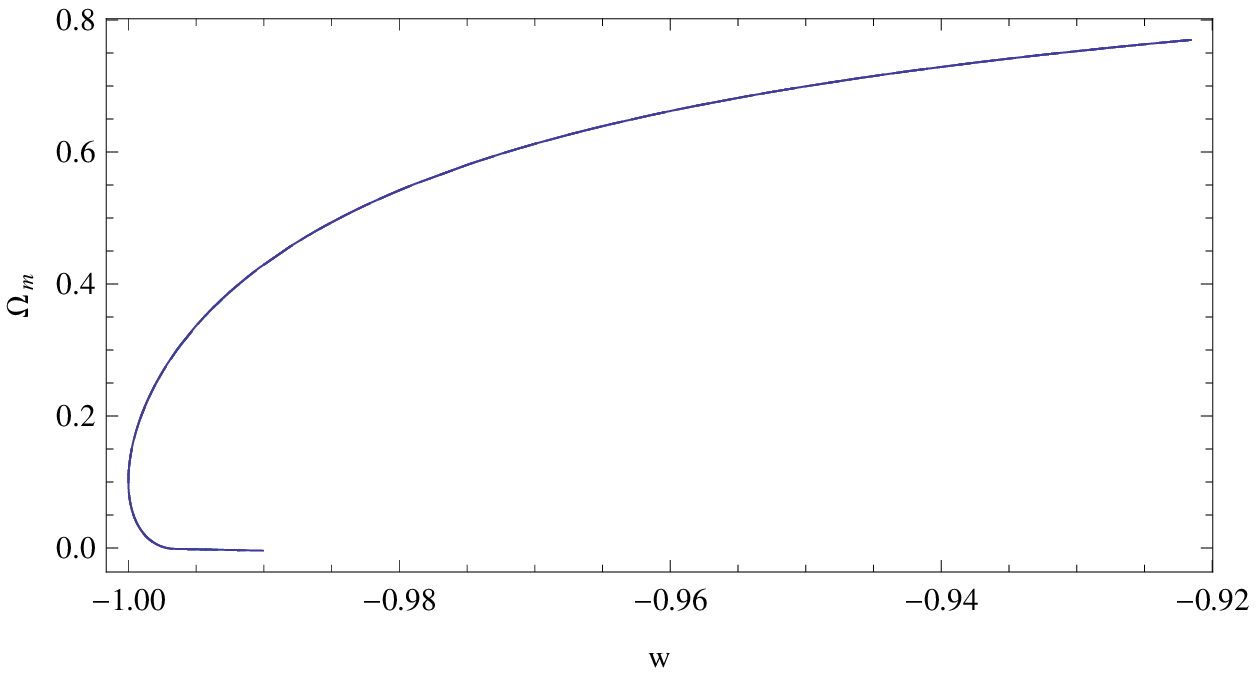}
\caption{ Variation of the density parameter of the tachyonic scalar field
with power law potential as a function of the density parameter of the dust
cosmological matter ($\protect\gamma _m=1$) (left figure) and of the
parameter $w$ of the total equation of state (right figure) for different
values of $\protect\alpha $: $\protect\alpha =0.8$ (solid curve), $\protect%
\alpha =1$ (dotted curve), $\protect\alpha =1.2$ (short dashed curve), and $%
\protect\alpha =1.4$ (dashed curve), respectively.}
\label{fig12}
\end{figure*}

In the following we assume again that the potential of the tachyonic scalar
field is power law type, with $V(\phi)=V_0\phi ^{\alpha}$, where $V_0$ and $%
\alpha $ are constants. This choice immediately gives $\Gamma
=(2/3)\left(\alpha -1\right)/\alpha $. The parametric dependence of the
density parameter $\Omega _{\phi}$ of the tachyonic scalar field with power
law potential on the matter energy density $\Omega _m$ is represented in the
left panel of Fig.~\ref{fig12}. The parametric variation of the matter
density parameter $\Omega _m$ as a function of the parameter of the total
equation of state of matter $w$ is shown in the right panel of Fig.~\ref%
{fig12}.

\begin{figure*}[tb]
\centering
\includegraphics[width=8.15cm]{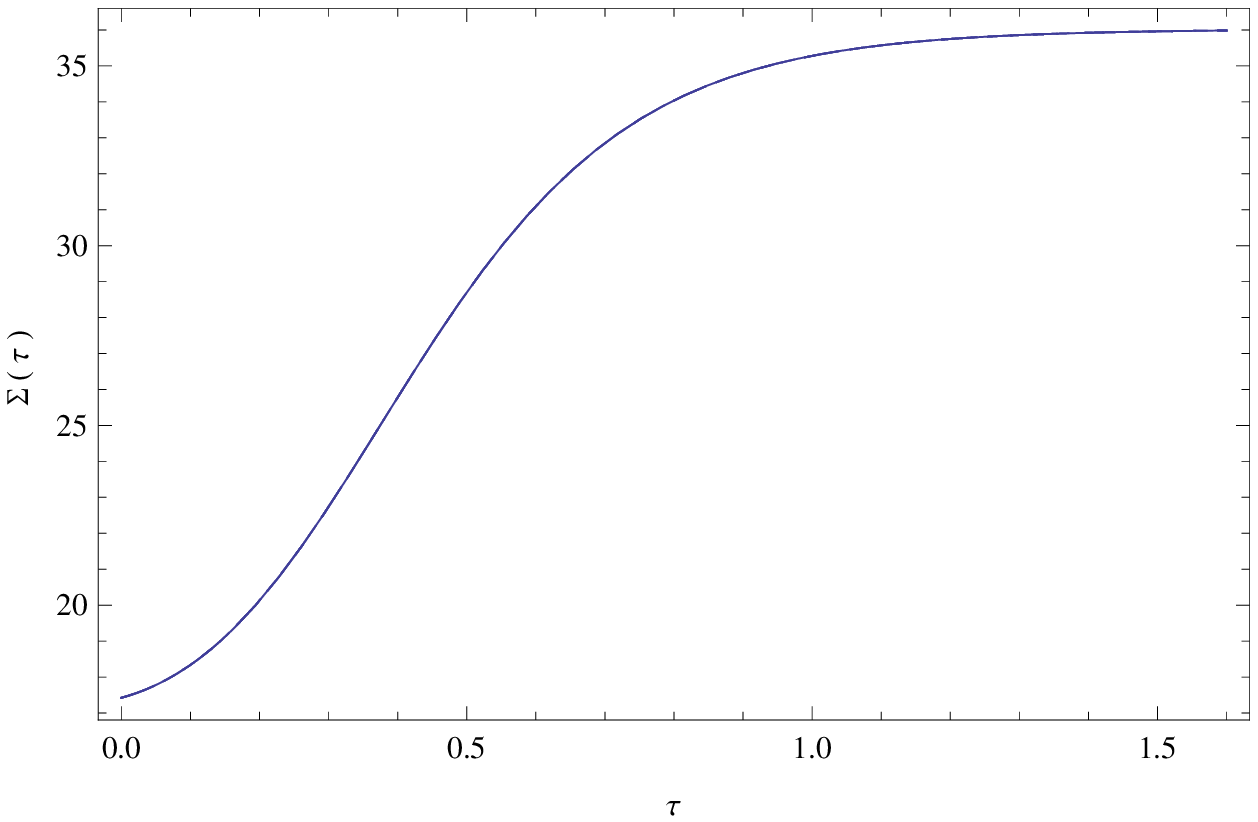} %
\includegraphics[width=8.15cm]{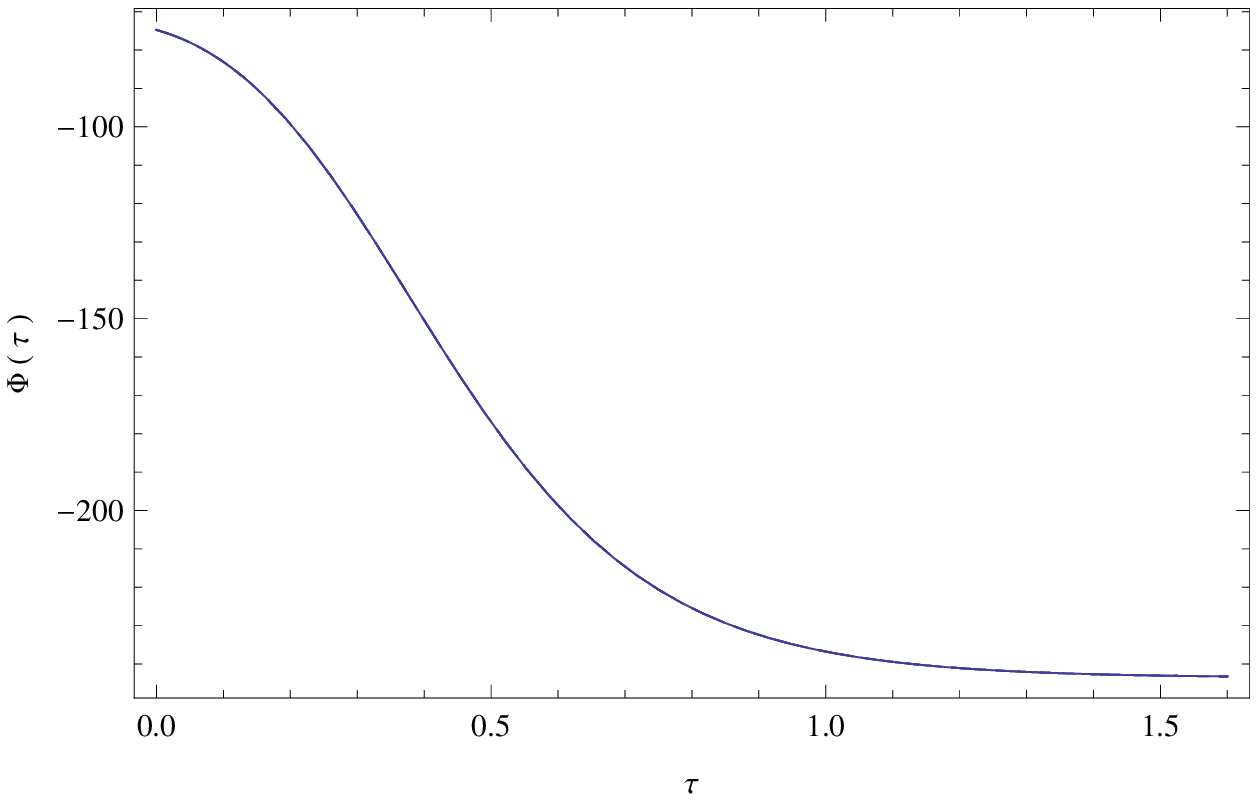} %
\includegraphics[width=8.15cm]{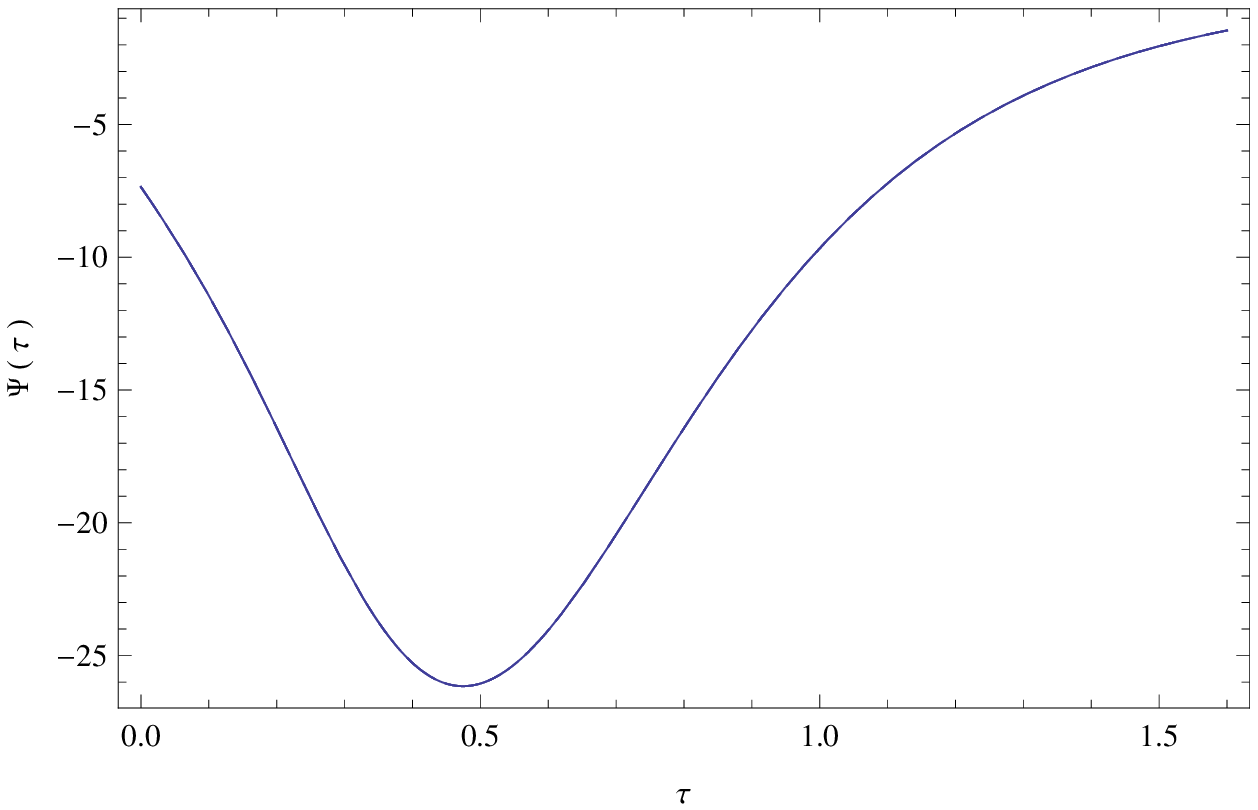} %
\includegraphics[width=8.15cm]{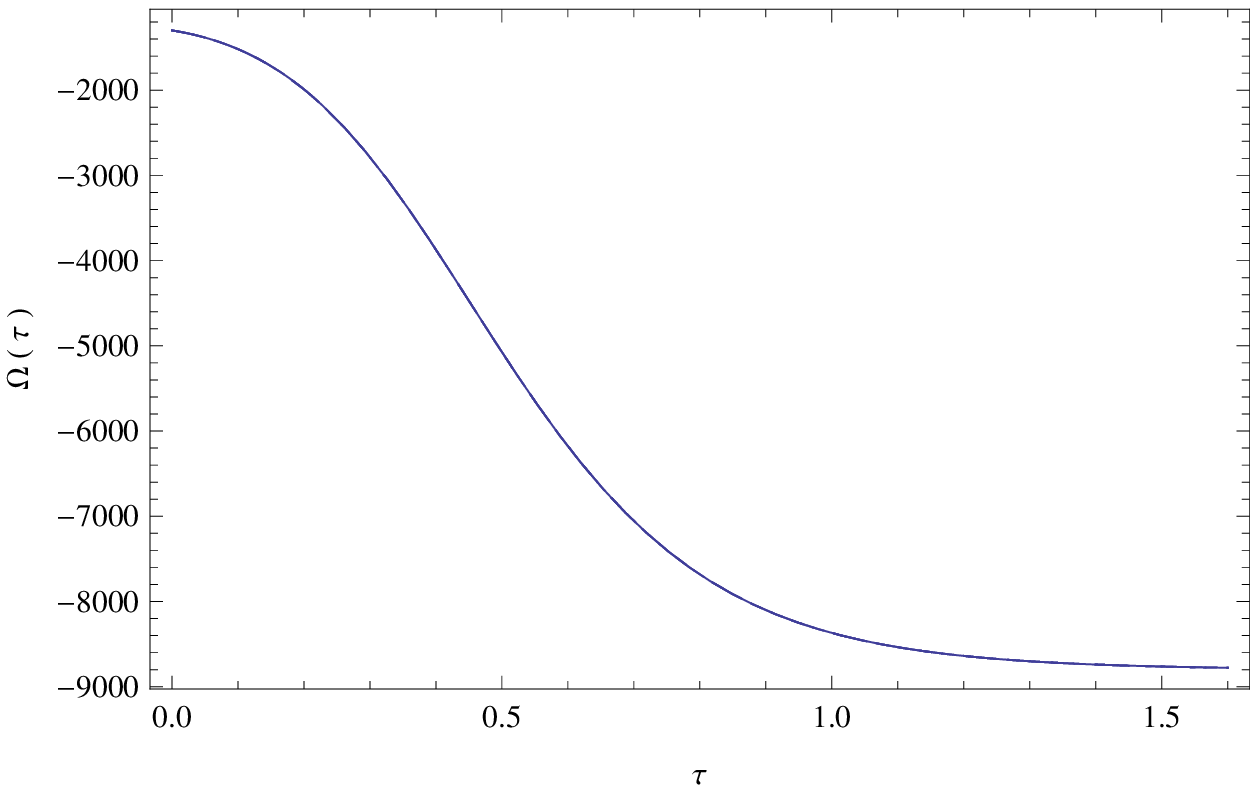}
\caption{ Time variation of the quantities $\left(\Sigma,
\Phi,\Psi,\Omega\right)$, giving the Jacobi stability of the tachyon scalar
field cosmological model with power law potential for different values of $%
\protect\alpha $: $\protect\alpha =0.8$ (solid curve), $\protect\alpha =1$
(dotted curve), $\protect\alpha =1.2$ (short dashed curve), and $\protect%
\alpha =1.4$ (dashed curve), respectively.}
\label{fig14}
\end{figure*}

In order to numerically integrate the dynamical system corresponding to the
tachyon scalar field model we have used the initial conditions $x(0)=0.28$, $%
y(0)=0.45$, and $z(0)=0.1$, respectively, and we have assumed that the
ordinary matter in the Universe is in the form of zero pressure dust, with $\gamma _{m}=1$. There is a linear relation between $\Omega _{\phi }$
and $\Omega _{m}$. The Universe starts its cosmological evolution in a
matter dominated phase, with $\Omega _{m}=1$, $\Omega _{\phi }=0$, with
decelerating expansion. The energy density of the tachyon field increases in
time, and the Universe enters in a de Sitter phase, with the matter density
becoming negligibly small. The parameter of the total equation of state
satisfies the condition $w<0$ during the entire cosmological evolution. It
is also interesting to note that the time evolution of this model is
basically independent on the numerical values of the exponent $\alpha $ in
the power law potential. The time variations of $\left( \Sigma ,\Phi ,\Psi
,\Omega \right) $ describing the Jacobi stability of the tachyon scalar
field cosmological model are represented in Fig.~\ref{fig14}.

As one can see from Fig.~(\ref{fig14}), the Jacobi stability condition $%
\Omega >0$ is not satisfied during the entire cosmological evolution period
of this scalar field model. Therefore it follows that tachyon scalar field
cosmological models with power law potential are \textit{Jacobi unstable}
for all time intervals. This result is independent on the numerical values
of the exponent $\alpha $ in the tachyon scalar field potential.

\section{Discussions and final remarks}

\label{sect5}

In the present paper we have investigated the Jacobi stability properties of
the scalar field cosmological models by using the KCC theory, which
represents a powerful mathematical method for the analysis of dynamical
systems. Scalar field cosmological models represent a non-trivial testing
object for studying non-linear effects in the framework of general
relativity. From a mathematical point of view the Jacobi (in)stability
represents a natural generalization of the (in)stability of the geodesic
flow on a differentiable manifold, endowed with a Riemannian or Finslerian
type metric to a \textit{non-metric setting}. The KCC theory can be applied
to scalar field cosmological models that can be formulated mathematically as
sets of second order ordinary non-linear differential equations. Then the
geometric invariants associated to this system (nonlinear and Berwald
connections), and the deviation curvature tensor, as well as its
eigenvalues, can be explicitly obtained. The time evolution of the
components of the deviation vector can also be obtained by explicitly
solving the geodesic deviation equations.

The Jacobi stability, and its theoretical foundation, the KCC theory, offers
an alternative approach to the "classical" Lyapunov approach, by
investigating \textit{the deviations of the entire trajectory} of the
cosmological evolution equations with respect to the nearby ones under the
effects of a small perturbation. In the framework of general relativity we
may call the applications of the KCC theory to the study of the
gravitational fields as a "second geometrization", in which already
geometric quantities are supplemented by additional geometric structures.
Hence general relativistic cosmological models can be described in geometric
terms originating from their dynamical system structure, with these new
geometric structure fully determined by the underlying Riemannian geometry,
and the physical properties of the scalar fields (their self-interaction
potential). The stability properties of the perturbations of a given
trajectory describing the cosmological evolution are determined by the
properties of the curvature deviation tensor, a geometric quantity
constructed from the connections (non-linear and Berwald) associated to the
dynamical system describing the cosmological evolution. It is important to
note that the KCC theory can be directly applied to systems of second order
differential equations, which can be interpreted geometrically as the paths
(or geodesics) associated to a semispray. In investigating the Jacobi
stability of cosmological models we have followed two approaches. Since the
cosmological evolution equations (the Friedmann equations) are second order
differential equations, the KCC theory can be naturally and directly applied
to study the stability of the cosmic evolution. As a first step one obtains
the two non-zero components of the non-linear connection, with the $N_2^1$
component depending on the product of the scale factor and of the time
derivative of the field, while the $N_2^2$ component depends on the energy
density of the scalar field, as well as of the scalar field potential. After
obtaining the components of the deviation curvature tensor we have
formulated the general condition of the stability of the scalar field
cosmological models, which is determined by two inequalities involving the
second and the first derivative of the scalar field potential, the energy
density of the field, and the time variation of the scalar field itself.

As an applications of the developed formalism we have considered two scalar
field models, both being relevant for the study of both the early and the
late stages of the cosmological evolution. The first case we did consider is
the scalar field with exponential potential. We have studied in detail the
KCC geometric properties of this model. It turns out that the Jacobi
stability condition, which can be expressed in terms of the components of
the deviation curvature tensor is not satisfied during the cosmological
evolution, and that the Universe described by the exponential potential
scalar field is in a Jacobi unstable state. This result is independent on
the numerical values of the parameter $\lambda $, describing the properties
of the potential, and can also be inferred from the behavior of the
components of the deviation vector $\xi ^i$, with $\xi ^1$ diverging
exponentially in time. As a second case we have considered the case of the
Higgs type potential. For this potential the KCC geometric quantities show a
complex behavior. After a period in which the scalar field and the potential
are almost constant, the field starts to oscillate, with the amplitude of
the oscillations decreasing in time. This behavior of the Higgs field is
also reflected in the behavior of the components of the deviation curvature
tensor, which are also some oscillating functions. The Jacobi stability of
this cosmological model strongly depends on the numerical value of the
parameter $\eta =\lambda /4M^2$. For small values of $\eta $, the Universe
evolves between successive Jacobi stable and unstable states. With the
increase of the numerical value of $\eta $ the time intervals in which the
Universe is Jacobi stable decrease quickly, and for large values of $\eta $
the Universe is in a Jacobi unstable state during its entire cosmological
evolution.

As a second approach for the study of the Jacobi stability of scalar field
cosmologies we have considered the first order dynamical system formulation
of scalar field evolution equations. In this approach by introducing a new
set of variables, expressed in terms of the square root of the potential,
the time derivative of the scalar field, and the Hubble function,
respectively, the Friedmann equations in the presence of scalar fields can
be reformulated as a first order dynamical system, consisting of three
highly non-linear ordinary differential equations. In order to apply the KCC
theory this dynamical system must be lifted to the tangent bundle, and
formulated as a second order differential system. We have analyzed, by using
this approach, two specific scalar field models, the phantom quintessence
and the tachyon scalar field with power law potentials. It turns out that
for this choice of the potential both scalar field models are Jacobi
unstable. The power law potential gives a very simple form for the function $%
\Gamma (z)$, which takes constant values during the cosmological evolution.
This situation is similar to the case of the exponential potential, and
leads to a significant simplification of the mathematical formalism.

We have started our study of the applications of the KCC theory to
cosmological problems with the investigation of the standard mater dominated
cosmological models in their dynamical system formulation. We have studied
the Jacobi stability of the \textit{critical points} of different models,
and we have shown that they are Jacobi unstable. A full comparison between
the Jacobi and Lyapunov properties of the critical points for second order
systems was given in \cite{Sa05} and \cite{Sa05a}, and hence we will not
discuss in detail this relation. However, this study of the critical points
of matter dominated cosmological models also shows the fundamental
differences between the Lyapunov stability and KCC theories: while Lypunov
stability is mostly restricted to the study of critical points, the KCC
theory has the potential of investigating the deviation of the full
trajectory during the entire period of the cosmological evolution. Therefore
we can consider a Lyapunov stability analysis of steady states (called the
linear analysis), and a "Lyapunov type" stability analysis of the whole
trajectory (the KCC or Jacobi stability analysis), and these two methods are
\textit{complementary} but \textit{distinct} to each other.

The KCC theory also introduces the first set of KCC invariants $\epsilon ^i$%
, $i=1,...,n$, giving the contravariant KCC derivative of the vector field $%
y^i$. The first KCC invariant can be interpreted as an external force. We
did not study in detail the time evolution of the first KCC invariant, since
its properties are not directly related to the stability issues that were
our main points of interest.

In the present paper we have performed a stability analysis of the scalar
field cosmological models, in which we have considered a description of the
deviations of the whole trajectories of the differential system describing
the cosmological dynamics, and we have provided some basic theoretical and
computational tools for this study. Further investigations of the Jacobi
stability properties of cosmological models may provide some methods for
discriminating between different evolutionary scenarios, as well as for the
better understanding of some other fundamental processes, like, for example,
structure formation, that played an essential role in the evolution of our
Universe.

\paragraph*{{\bf Conflict of interest}}

The authors declare that there is no conflict of interest regarding the publication of this paper.

\appendix

\section{The non-linear connection, the curvature deviation tensor and the
geodesic deviation equations for the phantom quintessence scalar field
cosmological model}

\label{appA}

The coefficients of the non-linear connection associated to the dynamical
system describing the phantom quintessence cosmological model are given by

\begin{eqnarray*}
N_{1}^{1}&=&-3 \gamma_m x^2+6 \zeta  x^2-3 \left(1-\zeta  x^2\right) +\\
&& \frac{3}{2} \gamma_m \left(-x^2-y^2+1\right),
\end{eqnarray*}
\begin{eqnarray*}
N_{1}^{2}=-3 \gamma_m x y-\sqrt{6} \zeta  y z,\;\; N_{1}^{3}=-\sqrt{\frac{3}{2%
}} \zeta y^2,
\end{eqnarray*}
\begin{eqnarray*}
N_{2}^{1}=y (6 \zeta x-3 \gamma_m x)+\sqrt{\frac{3}{2}} y z,
\end{eqnarray*}
\begin{eqnarray*}
N_{2}^{2}&=&3 \zeta x^2+\frac{3}{2} \gamma_m \left(-x^2-y^2+1\right) +\sqrt{%
\frac{3}{2}} x z-3 \gamma_m y^2,
\end{eqnarray*}
\begin{eqnarray*}
N_{2}^{3}=\sqrt{\frac{3}{2}} x y, N_{3}^{1}=\sqrt{6} z^2 \left[\Gamma (z)-1%
\right], N_{3}^{2}=0,
\end{eqnarray*}
\begin{eqnarray*}
N_{3}^{3}=\sqrt{6} x z^2 \Gamma ^{\prime }(z)+2 \sqrt{6} x z \left[\Gamma
(z)-1\right].
\end{eqnarray*}

The first KCC invariants $\epsilon ^i$, $i=1,2,3$ are given by
\begin{eqnarray*}
\epsilon^{1}&=&\frac{3}{2} \Bigg\{x' \left(\gamma_m-3 (\gamma_m-2 \zeta ) x^2-\gamma_m y^2-2\right) -\\
&& y \left(6 \gamma_m x y'+\sqrt{6} \zeta  \left(2 z y'+y z'\right)\right)\Bigg\},
\end{eqnarray*}
\begin{eqnarray*}
\epsilon^{2}&=&\frac{1}{2} \Bigg\{x \left(\sqrt{6} \left(z y'+y z'\right)-6 (\gamma_m-2 \zeta ) y x'\right) +\\
&& \sqrt{6} y z x'-3 (\gamma_m-2 \zeta ) x^2 y'-9 \gamma_m y^2 y'+3 \gamma_m y'\Bigg\},
\end{eqnarray*}
\begin{eqnarray*}
\epsilon^{3}=\sqrt{6} z \left((\Gamma (z)-1) \left(z x^{\prime }+2 x
z^{\prime }\right)+x z z^{\prime }\Gamma ^{\prime }(z)\right).
\end{eqnarray*}

All Berwald connection components are zero here. \newline
The components of the curvature deviation tensor are
\begin{eqnarray*}
P_{1}^{1}&=&\frac{1}{4} \Bigg\{9 \left(4 (\gamma_m-2 \zeta ) x x'+\left(\gamma_m-3 (\gamma_m-2 \zeta ) x^2-2\right)^2\right) + \\
&& 6 y^2 \Bigg(-3 (\gamma_m-2) \gamma_m+15 \gamma_m (\gamma_m-2 \zeta ) x^2 -\\
&& \sqrt{6} \left(-2 \gamma_m \zeta +\gamma_m+4 \zeta ^2\right) x z-2 \zeta  z^2 \Gamma (z)\Bigg) + \\
&& 12 \gamma_m y y'+9 \gamma_m^2 y^4\Bigg\},
\end{eqnarray*}
\begin{eqnarray*}
P_{1}^{2}&=&3 \gamma_m y x'+3 x \Bigg[\gamma_m y'+6 \gamma_m^2 y^3-y \Bigg(3 (\gamma_m-1) \gamma_m + \\
&& \zeta  z^2\Bigg)\Bigg]+18 \gamma_m (\gamma_m-2 \zeta ) x^3 y + \\
&& 3 \sqrt{\frac{3}{2}} \left(-4 \gamma_m \zeta +\gamma_m+8 \zeta ^2\right) x^2 y z +\\
&& \sqrt{6} \zeta  \left(z \left(y'+6 \gamma_m y^3-3 (\gamma_m-1) y\right)+y z'\right),
\end{eqnarray*}
\begin{eqnarray*}
P_{1}^{3}&=&\frac{1}{2} \sqrt{\frac{3}{2}} y \Bigg\{-y \Bigg[3 (\gamma_m-2) \zeta +\left(-9 \gamma_m \zeta +6 \gamma_m+18 \zeta ^2\right) \times \\
&& x^2+2 \sqrt{6} \zeta  x z \left(z \Gamma '(z)+2 \Gamma (z)-1\right)\Bigg]+ \\
&& 4 \zeta  y'+3 \gamma_m \zeta  y^3\Bigg\},
\end{eqnarray*}
\begin{eqnarray*}
P_{2}^{1}&=&\frac{1}{2} \Bigg\{y \Bigg[6 (\gamma_m-2 \zeta ) x'+36 (\gamma_m-2 \zeta )^2 x^3 - \\
&& 3 x \left(6 (\gamma_m-1) (\gamma_m-2 \zeta )+z^2 (1-2 \Gamma (z))\right)- \\
&& 9 \sqrt{6} (\gamma_m-2 \zeta ) x^2 z-\sqrt{6} z'+3 \sqrt{6} (\gamma_m-1) z\Bigg]+ \\
&& y' \left(6 (\gamma_m-2 \zeta ) x-\sqrt{6} z\right)+ \\
&& 6 \gamma_m y^3 \left(6 (\gamma_m-2 \zeta ) x-\sqrt{6} z\right)\Bigg\},
\end{eqnarray*}
\begin{eqnarray*}
P_{2}^{2}&=&\frac{1}{4} \Bigg\{2 x \Bigg[6 (\gamma_m-2 \zeta ) x'+3 \sqrt{6} z \times \\
&&  \left(\gamma_m+2 \left(\gamma_m (\zeta -2)-2 \zeta ^2\right) y^2\right)-\sqrt{6} z'\Bigg]-2 \sqrt{6} z x'+ \\
&& 9 (\gamma_m-2 \zeta )^2 x^4+6 x^2 \left(3 \gamma_m (\gamma_m-2 \zeta ) \left(5 y^2-1\right)+z^2\right)-\\
&& 6 \sqrt{6} (\gamma_m-2 \zeta ) x^3 z+9 \gamma_m \left(4 y y'+\gamma_m \left(1-3 y^2\right)^2\right)- \\
&& 12 \zeta  y^2 z^2\Bigg\},
\end{eqnarray*}
\begin{eqnarray*}
P_{2}^{3}&=&-\frac{1}{2} \sqrt{\frac{3}{2}} \Bigg\{(y \Bigg[2 x'+3 (\gamma_m-2 \zeta ) x^3-3 \gamma_m x+ \\
&& \sqrt{6} x^2 z \left(-2 z \Gamma '(z)-4 \Gamma (z)+3\right)\Bigg]+2 x y' \\
&& +y^3 \left(\left(\gamma_m (9-6 \zeta )+12 \zeta ^2\right) x+\sqrt{6} \zeta  z\right)\Bigg\},
\end{eqnarray*}
\begin{eqnarray*}
P_{3}^{1} &=&\sqrt{\frac{3}{2}} z \Bigg\{z \Bigg[-3 (\Gamma (z)-1) \Bigg(-\gamma_m+3 (\gamma_m-2 \zeta ) x^2 \\
&& +\gamma_m y^2+2\Bigg)-2 z' \Gamma '(z)\Bigg]+2 \sqrt{6} x z^3 \times \\
&& (\Gamma (z)-1) \Gamma '(z)+4 \sqrt{6} x z^2 (\Gamma (z)-1)^2- \\
&& 4 (\Gamma (z)-1) z'\Bigg\},
\end{eqnarray*}
\begin{eqnarray*}
P_{3}^{2}=-3 y z^2 (\Gamma (z)-1) \left(\sqrt{6} \gamma_m x+2 \zeta  z\right),
\end{eqnarray*}
\begin{eqnarray*}
P_{3}^{3}&=&-z \left(\sqrt{6} x' \left(z \Gamma '(z)+2 \Gamma (z)-2\right)+3 \zeta  y^2 z (\Gamma (z)-1)\right)-\\
&& \sqrt{6} x z' \left(z^2 \Gamma ''(z)+4 z \Gamma '(z)+2 \Gamma (z)-2\right) +\\
&& 6 x^2 z^2 \left(z \Gamma '(z)+2 \Gamma (z)-2\right)^2.
\end{eqnarray*}

The geodesic deviation equations are obtained in the form
\begin{eqnarray*}
&&\frac{d^2\xi^1}{d\tau ^2}+\xi^1 \left(-18 (\gamma_m-2 \zeta ) x x'-6 \gamma_m y y'\right)- \\
&& 2 \xi^2 \left(y \left(3 \gamma_m x'+\sqrt{6} \zeta  z'\right)+y' \left(3 \gamma_m x+\sqrt{6} \zeta  z\right)\right)+\\
&& \frac{\xi^1}{d\tau} \left(3 \gamma_m-9 (\gamma_m-2 \zeta ) x^2-3 \gamma_m y^2-6\right)- \\
&& 2 \frac{\xi^2}{d\tau} y \left(3 \gamma_m x+\sqrt{6} \zeta  z\right)-2 \sqrt{6} \zeta  \xi^3 y y'-\sqrt{6} \zeta  \frac{\xi^3}{d\tau} y^2=0,
\end{eqnarray*}

\begin{eqnarray*}
&&\frac{d^2\xi^2}{d\tau ^2}+\sqrt{6} \xi^3 \left(y x'+x y'\right)+\xi^1 \times \\
&&  \left(y \left(\sqrt{6} z'-6 (\gamma_m-2 \zeta ) x'\right)+y' \left(\sqrt{6} z-6 (\gamma_m-2 \zeta ) x\right)\right)+ \\
&& \xi^2 \left(x' \left(\sqrt{6} z-6 (\gamma_m-2 \zeta ) x\right)+\sqrt{6} x z'-18 \gamma_m y y'\right)+ \\
&& \sqrt{6} \frac{\xi^3}{d\tau} x y+\frac{\xi^1}{d\tau} y \left(\sqrt{6} z-6 (\gamma_m-2 \zeta ) x\right)+ \\
&& \frac{\xi^2}{d\tau} \left(3 \gamma_m-3 (\gamma_m-2 \zeta ) x^2+\sqrt{6} x z-9 \gamma_m y^2\right)=0,
\end{eqnarray*}

\begin{eqnarray*}
&&\frac{d^2\xi^3}{d\tau ^2}+2 \sqrt{6} \xi^3 \Bigg(z x' \left(z \Gamma '(z)+2 \Gamma (z)-2\right)+ \\
&& x z' \left(z^2 \Gamma ''(z)+4 z \Gamma '(z)+2 \Gamma (z)-2\right)\Bigg)+ \\
&& 2 \sqrt{6} \frac{\xi^3}{d\tau} x z \left(z \Gamma '(z)+2 \Gamma (z)-2\right)+2 \sqrt{6} \times \\
&&  \xi^1 z z' \left(z \Gamma '(z)+2 \Gamma (z)-2\right)+2 \sqrt{6} \frac{\xi^1}{d\tau} z^2 (\Gamma (z)-1)=0.
\end{eqnarray*}

\section{The non-linear connection, the curvature deviation tensor and the
geodesic deviation equations for the tachyon scalar field cosmological model}

\label{appB}

The coefficients of the non-linear connection associated to the dynamical
system describing the tachyon scalar field cosmological model are given by
\begin{eqnarray*}
N_{1}^{1}=2 \sqrt{3} x \left(\sqrt{3} x+y \right)-3 \left(1-x^2\right),
\end{eqnarray*}
\begin{eqnarray*}
N_{1}^{2}=-\sqrt{3} \left(1-x^2\right) z,\; N_{1}^{3}=-\sqrt{3}
\left(1-x^2\right) y,
\end{eqnarray*}
\begin{eqnarray*}
N_{2}^{1}=\frac{\sqrt{3}}{2} \left(3 y \left(2 x y^2-\frac{\gamma_m x y^2}{%
\left(1-x)^2\right)^{3/2}}\right)+y^2 z\right),
\end{eqnarray*}
\begin{eqnarray*}
N_{2}^{2}&=&\frac{\sqrt{3}}{2} \Bigg\{3 \left(2 x^2 y-\frac{2 \gamma_m y}{%
\sqrt{1-x^2}}\right)+  \notag \\
&&3 \Bigg[\gamma_m \left(1-\frac{y^2}{\sqrt{1-x^2}}\right) +x^2 y^2\Bigg]+2
x y z\Bigg\},
\end{eqnarray*}
\begin{eqnarray*}
N_{2}^{3}=\frac{1}{2} \sqrt{3} x y^2,\; N_{3}^{1}=\frac{3}{2} \sqrt{3} y z^2
(\Gamma (z)-1),
\end{eqnarray*}
\begin{eqnarray*}
N_{3}^{2}=\frac{3}{2} \sqrt{3} x z^2 (\Gamma (z)-1),
\end{eqnarray*}
\begin{eqnarray*}
N_{3}^{3}&=&\frac{3}{2} \sqrt{3} x y z^2 \Gamma ^{\prime }(z) +3 \sqrt{3} x
y z (\Gamma (z)-1).
\end{eqnarray*}

The first KCC invariants $\epsilon ^i$, $i=1,2,3$ are
\begin{eqnarray*}
\epsilon^{1}&=&x^{\prime }\left(9 x^2+2 \sqrt{3} x y z-3\right) +\sqrt{3}
\left(x^2-1\right) \left(z y^{\prime }+y z^{\prime }\right),
\end{eqnarray*}
\begin{eqnarray*}
\epsilon^{2}&=&\frac{1}{2 \left(x^2-1\right)^2}\Bigg\{\sqrt{3} \Bigg[-3
\gamma_m x \sqrt{1-x^2} y^3 x^{\prime }+ \\
&& \left(x^2-1\right)^2 y^2 x^{\prime }(6 x y+z)+9 \gamma_m x^2 \sqrt{1-x^2}
y^2 y^{\prime }- \\
&& 9 \gamma_m \sqrt{1-x^2} y^2 y^{\prime }+\left(x^2-1\right)^2 \times \\
&& \left(y^{\prime }(3 \gamma_m+x y (9 x y+2 z))+x y^2 z^{\prime }\right)%
\Bigg]\Bigg\},
\end{eqnarray*}
\begin{eqnarray*}
\epsilon^{3}=\frac{3}{2} \sqrt{3} z \left\{(\Gamma (z)-1) \left[y \left(z
x^{\prime }+2 x z^{\prime }\right)+x z y^{\prime }\right]+x y z z^{\prime
}\Gamma ^{\prime }(z)\right\}.
\end{eqnarray*}
All Berwald connection components are zero here. The components of the
curvature deviation tensor are
\begin{eqnarray*}
P_{1}^{1}&=&x \Bigg[-2 \left(9 x^{\prime }+\sqrt{3} z y^{\prime }\right)+%
\frac{9}{2} y^3 z \left(\frac{\gamma_m}{\sqrt{1-x^2}}-2\right) \\
&&-2 \sqrt{3} y \left(z^{\prime }+6 z\right)\Bigg]-\frac{1}{2} y z \left(4
\sqrt{3} x^{\prime }+3 y z (3 \Gamma (z)-2)\right) \\
&& +81 x^4+\frac{9}{2} x^2 \left(y^2 z^2 (\Gamma (z)+2)-12\right) \\
&& +9 x^3 y \left(y^2+4 \sqrt{3}\right) z+9,
\end{eqnarray*}
\begin{eqnarray*}
P_{1}^{2}&=&\frac{1}{2} \Bigg\{z \Bigg[-9 \gamma_m-4 \sqrt{3} x x^{\prime 2
}\left(-3 \gamma_m+9 y^2+8 \sqrt{3}\right) \\
&& +27 \gamma_m \sqrt{1-x^2} y^2+9 x^4 \left(3 y^2+2 \sqrt{3}\right)+6 \sqrt{%
3}\Bigg] \\
&& +9 x \left(x^2-1\right) y z^2 (\Gamma (z)+1)-2 \sqrt{3}
\left(x^2-1\right) z^{\prime }\Bigg\},
\end{eqnarray*}
\begin{eqnarray*}
P_{1}^{3}&=&\sqrt{3} y \left(-2 x x^{\prime 4}-12 x^2+3\right)-\sqrt{3}
\left(x^2-1\right) y^{\prime } \\
&& +\frac{3}{2} x \left(x^2-1\right) y^2 z \left(3 z \Gamma ^{\prime }(z)+6
\Gamma (z)-1\right),
\end{eqnarray*}
\begin{eqnarray*}
P_{2}^{1}&=&\frac{1}{4} y \Bigg\{6 \sqrt{3} y^2 \left(\frac{\gamma_m}{%
\left(1-x^2\right)^{5/2}}-2\right) x^{\prime 2 }y \\
&& \times \left(\frac{4 \sqrt{3} \gamma_m y x^{\prime }}{\left(1-x^2%
\right)^{5/2}}+\left(45 y^2+6 \sqrt{3}\right) z\right) \\
&& +\frac{1}{\left(1-x^2\right)^{5/2}} \Bigg[9 x^3 y \Bigg(18
\left(1-x^2\right)^{5/2} y^3+12 \sqrt{3} \\
&& \times \left(1-x^2\right)^{5/2} y-2 \sqrt{3} \gamma_m y^{\prime }+3
\gamma_m^2 y+27 \gamma_m y^3 \\
&& -8 \sqrt{3} \gamma_m y\Bigg) \Bigg]+9 x y \Bigg[2 \sqrt{3} \left(\frac{%
\gamma_m}{\left(1-x^2\right)^{5/2}}-2\right) y^{\prime } \\
&& +\frac{9 \gamma_m y^3 \left(\gamma_m \sqrt{1-x^2}-2\right)}{%
\left(1-x^2\right)^{5/2}} +y \\
&& \times \Bigg(\frac{\left(2 \sqrt{3}-3 \gamma_m\right) \left(\gamma_m-2
\left(1-x^2\right)^{5/2}\right)}{\left(1-x^2\right)^{5/2}} \\
&& +z^2 (\Gamma (z)+1)\Bigg)\Bigg]+\frac{27 \gamma_m x^5 y^2 \left(2 \sqrt{3}%
-3 y^2\right)}{\left(1-x^2\right)^{5/2}} \\
&& +\frac{27 \gamma_m x^4 y^3 z}{\left(1-x^2\right)^{5/2}}-\frac{27 \gamma_m
y^3 z}{\left(1-x^2\right)^{5/2}}-4 \sqrt{3} z y^{\prime } \\
&& +y \left(9 \gamma_m z-2 \sqrt{3} \left(z^{\prime }+3 z\right)\right)%
\Bigg\},
\end{eqnarray*}
\begin{eqnarray*}
P_{2}^{2}&=&\frac{1}{4 \left(x^2-1\right)^2}\Bigg\{-4 \left(x^2-1\right)^2 y %
\Bigg[z \left(\sqrt{3} x^{\prime }-9 \gamma_m x\right) \\
&& +\sqrt{3} x \left(9 x y^{\prime }+z^{\prime }\right)\Bigg]+18 \gamma_m x
\sqrt{3-3 x^2} y^2 x^{\prime } \\
&& +3 \left(x^2-1\right)^2 y^2 \Bigg[3 x \left(x \left(18 \gamma_m+z^2
(\Gamma (z)+1)\right)-4 \sqrt{3} x^{\prime }\right) \\
&& -2 z^2\Bigg]-36 \gamma_m x^2 \sqrt{3-3 x^2} y y^{\prime }+36 \gamma_m
\sqrt{3-3 x^2} y y^{\prime } \\
&& +\left(x^2-1\right)^2 \left(27 \gamma_m^2-4 \sqrt{3} x z y^{\prime
}\right) \\
&& -243 \left(x^2-1\right) y^4 \left(\gamma_m^2-x^6+x^4\right) \\
&& -162 \gamma_m^2 \sqrt{1-x^2} y^2+162 \gamma_m^2 x^2 \sqrt{1-x^2} y^2 \\
&& +486 \gamma_m x^4 \sqrt{1-x^2} y^4-486 \gamma_m x^2 \sqrt{1-x^2} y^4 \\
&& +90 \gamma_m x^3 \sqrt{1-x^2} y^3 z-90 \gamma_m x \sqrt{1-x^2} y^3 z \\
&& +36 x \left(x^2-1\right)^2 \left(4 x^2-1\right) y^3 z\Bigg\},
\end{eqnarray*}
\begin{eqnarray*}
P_{2}^{3}&=&\frac{1}{4} y \Bigg\{y \left(9 \gamma_m x-2 \sqrt{3} x^{\prime
}\right)-4 \sqrt{3} x y^{\prime } \\
&& -\frac{9 \gamma_m x y^3}{\sqrt{1-x^2}}+9 x \left(7 x^2-4\right) y^3 \\
&& +3 y^2 z \left(x^2 \left(3 z \Gamma ^{\prime }(z)+6 \Gamma
(z)-2\right)-2\right)\Bigg\},
\end{eqnarray*}
\begin{eqnarray*}
P_{3}^{1}&=&-\frac{1}{4 \left(1-x^2\right)^{3/2}}\Bigg\{3 z \Bigg[-2 \sqrt{%
3-3 x^2} \left(x^2-1\right) \\
&& \times z y^{\prime }(\Gamma (z)-1)+2 \sqrt{3-3 x^2} \left(x^2-1\right) y %
\Bigg((\Gamma (z)-1) \\
&& \times \left(\left(9 x^2-3\right) z-2 z^{\prime }\right)-z z^{\prime
}\Gamma ^{\prime }(z)\Bigg) \\
&& -3 x \left(1-x^2\right)^{3/2} y^2 z^2 (\Gamma (z)-1) \Bigg(3 z \Gamma
^{\prime }(z) \\
&& +6 \Gamma (z)-1\Bigg)+9 x^2 y^3 z \Bigg(\gamma_m+2 \sqrt{1-x^2} x^2 \\
&& -2 \sqrt{1-x^2}\Bigg) (\Gamma (z)-1)\Bigg]\Bigg\},
\end{eqnarray*}
\begin{eqnarray*}
P_{3}^{2}&=&\frac{1}{4 \sqrt{1-x^2}}\Bigg\{3 z \Bigg[-2 \sqrt{1-x^2} z
(\Gamma (z)-1) \\
&& \times \left(\sqrt{3} x^{\prime }+3 y z\right)+x \Bigg(z \Bigg(\sqrt{1-x^2%
} \\
&& \times \left(-9 \gamma_m-2 \sqrt{3} z^{\prime }\Gamma ^{\prime }(z)+9
\gamma_m \Gamma (z)\right) \\
&& -27 \gamma_m y^2 (\Gamma (z)-1)\Bigg)-4 \sqrt{3-3 x^2} (\Gamma (z)-1)
z^{\prime }\Bigg) \\
&& +3 \sqrt{1-x^2} x^2 y z^2 (\Gamma (z)-1) \left(3 z \Gamma ^{\prime }(z)+6
\Gamma (z)-2\right) \\
&& +27 \sqrt{1-x^2} x^3 y^2 z (\Gamma (z)-1)\Bigg]\Bigg\},
\end{eqnarray*}
\begin{eqnarray*}
P_{3}^{3}&=&\frac{3}{4} \Bigg\{-2 \sqrt{3} y \Bigg(2 z \left(x^{\prime
}(\Gamma (z)-1)+2 x z^{\prime }\Gamma ^{\prime }(z)\right) \\
&& +z^2 \left(x^{\prime }\Gamma ^{\prime }(z)+x z^{\prime }\Gamma ^{\prime
\prime }(z)\right)+2 x (\Gamma (z)-1) z^{\prime }\Bigg) \\
&& -2 \sqrt{3} x z y^{\prime }\left(z \Gamma ^{\prime }(z)+2 \Gamma
(z)-2\right)+3 y^2 z^2 \\
&& \times \Bigg[3 x^2 \Bigg(z^2 \Gamma ^{\prime 2}+4 z (\Gamma (z)-1) \Gamma
^{\prime }(z) \\
&& +4 \Gamma (z)^2-7 \Gamma (z)+3\Bigg)-2 \Gamma (z)+2\Bigg]\Bigg\}.
\end{eqnarray*}

The geodesic deviation equations are obtained in the form

\begin{eqnarray*}
&&\frac{d^2\xi ^1}{d\tau ^2}+2 \frac{d\xi ^1}{d\tau } \left(9 x^2+2 \sqrt{3}
x y z-3\right) + \\
&& 2 \sqrt{3} \left(x^2-1\right) y \frac{d\xi ^3}{d\tau}+2 \sqrt{3}
\left(x^2-1\right) z \frac{d\xi ^2}{d\tau} + \\
&& 4 \xi ^1 \left[x^{\prime }\left(9 x+\sqrt{3} y z\right)+\sqrt{3} x
\left(z y^{\prime }+y z^{\prime }\right)\right]+ \\
&& 2 \sqrt{3} \xi ^2 \left[\left(x^2-1\right) z^{\prime }+2 x z x^{\prime }%
\right]+ \\
&& 2 \sqrt{3} \xi ^3 \left[\left(x^2-1\right) y^{\prime }+2 x y x^{\prime }%
\right]=0,
\end{eqnarray*}
\begin{eqnarray*}
&&\frac{d^2\xi ^2}{d\tau ^2}+\frac{d\xi ^2}{d\tau} \left(3 \sqrt{3} \gamma_m-%
\frac{9 \sqrt{3} \gamma_m y^2}{\sqrt{1-x^2}}+9 \sqrt{3} x^2 y^2+2 \sqrt{3} x
y z\right)+ \\
&& 9 \sqrt{3} \xi ^1 x y^2 \left(2-\frac{\gamma_m}{\left(1-x^2\right)^{3/2}}%
\right) y^{\prime }+\sqrt{3} \xi_3 y^2 x^{\prime }+ \\
&& \sqrt{3} y^3 \Bigg[3 x \xi_1^{\prime }\left(2-\frac{\gamma_m}{%
\left(1-x^2\right)^{3/2}}\right) + \\
&& 3 \xi ^1 \left(-\frac{2 \gamma_m x^2}{\left(1-x^2\right)^{5/2}}-\frac{%
\gamma_m}{\left(1-x^2\right)^{5/2}}+2\right) x^{\prime }\Bigg]+ \\
&& \xi ^2 \Bigg[-\frac{18 \sqrt{3} \gamma_m y y^{\prime }}{\sqrt{1-x^2}}+18
\sqrt{3} x^2 y y^{\prime }+2 \sqrt{3} y z x^{\prime }+ \\
&& 9 \sqrt{3} x y^2 \left(2-\frac{\gamma_m}{\left(1-x^2\right)^{3/2}}\right)
x^{\prime }+2 \sqrt{3} x z y^{\prime }+2 \sqrt{3} x y z^{\prime }\Bigg]+ \\
&& \sqrt{3} x y^2 \frac{d\xi ^3}{d\tau }+2 \sqrt{3} \xi^3 x y y^{\prime }+%
\sqrt{3} \xi ^1 y^2 z^{\prime }+ \\
&&\sqrt{3} y^2 z \frac{d\xi ^1}{d\tau }+2 \sqrt{3} \xi ^1 y z y^{\prime }=0,
\end{eqnarray*}
\begin{eqnarray*}
&&\frac{d^2\xi ^3}{d\tau ^2}+\xi ^3 \Bigg[(3 \sqrt{3} y z^2 x^{\prime
}\Gamma ^{\prime }(z)+6 \sqrt{3} y z \Bigg(x^{\prime }(\Gamma (z)-1) + \\
&& 2 x z^{\prime }\Gamma ^{\prime }(z)\Bigg)+3 \sqrt{3} x z^2 y^{\prime
}\Gamma ^{\prime }(z)+6 \sqrt{3} x z y^{\prime }(\Gamma (z)-1)+ \\
&& 6 \sqrt{3} x y (\Gamma (z)-1) z^{\prime }+3 \sqrt{3} x y z^2 z^{\prime
}\Gamma ^{\prime \prime }(z)\Bigg]+ \\
&& 3 \sqrt{3} \xi^2 z^2 \left[x^{\prime }(\Gamma (z)-1)+x z^{\prime }\Gamma
^{\prime }(z)\right] + \\
&& \frac{d\xi ^3}{d\tau } \left[3 \sqrt{3} x y z^2 \Gamma ^{\prime }(z)+6
\sqrt{3} x y z (\Gamma (z)-1)\right] + \\
&& 3 \sqrt{3} x z^2 \frac{d\xi ^2}{d\tau} (\Gamma (z)-1)+6 \sqrt{3} \xi ^2 x
z (\Gamma (z)-1) z^{\prime }+ \\
&& 3 \sqrt{3} \xi ^1 z^2 \left[y^{\prime }(\Gamma (z)-1)+y z^{\prime }\Gamma
^{\prime }(z)\right] + \\
&& 3 \sqrt{3} y z^2 \frac{d\xi ^1}{d\tau} (\Gamma (z)-1)+6 \sqrt{3} \xi_1 y
z (\Gamma (z)-1) z^{\prime }=0.
\end{eqnarray*}

\end{document}